\documentclass[a4paper,11pt,longbibliography]{article}

\usepackage{jheppub} 
\usepackage{amsmath,graphicx}
\usepackage{color}
\usepackage[dvipsnames]{xcolor}
\usepackage{hyperref}

\definecolor{mygray}{gray}{0.6}

\def\F{{\mathcal F}}
\def\P{{\mathcal P}}

\def\L{{\mathcal L}}

\def\tR{\tau_R}
\def\p{{\boldsymbol p}}

\def\Eq#1{Eq.~(\ref{#1})}

\def\Fig#1{Fig.~\ref{#1}}

\def\Ref#1{Ref.~\cite{#1}}

\def\bra{\langle}
\def\ket{\rangle}

\newcommand{\be}{\begin{equation}}
\newcommand{\ee}{\end{equation}}

\newcommand{\beq}{\begin{eqnarray}}
\newcommand{\eeq}{\end{eqnarray}}
\newcommand{\nn}{\nonumber\\ }
\newcommand{\rme}{{\rm e}}
\newcommand{\rmd}{{\rm d}}
\newcommand{\del}{\partial}

\def\be{\begin{equation}}
\def\ee{\end{equation}}
\def\bea{\begin{eqnarray}}
\def\eea{\end{eqnarray}}

\begin{document}

\title{Emergence of hydrodynamical behavior in expanding quark-gluon plasmas}

\author[a]{Jean-Paul Blaizot}
\author[b]{and Li Yan}

\affiliation[a]{
	Institut de Physique Th{\'e}orique, Universit\'e Paris Saclay, 
        CEA, CNRS, 
	F-91191 Gif-sur-Yvette, France} 
\affiliation[b]{
 Institute of Modern Physics, 
Fudan University, 
Handan Road 220, 
Yangpu District, Shanghai, 200433, China
}
\date{\today}

\abstract{
We use a set of simple angular moments to solve the Boltzmann equation in the relaxation time approximation for a boost invariant longitudinally expanding gluonic plasma. The transition from the  free streaming regime at early time to the hydrodynamic regime at late time  is well captured by the first two-moments, corresponding to the monopole and quadrupole components of the momentum distribution, or equivalently to  the energy density and the difference between the longitudinal and the transverse pressures.  We relate this property to the existence of fixed points in the infinite hierarchy of equations satisfied by the moments. These fixed points are already present in the two-moment truncations and are only moderately affected by the coupling to higher moments.   Collisions contribute to a damping of all the non trivial moments. At late time, when the hydrodynamic regime is entered, only the monopole and quadrupole moments are significant and remain strongly coupled, the decay of the quadrupole moment being delayed by the expansion, causing in turn a delay in the full isotropization of the system. The two-moment truncation contains second order viscous hydrodynamics, in its various variants, and third order hydrodynamics, together with explicit values of the relevant transport coefficients, can be easily obtained from the three-moment truncation. }

\maketitle


\section{Introduction}

One of the striking features of relativistic heavy-ion experiments at RHIC and the LHC is the collective, fluid dynamical, behavior of matter produced in these collisions. Relativistic hydrodynamics has thus become an essential tool in the modeling of these collisions, and many bulk observables are well understood from  simulations based on such a framework (for recent reviews see for instance~\cite{Heinz:2013th,Gale:2013da,Yan:2017ivm}, and also \cite{Nagle:2018nvi} dealing with the special case of small colliding systems). These phenomenological studies have been accompanied by many theoretical developments, leading to a better understanding of the foundations of relativistic hydrodynamics, as well as  improved  numerical implementations of higher order viscous corrections (see e.g. the recent reviews  
\Ref{Romatschke:2017ejr,Florkowski:2017olj}). 

This success of hydrodynamics hides in fact a number of long-standing theoretical questions. 
Indeed, the reasons why hydrodynamics work so well are far from obvious. In the traditional view, hydrodynamics requires some form of local equilibrium, and usually applies where deviations from local equilibrium are small, and can be accounted for by so-called viscous corrections. The magnitude of such corrections can be measured by the size of the typical gradients in the system, or by a Knudsen number, the ratio of microscopic to macroscopic scales. The corrections are expected to be  small when the gradients, or the Knudsen number, are small. It is not clear whether such conditions are well  satisfied for all the systems studied, nor whether local equilibrium is attained on the  short time scales that are involved  in hydrodynamical simulations.

Recent developments, in particular those based on holography and strong coupling techniques \cite{Baier:2007ix,Bhattacharyya:2008jc}, suggest that viscous hydrodynamics may work even well before local equilibrium is achieved. As was first observed in  \cite{Heller:2011ju},  viscous hydrodynamics can indeed handle sizeable deviations to local equilibrium,  measured in \cite{Heller:2011ju} by the difference between the longitudinal and the transverse pressures. Similar results were obtained within kinetic theory (see e.g. \cite{Keegan:2015avk,Kurkela:2015qoa}). This apparent emergence of hydrodynamical behavior prior to reaching local thermal equilibrium is sometimes dubbed ``hydrodynamization''.

Further insight into this question came from the realization  that the late time dynamics in several settings is controlled by an attractor that drives the solution of the out-of-equilibrium equations of motion towards hydrodynamics \cite{Heller:2015dha}. This attractor has universal properties, such as the loss of memory of the initial conditions, and a relative independence of the pre-equilibrium microphysics that precedes the hydrodynamic regime. This behavior was observed both in strong coupling, based on gauge-fluid
duality~\cite{Heller:2011ju,Romatschke:2017vte}, and in weak coupling kinetic theory where attractor solutions have   been identified in  the case of  Bjorken expansion of
conformal plasmas ~\cite{Heller:2016rtz,Romatschke:2017vte,Denicol:2016bjh,Heller:2018qvh,Blaizot:2017ucy},  and  extended beyond this regime  (see e.g.\cite{Denicol:2018pak,Behtash:2017wqg,Romatschke:2017acs}). 
This has triggered a number of interesting mathematical developments on the nature of the gradient expansion, its possible resummation, as well as a detailed analysis of the asymptotic solutions of differential equations whose long time behavior admits an hydrodynamic regime (see e.g. \cite{Basar:2015ava}). 

Our goal in this paper is to shed light on some of these questions, starting from elementary physical considerations. To do so, we shall   exploit  the approach initiated in Refs.~\cite{Blaizot:2017lht,Blaizot:2017ucy}. This approach is based on kinetic theory, which serves as a model for the pre-equilibrium dynamics (limited here to free streaming with corrections due to collisions), and which allows for a smooth transition to hydrodynamics. It is presently limited to the specific context of a longitudinally expanding system with  boost invariance. It uses as basic degrees of freedom simple angular moments of the distribution function. Using moments is a standard strategy in the context of kinetic theory. They have the advantage of averaging away much of  the superfluous information contained in the distribution function, and offer a simple way to realize the transition from the kinetic to the hydrodynamic regimes. For recent applications in this context see e.g. \cite{Denicol:2012cn,Behtash:2019txb,Strickland:2018ayk}.
The moments that we are using are not general moments though, and their knowledge does not allow us to reconstruct the full momentum distribution. However they are enough to describe accurately the angular dynamics, and in particular capture the physics of isotropisation. They   constitute the basic degrees of freedom in the present discussion. 

The plan of this paper is as follows. The next section gathers well known results concerning the simple setting that we consider: a system of gluons undergoing a boost invariant longitudinal expansion, its kinetic description by a Boltzmann equation whose collision term is treated in the relaxation time approximation. We also recall there the definition of the moments that were introduced in \cite{Blaizot:2017lht} as well as the infinite hierarchy of equations that they satisfy \cite{Blaizot:2017ucy}. We end this section by showing that this hierarchy can be truncated, and that even the lowest non trivial truncation, the two-moment truncation, yields accurate results in the calculation of the energy density and the pressures, and predicts correctly the transition to hydrodynamics. Most of the rest of the paper is devoted to the explanation of these results. We start with a discussion of the free-streaming regime and show that already there the two-moment truncation captures the main qualitative features. We attribute this surprising fact to the existence of fixed points that are already present in the two-moment truncation, and whose locations are only slightly modified by the couplings to the higher moments. Then we move to the hydrodynamic regime, controlled by a fixed point of a different nature that we analyse in details. We discuss the gradient expansion, and the attractor solution which we define here as the particular solution of the kinetic equation that joins the stable free-streaming fixed point at short time to the hydrodynamic fixed point at late time. The following section is devoted to a detailed study of the two-moment truncation, where many features can be analysed in great detail, using semi-analytical techniques. This section ends with a discussion of the role of the higher moments which are left out of the two-moment truncation. We show how, in many cases, the main effect of these moments can be accounted for by a  renormalization of the equations of the two-moment truncation. Finally, in Sec.~\ref{sec:hydro} we revisit the various versions of viscous hydrodynamics from the perspective of the truncated moment equations. Ambiguities that arise in second, and higher orders, are made apparent, and the numerical values of the corresponding transport coefficients are obtained. The paper ends with a conclusion section. Several Appendices gather technical material that complements various discussions of the main text. 

\section{Pre-eqilibrium expansion with Bjorken symmetry}

In this paper, we consider an expanding system with Bjorken symmetry, i.e.,  translationally invariant in the transverse plane ($xy$-plane), and boost invariant along the collision 
axis ($z$-axis). As a result of this symmetry, physical quantities at any space time point are functions only of the proper time $\tau = \sqrt{t^2-z^2}$, and they can be deduced from the corresponding quantities in a small slice centered around $z=0$ \cite{Bjorken:1982qr}.

\subsection{A simple kinetic equation}

The kinetic description is based on a single particle  
 phase-space distribution function $f(\tau,\p)$ which depends on the momentum $\p$ of the particles, and on space-time coordinates solely via the proper time $\tau$. This distribution function obeys a kinetic equation which, in the $z=0$ slice,  reads
\be
\label{eq:trans0}
\left(\frac{\partial}{\partial\tau} - \frac{p_z}{\tau} \frac{\partial}{\partial p_z}\right)f(\tau,{\bf p})
=\mathcal{C}[f(\tau, {\bf p})],
\ee
where $\mathcal{C}[f]$ denotes the collision integral. In this work collisions are treated in the relaxation time approximation, that is we write Eq.~(\ref{eq:trans0}) as 
\be
\label{eq:trans1}
\left(\frac{\partial}{\partial\tau} - \frac{p_z}{\tau} \frac{\partial}{\partial p_z}\right)f(\tau,{\bf p})
=-\frac{f(\tau,{\bf p})-f_{\rm eq}(p/T)}{\tR},
\ee 
where $\tau_R$ denotes the relaxation time. 
This equation has been solved long ago \cite{Baym:1984np} in the case where  $\tR$ is constant. The solution has the following form      
\beq\label{solconstanttauR}
f(\tau, \p_\perp, p_z)={\rm e}^{-(\tau-\tau_0)/\tau_R}f_0(\p_\perp, p_z\tau/\tau_0)
+\int_{\tau_0}^\tau\frac{d\tau'}{\tau_R} {\rm e}^{-(\tau-\tau')/\tau_R}\, f_{\rm eq}\left(\sqrt{p_\perp^2+(p_z\tau/\tau')^2}/T(\tau')\right).\cr
\eeq
In \Eq{eq:trans1} and \Eq{solconstanttauR}, $f_{\rm eq}$ is the local equilibrium distribution, a function of the energy $E_p$ of the particles. For massless particles, the case considered in this work, $E_p=p$, with $p$ denoting the modulus of the momentum $\p$. The local equilibrium distribution function depends on a temperature $T(\tau)$ which is fixed through the requirement that, at each time $\tau$,  the energy density $\varepsilon$ be the same when calculated from the local equilibrium distribution and from the actual distribution, that is\footnote{Here, and throughout
\beq
\int_\p \equiv \int \frac{d^3 \p}{(2\pi)^3}
\eeq
}
\beq\label{Landaumatching}
\varepsilon=\int_\p p f(\p)=\int_\p p f_{\rm eq}(p).
\eeq
This condition is often referred to as the Landau matching condition. 
Once this condition is satisfied, a temperature $T$ is defined through its equilibrium relation  to the energy density, i.e.,    
$\varepsilon\propto T^4$.

The quantity $f_0(\p_\perp, p_z \tau/\tau_0)$ in the first term in the right-hand side of \Eq{solconstanttauR} is the 
free-streaming solution, that is, the solution of the kinetic equation in the absence of collisions:
\beq
\label{eq:fs}
\left(\frac{\del}{\partial\tau} - \frac{p_z}{\tau}\frac{\del}{\partial p_z}\right) f(\tau,\p) = 0 .     
\eeq   
This  solution  is indeed of the form $
f(\tau,\p)=f_0(\p_\perp, p_z \tau/\tau_0),
$
with $f_0(\p)$ the distribution at the initial time $\tau_0$. Free-streaming tends to drive the momentum distribution to a very flat distribution along the $z$
direction, an effect reflecting  the fast longitudinal expansion of the system.

So far we have considered a constant relaxation time $\tau_R$. More generally, the relaxation time may depend on momentum, a possibility that we shall not consider here 
(for a discussion of the effect of such a dependence see e.g. \cite{Dusling:2009df,Blaizot:2017lht}).      
The relaxation time may also depend on time. 
This occurs in particular when one enforces scale invariance, and measure $\tau_R$  
in units of the inverse temperature, the only available parameter with the relevant dimension. Since 
the temperature (defined through the Landau matching condition (\ref{Landaumatching})) 
depends on time, so does $\tau_R$, the product $\tau_R T$ being kept  constant: $\tau_R T=5\eta/s$, where $\eta$ is the shear viscosity and 
$s$ the entropy density (related to the temperature by the usual equilibrium relation, i.e., $s\propto T^3$).  

The solution (\ref{solconstanttauR}) of the kinetic equation easily generalizes to the case of a time-dependent relaxation time \cite{Florkowski:2013lya} 
\beq
\label{eq:exactf}
f(\tau, {\bf p}_\perp, p_z)=D(\tau,\tau_0)f_0({\bf p}_\perp, p_z\tau/\tau_0)+\int_{\tau_0}^\tau \frac{\rmd\tau'}{\tau_R(\tau')} D(\tau,\tau')
f_{\rm eq}(\sqrt{p_\perp^2+(p_z\tau/\tau')^2}/T(\tau'))\cr
\eeq
where
\beq
D(\tau_2,\tau_1)\equiv \exp\left[-\int_{\tau_1}^{\tau_2} \rmd \tau' \tR^{-1}(\tau')\right].
\eeq

At late time, we expect the collisions to isotropize the momentum distribution, and eventually drive the system to a state of local equilibrium, describable by hydrodynamical equations. The dynamical variables in hydrodynamics are the fluid velocity and the components of the energy momentum tensor,\footnote{We do not impose here particule number conservation, so that the only conservation laws on which hydrodynamics is built are that of energy and momentum.} which can be obtained from the single particle distribution function as
\beq
\label{eq:Tmunu}
T^{\mu\nu} = \int_\p \frac{p^\mu p^\nu}{p} f(\tau,\p)\,.
\eeq
Because of Bjorken symmetry, this tensor has only three independent components, the energy density $
\varepsilon(\tau)$ and the longitudinal ($\P_L$) and the transverse ($\P_T$) pressures
\beq
\label{eq:Pres}
\P_L(\tau) = \int_\p \frac{p_z^2}{p} f(\tau, \p),
\qquad
\P_T(\tau) = \frac{1}{2}\int_\p \frac{p_\perp^2}{p} f(\tau, \p).
\eeq
As already stated, we consider in this paper only massless particles, in which case the trace of the energy-momentum tensor vanishes, so that 
\beq\label{eqofstate}
\varepsilon=\P_L+2\P_T.
\eeq
Because of this relation, there subsists only two independent components of $T^{\mu\nu}$, which we may choose to be either the energy density and the longitudinal pressure, or the difference of pressures $\P_L-\P_T$. 

The local conservation of energy and momentum, $\del_\mu T^{\mu\nu}=0$,  translates then into an equation that relates these two independent components, and takes the following forms depending on the choice of the independent variables:
\beq\label{deltauepsPL}    
\frac{\rmd (\tau\varepsilon)}{\rmd\tau} +\P_L=0,\qquad \tau\frac{\rmd \varepsilon}{\rmd\tau}+\frac{4}{3}\varepsilon+\frac{2}{3}\left( \P_L-\P_T  \right)=0.
\eeq
The equation of motion above  is usually closed by relating the pressure to the energy density via an equation of state, or more generally by writing a constitutive equation for $\P_L-\P_T$. We shall return to this issue shortly. We just note here that once the energy density is known, one can calculate the pressures from the relations     
\beq\label{relationPLPTeps}
{\cal P}_L=-\frac{\rmd (\tau\epsilon)}{\rmd \tau},\qquad {\cal P}_T=\frac{1}{2}(\epsilon-{\cal P}_L)=\frac{1}{2\tau}\frac{\rmd (\tau^2\epsilon)}{\rmd \tau}.
\eeq

\subsection{The approach to the hydrodynamic regime and a set of special moments}\label{sec:approachhydro}

At late time, when the system has reached local equilibrium,  the momentum distribution is 
isotropic and the longitudinal and transverse pressures are equal, $\P_L=\P_T=\P$. 
Then, the equation of state is simply $\varepsilon=3\P$, and the equation (\ref{deltauepsPL}) 
for the energy density becomes a closed equation
\beq
\frac{\rmd\varepsilon}{\rmd \tau}=-\frac{4}{3}\frac{\varepsilon}{\tau}.
\eeq
This is the ideal hydrodynamic regime.

Corrections to ideal hydrodynamics are generally implemented as viscous corrections to the energy-momentum tensor. These are derived by writing so-called constitutive equations for the pressure difference, in the form of a gradient expansion. Thus for instance
\beq
\label{eq:pressA_hydro}
\P_L-\P_T = -\frac{2\eta}{\tau} + \frac{4}{3\tau^2} (\lambda_1-\eta \tau_\pi) + O\left(\frac{1}{\tau^3}\right).
\eeq   
In this equation, the gradients appear as powers of $1/\tau$, 
a consequence of the boost invariance, as we shall discuss in more detail later (see also Appendix~\ref{Chapman}). The dominant contribution to the  pressure anisotropy involves the shear viscosity $\eta$. The next correction in Eq.~(\ref{eq:pressA_hydro}) involves the transport coefficients $\lambda_1$ and $\eta\tau_\pi$, where we
consider here a conformal fluid and use the notation from Ref.~\cite{Baier:2007ix}.

The pressure difference $\P_L-\P_T$ can be expressed as a special moment of the distribution function. We have indeed
\beq
\label{eq:L1def}
\P_L-\P_T=\int_\p p\, P_{2}(\cos\theta)\, f(\tau, \p_\perp, p_z),
\qquad \cos\theta=p_z/p,
\eeq
where $P_{2}(x)=(3x^2-1)/2$ is a Legendre polynomial. 
More generally, we define the following set of moments, to be referred to as the 
$\L$-moments~\cite{Blaizot:2017lht},
\beq
\label{eq:Ldef}
\L_n = \int_\p p\, P_{2n}(\cos\theta) \,f(\tau, \p_\perp, p_z)\,, \qquad n=0,1,2,\ldots\,,
\eeq
where $P_{2n}$ is the Legendre polynomial of order $2n$. 
Note that odd order moments vanish as a consequence
of the invariance of the distribution function under parity (or under reflection with respect to the $z=0$ plane, i.e. $p_z\to -p_z$ and $\theta\to \pi-\theta$). Clearly, the energy-momentum tensor is entirely expressible in terms of the first two moments, 
\beq
\varepsilon=\L_0,\qquad  \P_L-\P_T=\L_1.
\eeq

The $\L$-moments allow to treat the approach to isotropy keeping along only the required minimal information on the distribution function. Note that these $\L$-moments do not allow us to reconstruct the full momentum distribution. 
This is because a single  
powers of $p$ is involved in their definition (and no higher powers as is usually the case -- see Ref.~\cite{Denicol:2012cn,Behtash:2019txb} for  recent studies involving more complete sets of moments). In other words, theses moments carry only information on the  rms radius of the radial momentum distribution. With this particular definition, all the moments have the same dimension, that of the energy-momentum tensor. They provide an intermediate description between the full kinetic theory dealing with the complete distribution function, and the hydrodynamics where only the first couple of moments are directly involved. As we shall see, these moments provide a simple picture of the isotropization of the momentum distribution, and the approach to hydrodynamics. 

The time dependence  of the $\L$-moments can be 
deduced from the formal solution in \Eq{eq:exactf}
\beq
\label{eq:rta_0a}
\L_n(\tau)=D(\tau,\tau_0) \L_n^{(0)}(\tau) + \int_{\tau_0}^\tau 
\frac{d\tau'}{\tR(\tau')} D(\tau,\tau') \L_0(\tau')(\tau'/\tau) \F_n(\tau'/\tau),
\eeq
where the function ${\cal F}_n$ is defined by
\be\label{calFdef}
\mathcal{F}_{n}(x)\equiv\frac{1}{2}\int_{-1}^1 dy \left[1-(1-x^2)y^2\right]^{1/2}
P_{2n}\left(\frac{xy}{\left[1-(1-x^2)y^2\right]^{1/2}}\right).
\ee
A detailed study of the function $\F_n(x)$ is presented in Appendix~\ref{App:functionFn}. 
The first term in \Eq{eq:rta_0a} contains the free-streaming moment $\L_n^{(0)}(\tau)$, which is given by
\be\label{FSLnISO}
\L_n^{(0)}(\tau)=\varepsilon_0 \,\frac{\tau_0}{\tau}\, \F_n\left( \frac{\tau_0}{\tau}\right),
\ee 
with $\varepsilon_0$ the initial energy density. Thus, Eq.~(\ref{eq:rta_0a}) allows the calculation of all the moments, once $\L_0$, that is, the energy density, is known (this may be seen as a generalization of Eqs.~(\ref{relationPLPTeps}) that allow the calculation of $\P_L$ and $\P_T$ from $\varepsilon(\tau)$). Thus, except for the case of the energy density, for which  Eq.~(\ref{eq:rta_0a}) is truly an equation to be solved to determine $\L_0$, for all $n>0$  Eq.~(\ref{eq:rta_0a}) is simply an integral representation of the various moments. This representation is exact if $\L_0$ is exactly calculated. 

\subsection{Initial conditions and relevant parameters}\label{initialconditions}           

In order to solve the kinetic equation, we need to specify the initial condition. We shall, in mainly cases, consider isotropic initial conditions, for which all moments vanish, except $\L_0$. But we shall also consider  flat initial distributions for which all moments take finite values. Such flat distributions naturally emerge as one lets the system free stream before switching on the effects of collisions, as we shall see shortly. They also naturally appear in microscopic determination of the energy momentum tensor in the early stage of a heavy ion collision (see e.g. \cite{Gelis:2013rba}). It is convenient to characterize these various initial conditions by a single parameter $\xi$ that expresses the ``deformation'' of the initial distribution, with $\xi=1$ corresponding to isotropy, and $\xi\to\infty$ to a flat distribution, and write the initial distribution as $f_0(p_T,p_z)=f_0\left(\sqrt{p_T^2+\xi^2 p_z^2}\right)$ \cite{Romatschke:2003ms}. Since the longitudinal pressure equals the transverse pressure for an isotropic distribution and vanishes for a flat distribution, one may as well characterize the deformation of the initial distribution by the ratio
\beq\label{PLoverPT}     
\frac{\P_L}{\P_T}=\frac{\L_0+2\L_1}{\L_0-\L_1}
=\frac{1+2\Lambda_0}{1-\Lambda_0},
\eeq
or equivalently by the ratio $\Lambda_0\equiv\L_1/\L_0$. These ratios are decreasing  functions of $\xi$, with $\L_1=0$ when $\xi=1$ and  $\L_1\to -\L_0/2$ as $\xi\to\infty$.

As we just mentioned, an anisotropic initial condition of the form $f_0\left(\sqrt{p_T^2+\xi^2 p_z^2}\right)$ can be reached from an isotropic initial condition that one lets evolve by free streaming from an earlier time. Indeed, given the function $f_0(p_T,p_z)$ at time $\tau_0$, the free streaming solution at time $\tau$ reads $f_0(p_T,p_z\tau/\tau_0)$.  This suggests setting
$\xi=\tilde\tau_0/\tau_0$ 
and interpreting the function $f_0\left(\sqrt{p_T^2+\xi^2 p_z^2}\right)$ as the solution of the free streaming equation obtained from an isotropic distribution at time $\tilde\tau_0$.  
 It is then straightforward to obtain the free streaming solution corresponding to this initial condition. In particular
 \beq\label{barepsilon0}
 \varepsilon_0^{(0)}(\tau)=\tilde\varepsilon_0\frac{\tau_0}{\tau}\frac{1}{\xi} \F_0\left(\frac{\tau_0}{\tau}\frac{1}{\xi} \right),\qquad \varepsilon_0=\varepsilon_0^{(0)}(\tau_0)=\tilde\varepsilon_0\frac{1}{\xi} \F_0\left(\frac{1}{\xi}\right),
 \eeq
 where $\tilde\varepsilon_0$ is the energy density at time $\tilde\tau_0$ and $\varepsilon_0$ that at the true initial time $\tau_0$.    
  Similarly, the general (free streaming) moment of the anisotropic distribution takes the form 
  \beq\label{FSLnanisotrop}
 \L_n^{(0)}(\tau)=\tilde\varepsilon_0\frac{\tau_0}{\tau}\frac{1}{\xi} \F_n\left(\frac{\tau_0}{\tau}\frac{1}{\xi} \right)=\varepsilon_0\frac{\tau_0}{\tau}\frac{\F_n\left(\frac{\tau_0}{\tau}\frac{1}{\xi} \right)}{\F_0\left(\frac{1}{\xi}\right)}.
 \eeq
 Note that because the integral in Eq.~(\ref{eq:rta_0a}) vanishes when $\tau=\tau_0$, all the information about $\xi$ is carried by the initial values of the free streaming moments. 
 
The initial energy density $\varepsilon_0$  plays no essential role in the discussion. When $\tau_R$ is constant,  the equation of motion is linear, and all the $\L$-moments are proportional to $\varepsilon_0$ and to a dimensionless function of $\tau/\tau_0$. This structure is explicit in the general expression (\ref{FSLnanisotrop}) of the free streaming moments. When collisions are present, the moments acquire a parametric dependence on $r_0\equiv \tau_0/\tau_R$, that is, on the ratio between the collision rate and the expansion rate at the initial time. 
 In the case of conformal fluids $\tau_R$ depends on time, with $\tau_R T$ constant. Although it can  be determined from the initial energy density and the ratio $\eta/s$, the value of this constant is actually irrelevant if the moments are written as functions of $\tau/\tau_R$ (and $r_0=\tau_0/\tau_R(\tau_0)$).       
 
\subsection{The equations for the $\L$-moments and their truncations}

By using well known relations among the Legendre polynomials, one can recast Eq.~(\ref{eq:trans0}) into the following hierarchy of coupled equations \cite{Blaizot:2017ucy}
\begin{align}
\label{eq:eomL}
\frac{\partial \L_n}{\partial \tau} =& -\frac{1}{\tau}\left[a_n\L_n
+b_n\L_{n-1}+c_n\L_{n+1}\right]-\frac{(1-\delta_{n0})\L_n}{\tR}\,,
\qquad
n=0,1,2,\ldots.
\end{align}
where the coefficients $a_n,b_n,c_n$ are pure numbers given by 
\beq
a_n &=&
\frac{2(14 n^2+7n-2)}{(4n-1)(4n+3)}\simeq \frac{7}{4}+\frac{5}{64 n^2}-\frac{5}{128 n^3}+O\left(\frac{1}{n}\right)^4\\
b_n&=&\frac{(2n-1)2n(2n+2)}{(4n-1)(4n+1)}\simeq \frac{n}{2}+\frac{1}{4}-\frac{7}{32 n}+\frac{1}{64 n^2}-\frac{7}{512
   n^3}+O\left(\frac{1}{n}\right)^4\\
c_n&=&\frac{(1-2n)(2n+1)(2n+2)}{(4n+1)(4n+3)}\simeq -\frac{n}{2}+\frac{7}{32 n}-\frac{3}{32 n^2}+\frac{27}{512
   n^3}+O\left(\frac{1}{n}\right)^4\,.
\eeq
As we shall see later, the transport coefficients are simple functions of these coefficients. It is to be observed that they are entirely determined by the part of the equation that describes free streaming, i.e., they are independent of the collision kernel. 

Note the relation 
\beq\label{identity2}
a_n+b_n+c_n=2
\eeq
valid for any $n$, as a simple calculation reveals. Note also the asymptotic value $a_n\simeq 7/4$ at large $n$, and the values of the first few coefficients  
\beq
a_0=4/3,\quad a_1=38/21,\quad b_1=8/15,\quad b_2= 8/7, \quad c_0=2/3,\quad c_1=-12/35. 
\eeq
These will be useful later in our discussion. 

The collision kernel in Eq.~(\ref{eq:eomL}) leads to
a damping of all the $\L$-moments, except ${\cal L}_0$ which is not directly affected by the collisions. The latter property is of course a consequence of  energy conservation  and the Landau matching 
condition. 

The advantage of transforming the simple kinetic equation (\ref{eq:trans0}) into an infinite hierarchy of equations for the $\L$-moments is that it suggests new approximations, in particular the truncation of the hierarchy in which a limited set of moments is kept, all the others being set equal to zero. We are indeed not really interested in all the moments, but mainly in the lowest ones, essentially $\L_0$ and $\L_1$ directly related to the hydrodynamical quantities.  A natural question is of course that of the convergence of the procedure. This will be much discussed in the rest of this paper. At this point, we shall just make a few general comments, and show some numerical results indicating that indeed selecting a few moments does provide a good description of the dynamics, at least if one is only interested in the time dependence of the energy-momentum tensor, i.e., in the first few moments. 

An important truncation, to be referred to as the \textit{two-moment truncation}, consists in keeping the first two moments $\L_0$ and $\L_1$ and setting $\L_{n\ge 2}=0$. 
This truncation results in two coupled equations
for $\L_0$ and $\L_1$,
\begin{subequations}
\label{eq:l0l1}
\begin{align}
\partial_\tau \L_0 + \frac{1}{\tau}(a_0 \L_0 + c_0 \L_1) =&\; 0\,,\\
\partial_\tau \L_1 + \frac{1}{\tau}(b_1 \L_0 + a_1 \L_1) =&\; -\frac{\L_1}{\tR}\,.
\end{align}
\end{subequations}
Note that the first equation (\ref{eq:l0l1}) is identical to \Eq{deltauepsPL} 
 since $a_0=4/3$ and $c_0=2/3$. As we shall see these two coupled equations provide an accurate description of the dynamics of the energy-momentum tensor, 
with the higher moments contributing to quantitative renormalizations, but no major qualitative modifications. 

\begin{figure}
\begin{center}
\includegraphics[width=0.75\textwidth] {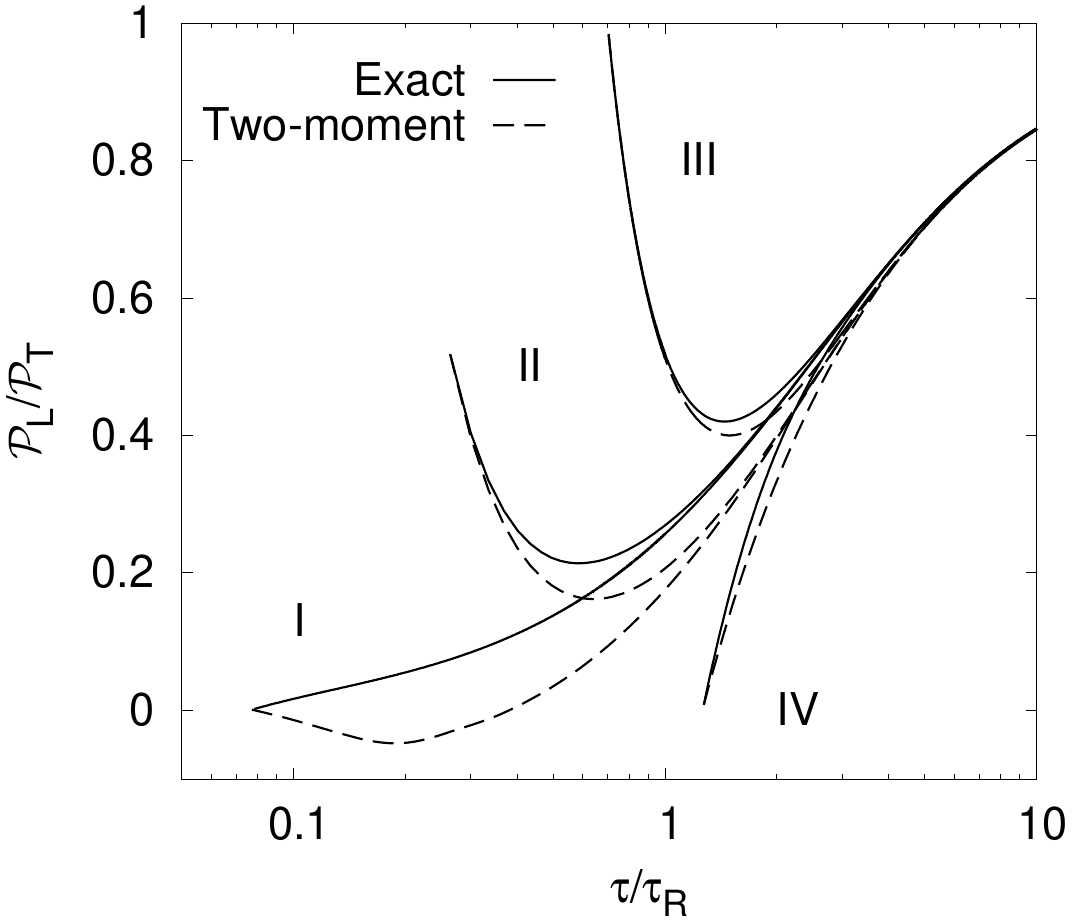}
\caption{Comparison between the exact solution of the kinetic equation (solid lines) and the solution of the two-moment truncation (dashed lines).  
Four sets of initial condition are chosen which lead to different time evolutions of the pressure anisotropy. 
 \label{fig:exactI}
}
\end{center}
\end{figure}

To demonstrate the validity of the aforementioned truncation scheme, we compare in Fig.~\ref{fig:exactI} the pressure ratio $\P_L/\P_T$ obtained form the exact solution of the kinetic equation to that obtained from the two-moment truncation. Four different initial conditions are considered, specified by different choices of $\P_L/\P_T$ and $r_0=\tau_0/\tau_R$. The choices 
(I) and (IV) correspond to a very small value of $\P_L/\P_T$, and respectively $r_0=0.07$ (I) or $r_0=1.3$ (IV). Case (III) corresponds to isotropic initial conditions and $r_0=0.7$, while in case (II) $\P_L/\P_T=0.5$ and $r_0=0.26$.  
These initial conditions cover most typical situations. The first observation is that the two-moment truncation is in good agreement with the exact solution, for nearly all initial conditions. The largest deviations occur in case (I) where the collision rate is small compared to the expansion rate and the initial longitudinal pressure is small. In this case, free-streaming dominates at early time, and drives the longitudinal pressure to negative values. This is an artefact of the two-moment truncation that we shall discuss further later. Note however that as soon as the collision rate ceases to be negligible (as in case IV for instance) this unphysical feature disappears. 

Fig.~\ref{fig:exactI} contains another important message. When $\tau$ exceeds a few times $\tau_R$, the solutions corresponding to different initial conditions merge into a single curve. That is, at that time, the memory of the details of the initial state is lost, and some universal behavior emerges. As we shall discuss at length later, this reflects the emergence of the hydrodynamic behavior. This regime sets in while the pressure anisotropy is still significant, i.e. for $\P_L/\P_T\gtrsim 0.6$. Note that the two-moment truncation describes accurately this regime, as well as the pre-equlibirum regime which is very sensitive to the initial conditions.
The value $\P_L/\P_T\simeq 0.6$ is often considered as an indication of a large anisotropy. Note however that, according to Eq.~(\ref{PLoverPT}), this value translates into a smaller ratio of the first two moments, $\L_1/\L_0\simeq 0.15$. Since the approach to local equilibrium, or at least the isotropisation of the system, is characterized by the decay of the non trivial moments of the distribution function, it may not be too surprising that viscous hydrodynamics start to work when the largest non trivial moment represents a $15\%$ correction.

\section{The free streaming regime}\label{sec:freestreaming}

In this section we study the free streaming regime from the point of view of the moments of the kinetic equation. A priori this may look as an unnecessary complication, since the explicit solution of the free streaming kinetic equation is indeed trivial. However, in doing so, we prepare the ground for the more complete discussion of the kinetic equation in the presence of collisions. Besides, this study of the free streaming moments is interesting in its own sake, as it illustrates some important features that are not immediately visible in the exact solution.

\subsection{The exact solution}

  The exact expression of the free streaming moments have already been given in the previous section,  for isotropic (Eq.~(\ref{FSLnISO}))  and anisotropic (Eq.~(\ref{FSLnanisotrop})) initial conditions. 
 These involve the function  $\F_n(x)$ defined in
\Eq{calFdef}, with here $x=\tau_0/\tau$. This function has the following limits:     
\beq
\mathcal{F}_n(x\to 0)\to \frac{\pi}{4} P_{2n}(0),\quad  \mathcal{F}_{n\ne 0}(x\to 1)\to 0,\quad \mathcal{F}_{0}(x\to 1)\to 1.
\eeq
It follows in particular that, at late time ($x\to 0$), 
\beq
{\cal L}_n^{(0)}(\tau)\sim  \frac{ \tau_0}{\tau}  \frac{\pi}{4} P_{2n}(0), \qquad (\tau\gg \tau_0)
\eeq   
i.e., all moments decay as $1/\tau$ and are proportional to each other. We set  ${\cal L}_n(\tau)=A_n{\cal L}_0(\tau)$, where the dimensionless constants $A_n$ characterize the moments of a distribution that is flat in the $p_z$ direction \cite{Blaizot:2017lht}
\beq\label{Ans}
A_n=P_{2n}(0)=(-1)^n\,\frac{(2n-1)!!}{(2n)!!},\qquad A_1=-1/2, \qquad A_2=3/8.
\eeq
Note that $\L_1(\tau)/\L_0(\tau)=A_1=-1/2$  corresponds to a vanishing longitudinal pressure. 
As for the factor $1/\tau$,
 it reflects the conservation of the energy in the increasing comoving volume ($\tau \varepsilon(\tau)={\rm cste}$) in the absence of longitudinal pressure (see Eq.~(\ref{deltauepsPL})): when the longitudinal pressure vanishes, we have indeed $\varepsilon=2\P_T$, so that $(\P_L-\P_T)/\varepsilon=-1/2$.     
 \begin{figure}
\begin{center}
\includegraphics[width=0.80\textwidth] {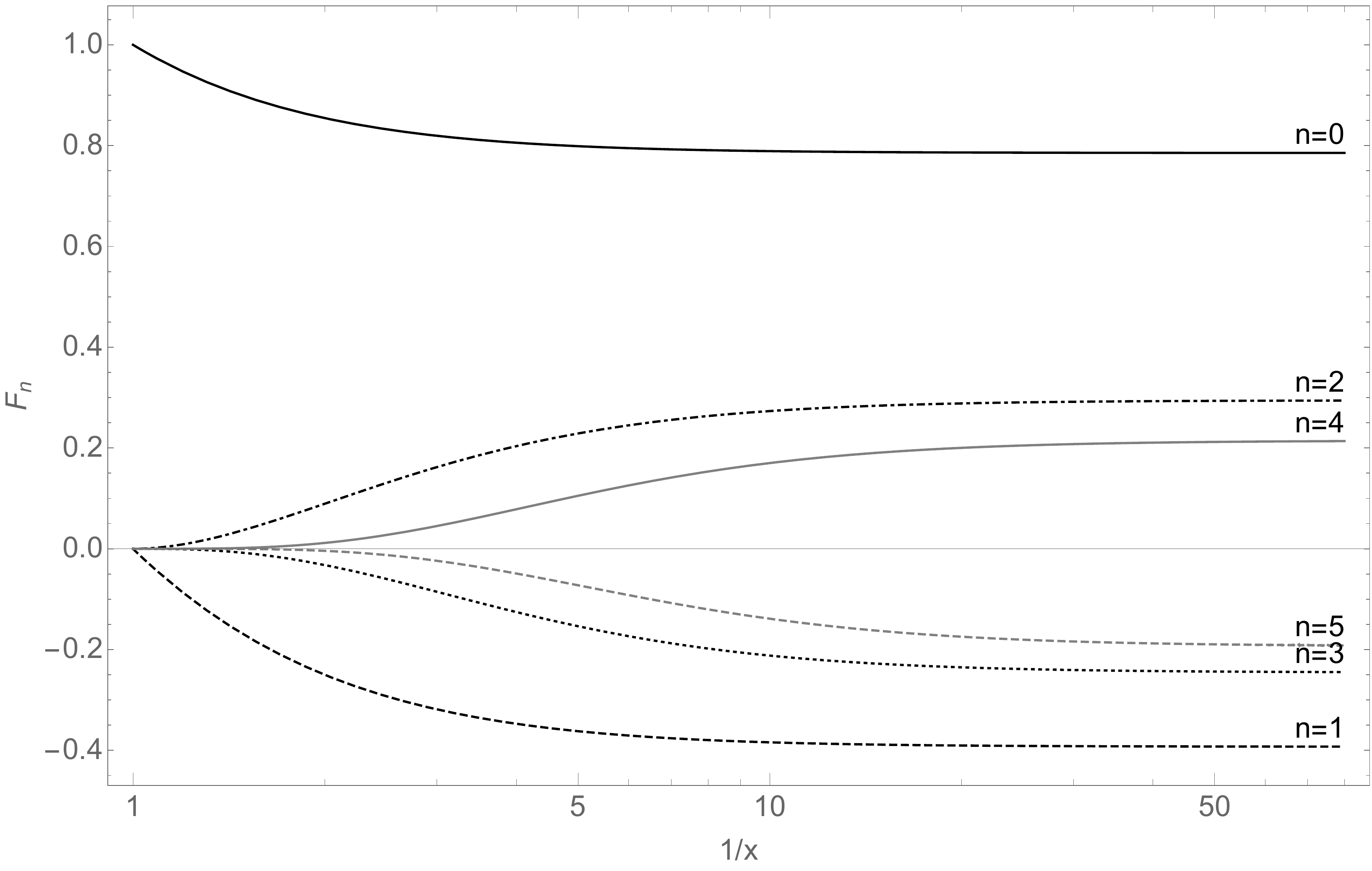}
\caption{
Time evolution of the first few moments shown through the function ${\cal F}_n(x)$, as a function of $\tau/\tau_0 (=1/x)$ on a logarithmic scale.
\label{fig:moments_evol_relax}
}
\end{center}
\end{figure}

The coefficient $A_n$ is a very slowly decreasing function of $n$: it takes many moments to describe the flat distribution. This may cast doubt on any attempt to solve the kinetic equation in terms of a finite set of moments, as we shall do in the next subsection. Note however, that starting from an isotropic distribution, for which all moments vanish, the  higher moments develop very slowly in time, since  all derivatives of ${\cal F}_n$ vanish up to order $n-1$:  ${\cal F}_n(\tau)\sim (\tau-\tau_0)^n$ (see Eq.~(\ref{shorttimemoments}), and Fig.~\ref{fig:moments_evol_relax}).   Thus it takes time for higher moments to develop,  and at late time they are damped by the expansion (recall that $\L_n(x)\sim x{\cal F}_n(x)$). As a result, at least for isotropic initial conditions, moments of rank $n\ge 2$  do not affect significantly the evolution of the lowest two moments  $\L_0$ and $\L_1$. In Sec.~\ref{FSfpt} we shall present a deeper argument for why truncations work.

\begin{figure}[h]
\begin{center}
\includegraphics[width=0.45\textwidth] {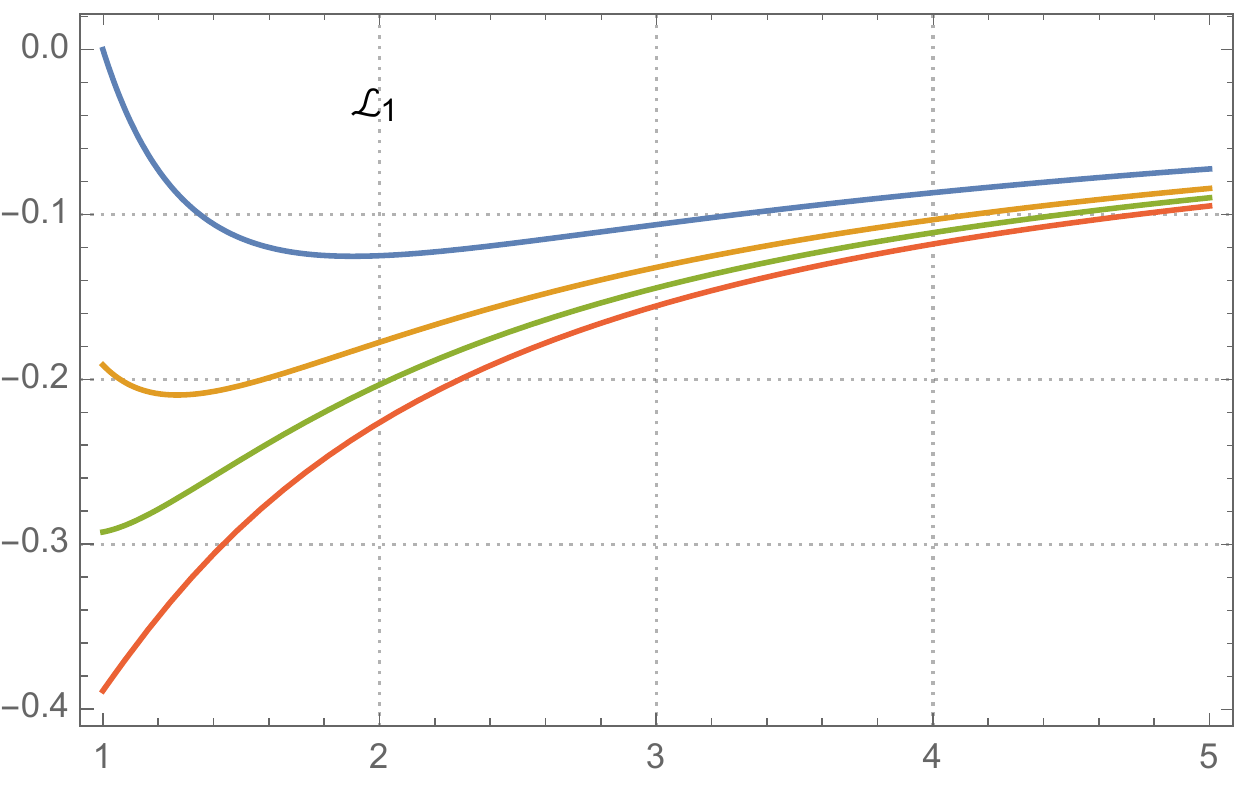}\quad\includegraphics[width=0.45\textwidth] {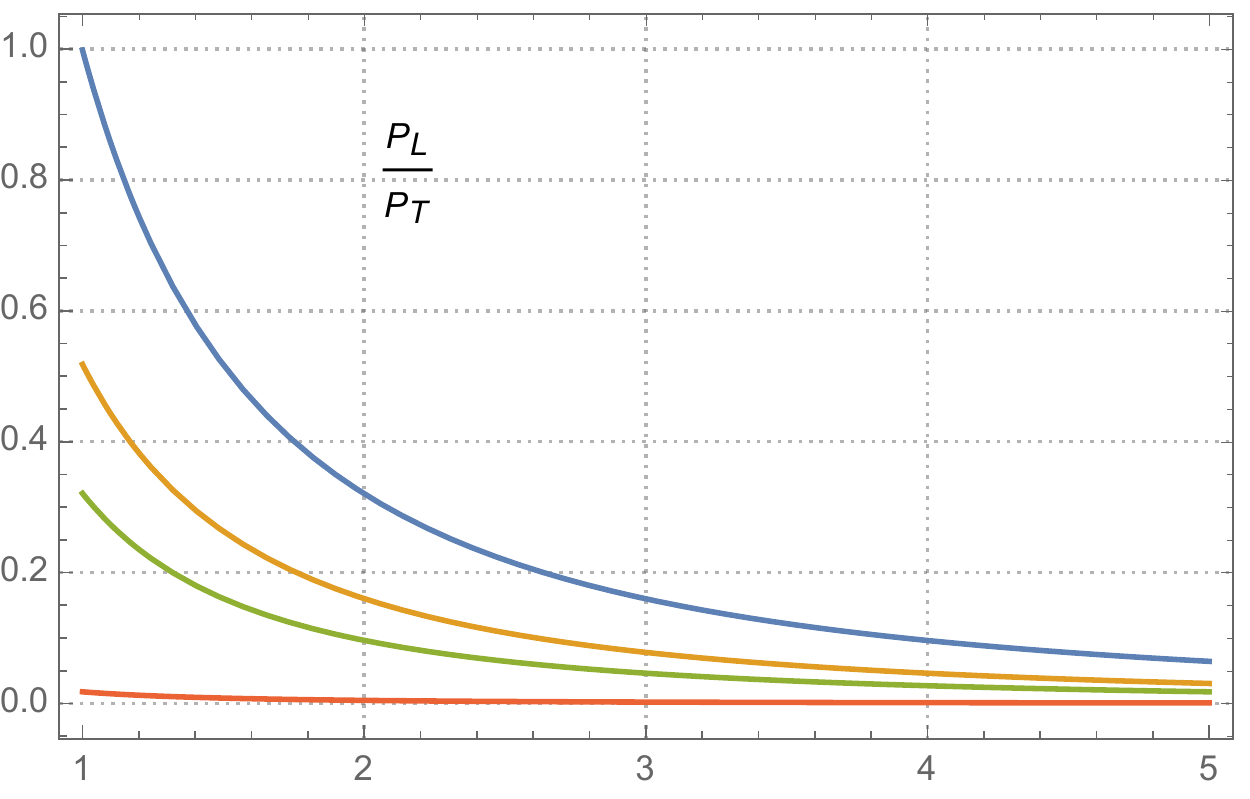}
\caption{(Color online.) The moment $\L_1$ (left) and the pressure ratio $\P_L/\P_T$ (right) as a function of $\tau/\tau_0$ for various values of $\xi$, covering a range of distributions from isotropic to flat: for $\L_1$, $\xi=1,1.5,2,3$ from top to bottom; for $\P_L/\P_T$, $\xi=1,1.5,2,10$ from top to bottom. 
\label{fig:FSl1a0105a}
}
\end{center}
\end{figure}

The figure \ref{fig:FSl1a0105a} illustrates the behavior of quantities that we shall be discussing many times in this paper, namely the first   moment $\L_1$, and the ratio $\P_L/\P_T$, for various anisotropic initial conditions characterized by the parameter $\xi$ introduced in Sect.~\ref{initialconditions} (the moment $\L_0$ is a smoothly decreasing function of $\tau$, and is shown for instance in Fig.~\ref{fig:ell0ell1}). Noteworthy is the change of slope at the origin of $\L_1$ as the initial anisotropy increases. This is easily understood by recalling how these various curves can be deduced from that corresponding to the isotropic initial condition (namely a shift of time and a rescaling, according to Eq.~(\ref{FSLnanisotrop})).  Note also the smooth decrease of $\P_L/\P_T$, related to the  regular drop of the longitudinal pressure as the initial distribution goes from an isotropic distribution to a flat distribution. These behaviors are like those in cases II and III in Fig.~\ref{fig:exactI}, for which indeed collisions  play a minor role at short time. When the initial distribution is a flat distribution, the longitudinal pressure vanishes initially and remains so at all times.     

\subsection{Truncating the moment equations}

We now turn to the hierarchy of equations (\ref{eq:eomL})  for the $\L$-moments, and ignore the effect of collisions (e.g. $\tau_R\to \infty$). We set $t\equiv \log \tau/\tau_0$ 
and consider $\vec {\cal L}=\L_1,\cdots, \L_n,\cdots$ as a vector (in an infinite dimensional space), and write Eq.~(\ref{eq:eomL}) as a matrix equation
\beq\label{linearsystem}
\del_t \vec {\cal L} =-M\vec {\cal L},
\eeq
where $M$ is a tridiagonal matrix, with constant elements. The truncations amount to restrict the size of this infinite dimensional linear problem to a finite dimensional one, which can then be solved by elementary linear algebra   techniques.

The simplest truncation corresponds to all moments vanishing except $\L_0$. It yields
\beq
\frac{\del \L_0}{\del \tau}=-\frac{a_0}{\tau}\L_0, \qquad \L_0(\tau)=\left( \frac{\tau_0}{\tau} \right)^{a_0}= x^{4/3},
\eeq
where we have set $x\equiv \tau_0/\tau$ and used $a_0=4/3$. We recognize here the ideal hydrodynamic behavior. We note that at small $\tau$, that is near $x=1$, the behaviors of the exact and approximate solutions are remarkably similar. In fact from the expansion of $\F_0(x)$ near $x=1$ given in Eq.~(\ref{shorttimemoments}), and recalling that $\L_0(x)=x\F_0(x)$, we get 
\beq
\L_0(x)\simeq 1+\frac{4}{3} (x-1)+O(x-1)^2.
\eeq
Physically, this corresponds to the fact that, at small time, the evolution of the system (as given by the exact solution) is dominated by the lowest moment (assuming isotropic initial condition): as already emphasized, it takes time for the higher order moments to build up and modify the evolution. Thus, at small time, and when the initial conditions are isotropic, the energy density $\L_0$  behaves as in ideal hydrodynamics. It is only through its interaction with $\L_1$ (and, through $\L_1$, with higher moments) that it will eventually reach the asymptotic  
behavior $\L_0(\tau)\sim 1/\tau$, corresponding to energy conservation in the absence of longitudinal pressure.

\begin{figure}
\begin{center}
\includegraphics[width=0.46\textwidth] {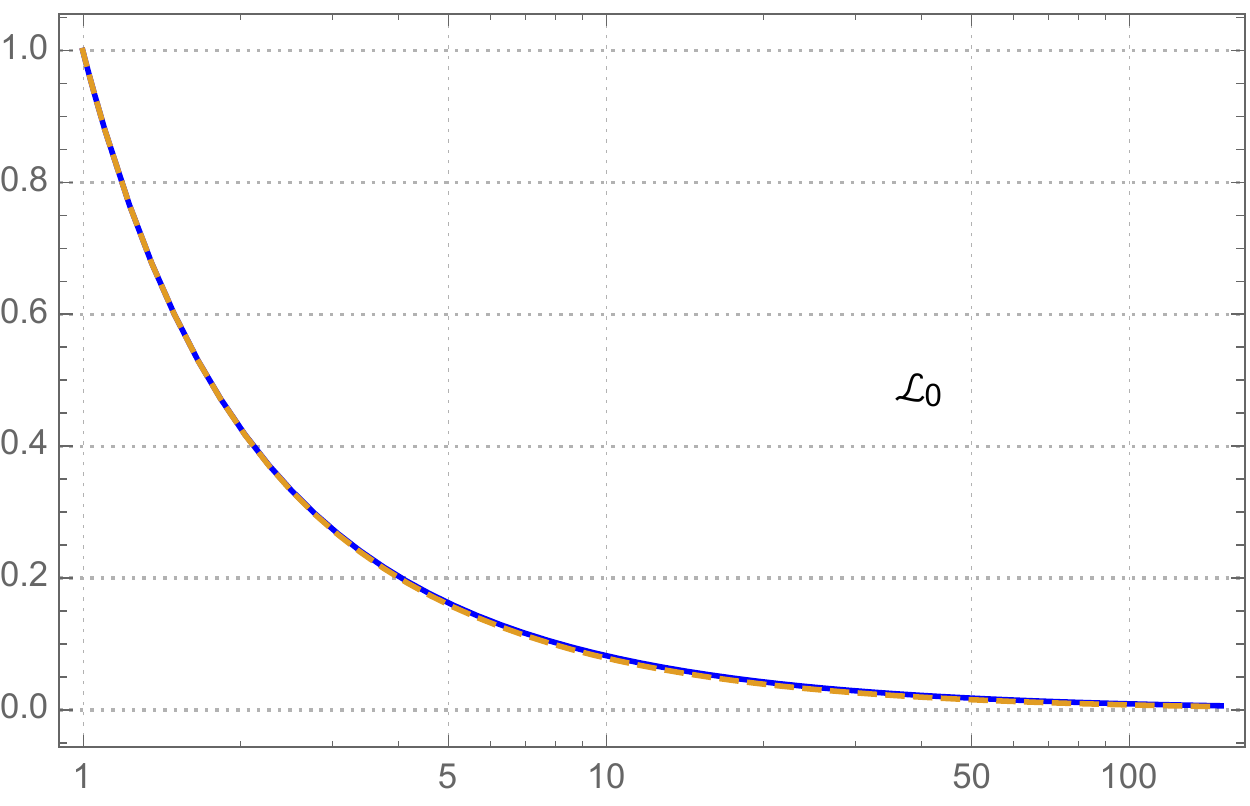}
\includegraphics[width=0.475\textwidth] {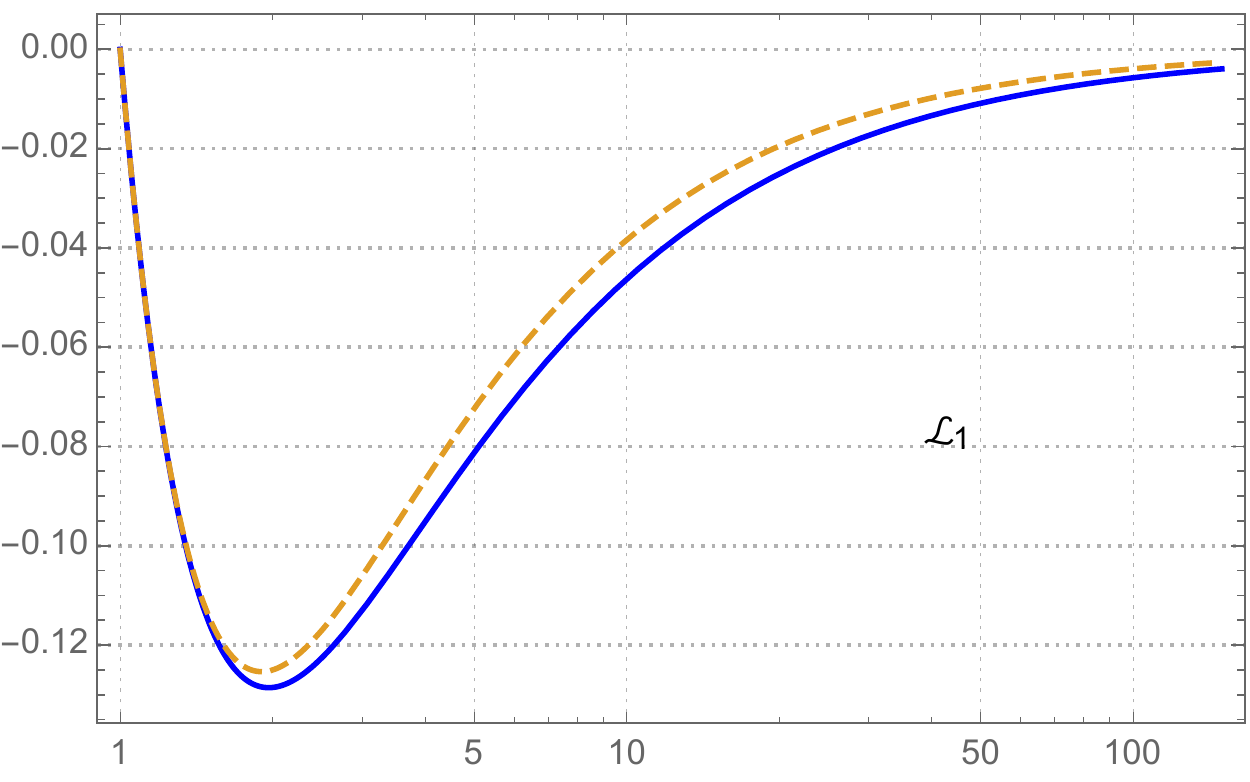}
\caption{(Color online.) The function $\L_0(\tau_0/\tau)$ (left) and $\L_1(\tau_0/\tau)$ (right), as a function of $\tau/\tau_0$ on a logarithmic scale, for isotropic initial conditions. The full lines represent the solution of the linear system of equations (\ref{FS}), the dashed line is the exact free streaming solution. In the case of $\L_0$ the two solutions are indistinguishable on the plot.   
\label{fig:ell0ell1}
}
\end{center}
\end{figure}

The next truncation is the two-moment truncation. It involves the two moments $\L_0$ and $\L_1$, with the  corresponding equations given by 
\beq\label{FS}
\frac{\partial}{\partial t}
\left(
\begin{array}{c}
 {\cal L}_0 \\
  {\cal L}_1
\end{array}
\right)
=
-\left(
\begin{array}{cc}
  a_0& c_0   \\
b_1  &  a_1  \end{array}
\right)
\left(
\begin{array}{c}
 {\cal L}_0 \\
{\cal L}_1
\end{array}
\right).
\eeq
The eigenvalues of the matrix $M$ are
 $\lambda_0\simeq 0.93$ and $\lambda_1 \simeq 2.21 $. Note that they are both positive so that the two modes are damped, and for isotropic initial conditions are given explicitly by:\beq\label{eigenmodes2}
\L_0(t)
&=&0.69\, \rme^{-0.93 \,t}+0.31  \,\rme^{-2.21 \, t} \nn
\L_1(t)
&=&-0.41\,  \rme^{-0.93\, t}+0.41\,\rme^{-2.21\,t}.
\eeq
At small $t$, $\L_0(t)\simeq 1-(4/3)t$, so that the ideal hydro behavior at short time is maintained. 
The expression of $\L_1(t)$ in Eq.~(\ref{eigenmodes2}) provides an analytical understanding of the behavior illustrated in Fig.~\ref{fig:ell0ell1}. At short time, the second exponential drops rapidly, making $\L_1$ negative as it tends to its fixed point value (see later). Then the first exponential takes over and causes a decay of the magnitude of $\L_1$. 

As can be seen in Fig.~\ref{fig:ell0ell1}, 
the agreement between the approximate solution and the exact one is excellent, in particular for the moment $\L_0$. There are deviations from the exact solution though. For instance, the dominant term at late time is not $\sim \tau^{-1}$ but $\sim \tau^{-0.9}$. As discussed in the appendix, the lowest eigenvalue of the linear system (\ref{linearsystem}) converges toward -1 rather slowly, with the first few values being -1.33, -0.93, -1,06, -0.97, etc (see Fig.~\ref{fig:attractorg0} in Appendix~\ref{truncationFS}). Similarly the coefficient of the dominant power in $\L_0$ is not $\pi/4\approx 0.785$, but 1 and 0.68 in the first two truncations, and 0.96, 0.71 in truncations with 3 and 4 moments, respectively. Also, at late time, the ratio $\L_1/\L_0$ takes the successive values 0,-0.61,-0.4,-0.55. These numbers converge slowly to the exact ratio $-1/2$. However, these quantitative aspects do not alter significantly the general picture.  

However, we should emphasize here an unphysical feature of the two-moment truncation. From the relation (\ref{PLoverPT}) and the positiveness of $\P_L$ and $\P_T$ in
kinetic theory, one easily derives the following bounds on $\L_1/\L_0$:    
\beq
\label{eq:bound_A}  
-0.5 \leq \frac{\L_1}{\L_0}\leq 1.
\eeq
The lower bound corresponds to $\P_L=0$, while the upper bound corresponds to $\P_T=0$. 
These two bounds are violated in the two-moment truncation. In particular, at late time, as we have seen, $\L_1/\L_0\simeq -0.6$, which corresponds to a negative longitudinal pressure (see case I in Fig.~\ref{fig:exactI}). This unphysical feature is an artefact of the two-moment truncation, and would eventually become negligible in truncations of sufficiently high order since, as we have mentioned above, the ratio $\L_1/\L_0$ converges (slowly) toward its exact value $A_1=-1/2$. We shall see later that once collisions are taken into account this unphysical feature becomes insignificant.

Further tests of convergence are presented in Appendix~\ref{truncationFS}, for isotropic initial conditions. A general pattern emerges from the higher order  truncations that is also discussed in this Appendix: i) The small time behavior is preserved at any order. For an initial isotropic distribution, the energy density at small time behaves as in ideal hydrodynamics. ii) The dominant eigenvalue of $-M$ converges (slowly) toward -1. iii) The ratios of moments at late time become constant. This is related to the exponential decays (in the variable $t$) of the various components of the moments. The ratios converge slowly toward the values $A_n$ corresponding to a flat distribution. iv) We find three types of eigenvalues. Two real ones, near -1 and -2, which can be associated with two fixed points to be discussed next, and a set of complex eigenvalues whose real part is close to 7/4 and whose imaginary part increases (roughly linearly) with $n$. These imaginary parts yield oscillating contributions to the moments. These, however, are strongly damped, and not visible in any of the plots displayed in this paper (see Appendix~\ref{truncationFS} for more details). 
  
\begin{figure}[h]    
\begin{center}
\includegraphics[width=0.45\textwidth] {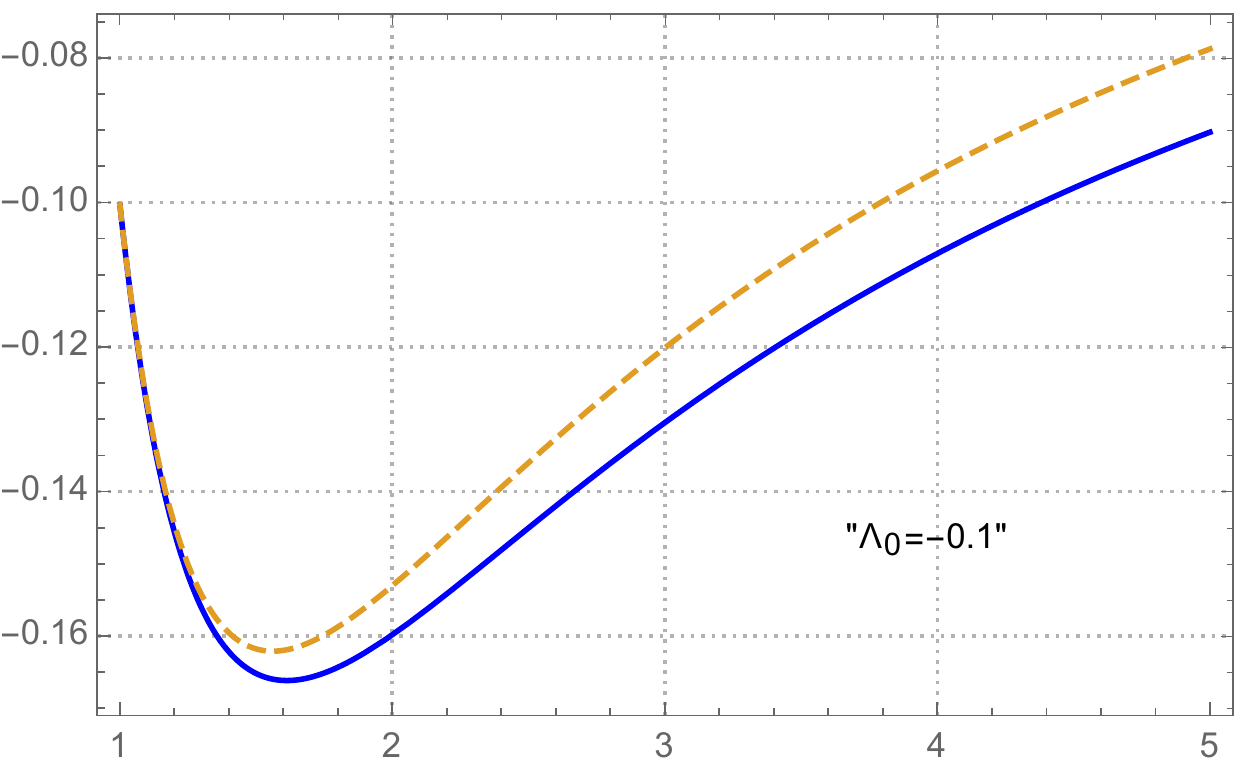}\includegraphics[width=0.45\textwidth] {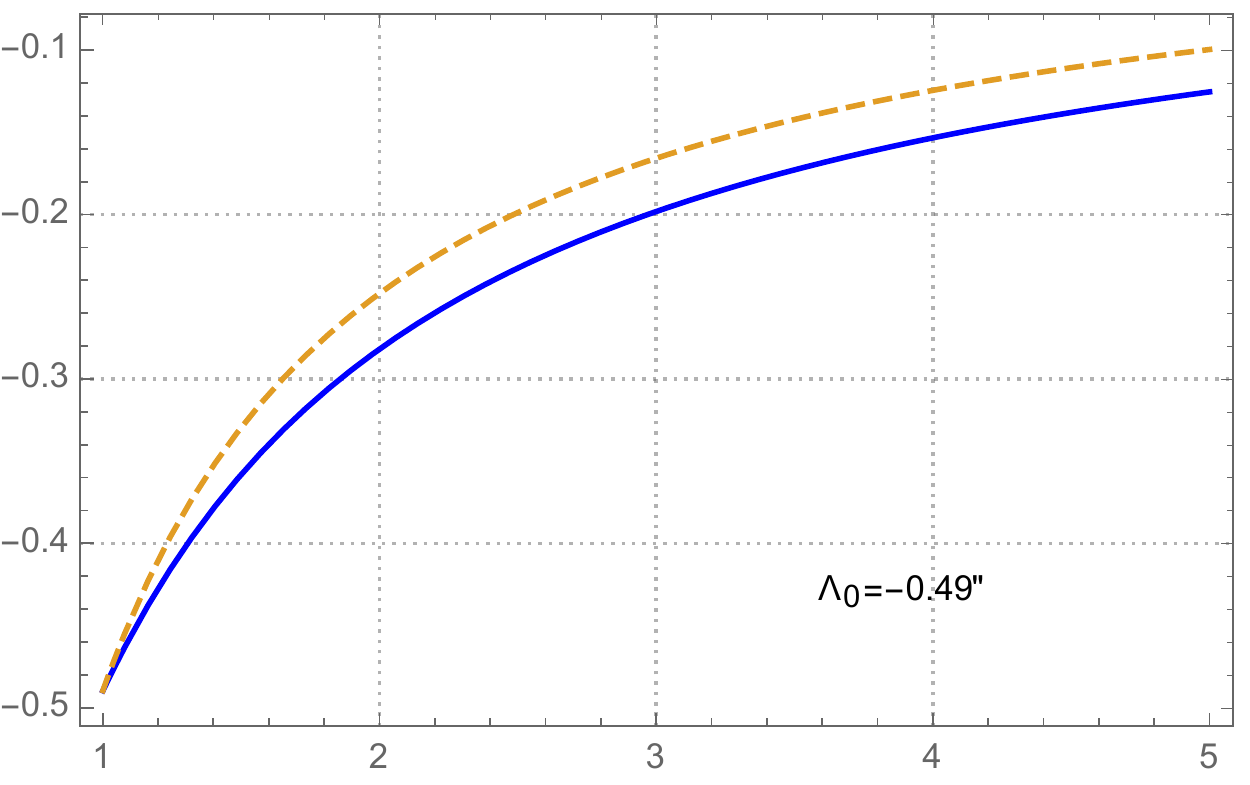}
\caption{Comparison of the function $\L_1(\tau)$ obtained from the two-moment truncation (blue line) and the rescaled exact solution (dashed line), as a function of $\tau/\tau_0$ for the values $\Lambda_0$:  -0.1  (left),  -0.49 (right) with $\Lambda_0=\L_1/\L_0$ at $\tau=\tau_0$.  
\label{fig:FSl1a0105}
}
\end{center}
\end{figure}

Finally, we consider briefly the case of anisotropic initial conditions. We expect of course similar convergence properties, in view of the close connection between the corresponding solution and that for isotropic initial conditions. This is indeed the case, as can be seen for instance in  Fig.~\ref{fig:FSl1a0105} where one compares the moment $\L_1$ calculated from the two-moment truncation and that obtained from the exact solution.   
       
\subsection{Fixed point analysis}\label{FSfpt}

We shall now elucidate the reasons why the two-moment truncation suffices to capture the essential features of the time evolution of the first two moments in the free streaming regime. We shall argue  that this results from the existence of fixed points of the general solution of the coupled systems of equations. These fixed points  are already present in the two-moment truncation, and are only moderately affected by the higher moments. 

The fixed points have already been identified, at least one of them. Indeed we have already noticed that at late time, all the moments become proportional to the energy density, $\L_n=A_n \L_0$, with $A_n$ given by Eq.~(\ref{Ans}). That this is a solution follows  form the identity
\beq
(a_n-1) A_n+b_n A_{n-1}+c_n A_{n+1}=0,
\eeq
which is easily checked (note that $A_0=1$, $b_0=0$). Another identity has also been mentioned, namely $a_n+b_n+c_n=2$, Eq.~(\ref{identity2}).  It follows from this identity that another fixed point exists, where all moments are equal, and decay as $\tau^{-2}$. Clearly the two fixed point that we have just identified correspond to the two eigenvalues 1 and 2 of the matrix $M$ in Eq.~(\ref{linearsystem}).     

\begin{figure}
\begin{center}
\includegraphics[width=.75\textwidth] {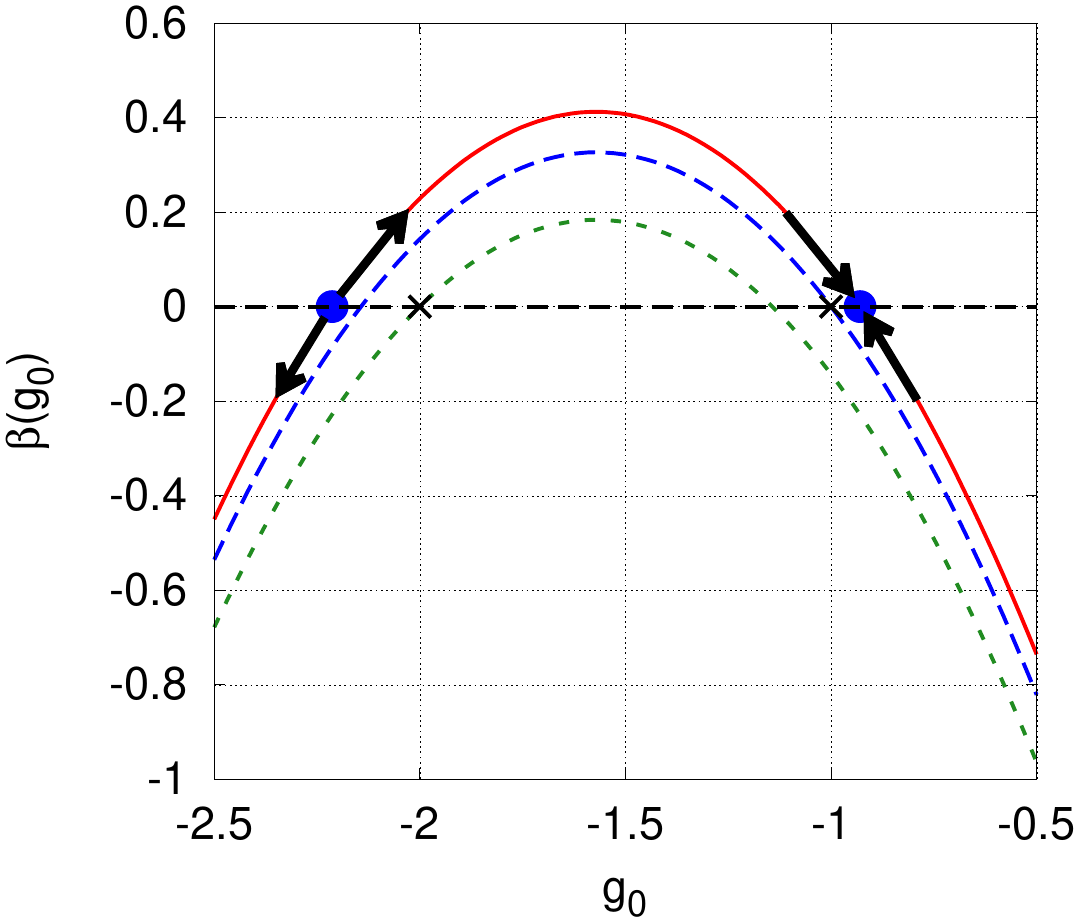}
\caption{ The function $\beta(g_0)$ in Eq.~(\ref{eqg0FSfp}). The full line corresponds to Eq.~(\ref{eqg0FSfp}), and the dots locate the two approximate fixed points, with the arrows indicating their stability or instability. The two dashed lines correspond to the shifts proportional to $\L_2$ in Eq.~(\ref{shiftfixpt}),  which induce minor displacements of the approximate fixed points and brings them to their exact (respective) locations.    \label{fig:fixed-points_FS}
}
\end{center}
\end{figure}

Since the moments continue to evolve at large time, it is convenient to consider their logarithmic derivatives
\be\label{logderivative}
g_n(\tau)\equiv
\tau\partial_\tau\ln \L_n, 
\ee 
which indeed go to constant values at late time.     
The quantity $g_0$ obeys an equation of motion that is easily obtained  in the two-moment truncation, by eliminating the moment $\L_1$ (assuming that $\L_2$ vanishes). One gets   
\beq\label{eqg0FSfp}
\tau\frac{\rmd g_0}{\rmd \tau}=\beta(g_0),\qquad \beta(g_0)=-g_0^2-(a_0+a_1) g_0-a_0a_1+c_0b_1.
\eeq 
 A plot of the function $\beta(g_0)$ is given in Fig.~\ref{fig:fixed-points_FS}. The fixed points correspond to the zeros of $\beta(g_0)$.  It is easy to verify that these coincide with the two eigenvalues of the matrix $-M$, the one close to -1 ($\lambda_0=0.93$), the other close to -2 ($\lambda_1=2.21 $).

Consider then small deviations away from the fixed points, and set $g_0(t)=\bar g_0+f(t)$, 
with $\tau=\tau_0\rme^t$ 
and $\bar g_0$  the value of $g_0$ at a fixed point. In linear order in  $f$ we get
\beq\label{stabilityg0}
\frac{\rmd f}{\rmd t}+2\bar g_0 f+(a_0+a_1) f=0, \qquad f(t)=f(0)\rme^{-(2\bar g_0+a_0+a_1)t}.
\eeq
Now, recall that $a_0=4/3$ and $a_1=38/21$, so that $a_0+a_1\simeq 3$. Thus when $\bar g_0\approx -1$, $2\bar g_0+a_0+a_1>0$, corresponding to a stable fixed point. When $\bar g_0\approx -2$, $2\bar g_0+a_0+a_1<0$, corresponding to an unstable fixed point.

One can  verify that these fixed points are also those of $g_1$. 
This is natural since in the vicinity of the stable fixed point, $\L_1$ and $\L_0$ are proportional, $\L_1=A_1\L_0$, while $\L_1=\L_0$ at the unstable fixed point. Thus in both cases, $g_0=g_1$ at the fixed point.  In fact this reasoning extends trivially to all the $g_n$'s: at the stable fixed point for instance, $g_n=-1$ for all $n$.

This basic structure is not changed when one takes into account higher moments in the truncation. The equation for $g_0$ that one obtains by keeping the moment $\L_2$ reads
\beq\label{shiftfixpt}
\tau\frac{\rmd g_0}{\rmd \tau}=- \left[g_0^2+(a_0+a_1) g_0+a_0a_1-c_0b_1\right] +c_0c_1 \frac{\L_2}{\L_0}.
\eeq
This equation is now exact. Of course, at this point $\L_2/\L_0$ is unknown, and can only be determined by solving the hierarchy of equations. But we can argue  that the effect of $\L_2$ is modest. This can be easily demonstrated since we know the values of $\L_2/\L_0$ in the vicinity of the two fixed points. Indeed, the effect of $\L_2/\L_0$ is simply to shift down the function $\beta(g_0)$  in Fig.~\ref{fig:fixed-points_FS} by the amount $c_0c_1 \frac{\L_2}{\L_0}$ ($c_1=-12/35$, $c_0=2/3$). This results in a decrease of the value of $\bar g_0$ at the stable fixed point, and an increase of $\bar g_0$ at the unstable fixed point. In fact, $\L_2/\L_0$ is a constant in the vicinity of each fixed point, equal to 1 near the unstable fixed points and to $A_2=3/8$ in the vicinity of the stable fixed point. If one inserts in Eq.~(\ref{shiftfixpt}) the exact value of $\frac{\L_2}{\L_0}$, $A_2=3/8$, the fixed point is moved from $-0.929366$ to exactly -1.  Similarly, injecting the value corresponding to the unstable fixed point, namely $\L_2/\L_0=1$, one moves $\bar g_0$ from  $-2.21349$ to -2.    

The existence of these two fixed points whose location is only moderately affected by the higher moments is the main reason why the two-moment truncation suffices to capture the main features of the free streaming.

                     
\section{ The gradient expansion and the hydrodynamic regime}\label{hydrogradients}

As we have seen in the previous section, starting from an isotropic momentum distribution, free streaming drives the system to a very anisotropic state, with an infinite number of moments being populated as time goes on, and the longitudinal pressure decreasing. We have seen also that, in spite of the fact that many moments are populated, even the simple two-moment truncation gives a fair account of the time dependence of the two lowest moments $\L_0$ and $\L_1$. We have argued that this can be understood from the existence of two fixed points which control the time evolution of these two lowest moments, and whose locations are only moderately affected by the higher moments.

We expect the truncations to become even more accurate once collisions are included.  The main effect of the collisions is indeed to wash out the anisotropy of the momentum distribution, a process which, in the absence of expansion, is exponentially fast, 
$\sim \exp{(-\tau/\tau_R)}$. In fact, a profound change takes place at late time, with the solution of the kinetic equation acquiring a simple representation as an expansion in powers of $1/\tau$. This corresponds to the fact that the late time behaviour is controlled by a different fixed point of the kinetic equation,  the hydrodynamic fixed point. It is the purpose of this section to analyse the main characteristics of this new fixed point.

\subsection{Gradient expansion and constitutive equations}

As recalled in Sec.~\ref{sec:approachhydro}, in the hydrodynamic regime, the energy momentum tensor admits an expansion in gradients, which, in the present setting with Bjorken symmetry, appears as an expansion in powers of $1/\tau$. The first couple of terms of this expansion are  displayed in  Eq.~(\ref{eq:pressA_hydro}).   
 We shall assume here that not only $\L_1$, but all moments $\L_n$  admit a gradiennt expansion and  we write
\beq
\label{gradientL10}
\L_n(\tau) =\sum_{m=n}^\infty \frac{\alpha_n^{(m)}}{\tau^m}.
\eeq
Such a structure follows for instance from the Chapman-Enskog expansion presented in Appendix~\ref{Chapman}. In the case of $\L_1$, which is  directly related to the energy-momentum tensor, the coefficients $\alpha_1^{(n)}$ are related to usual transport coefficients (such as the shear viscosity $\eta$ in Eq.~(\ref{eq:pressA_hydro})). The coefficients of higher moments are not directly related to usual transport coefficients, even though their contributions may affect dynamically the coefficients of the various gradients in the energy momentum tensor. 

The Chapman-Enskog expansion is an expansion for the deviation $\delta f=f-f_{\rm eq}$, in powers of the relaxation time $\tau_R$, as well as in Legendre polynomials (see Appendix~\ref{Chapman}). Each power of $\tau_R$ is accompanied by a gradient, i.e, in the present context, by a power of $1/\tau$, so that the expansion is an expansion in powers of $1/w\equiv\tau_R/\tau$ (see Sec.~\ref{sec:gradexpansion} for more on this variable $w$). The expansion is such that in $\L_n$ the leading term is of order $1/w^n$, as indicated in Eq.~(\ref{gradientL10}). 

In the expanding case, the coefficients $\alpha_n^{(m)}$ acquire a time dependence. In fact, for dimensional reason, $\L_n$ is  proportional to the energy density $\varepsilon$, a function of $\tau$. Also $\tau_R$ may depend on $\tau$. We may then rewrite Eq.~(\ref{gradientL10}) as follows
\beq
\label{gradientL10b}
\L_n(\tau) =\sum_{m=n}^\infty \frac{B_n^{(m)}\varepsilon \tau_R^m}{\tau^m}=\varepsilon \sum_{m=n}^\infty \frac{B_n^{(m)} }{w^m},
\eeq      
where $B_n^{(m)}$ is a dimensionless
constant number, and $\alpha_n^{(m)}=B_n^{(m)}\varepsilon \tau_R^m$. This observation is enough to determine the asymptotic behaviour of the moments in the hydrodynamic regime.  Note that the coefficients $B_n^{(m)}$ are entirely determined by the coefficients $a_n, b_n, c_n$, the effects of collisions being factored out in the explicit dependence on $\tau_R$.  
     
\subsection{The hydrodynamic fixed point}

From the remark above we have, for the leading term of each moment
\beq
\alpha_n^{(n)}=B_n^{(n)}\varepsilon \tau_R^n.
\eeq
We shall assume that the energy density behaves, in leading order, as in ideal hydrodynamics, i.e., $\varepsilon\sim \tau^{-4/3}\sim T^4$.
For constant $\tau_R$, the time dependence of $\alpha_n^{(n)}$ is just that of the energy density. In the conformal case, where $\tau_R T={\rm cste}$, we have instead
\beq
\alpha_n^{(n)}= B_n^{(n)}\frac{\varepsilon}{T^n} \left(\tau_R T\right)^n\sim \tau^{-4/3+n/3}.
\eeq   
 It follows that     
\beq\label{gninfinity}
\L_n(\tau)\sim \tau^{-(4/3+2n/3)} \quad(\tau_R T={\rm cste}),\qquad \L_n(\tau)\sim \tau^{-(4/3+n)} \quad(\tau_R ={\rm cste}).
\eeq
We can rewrite these relations as
\beq
\frac{\L_n(\tau)}{\L_0(\tau)}\sim \tau^{-2n/3}\quad(\tau_R T={\rm cste}),\qquad \frac{\L_n(\tau)}{\L_0(\tau)}\sim \tau^{-n}\quad(\tau_R ={\rm cste}),
\eeq
or in terms of the logarithmic derivatives (\ref{logderivative})
\beq
\label{eq:gninf}
g_n(\infty)=-\frac{4}{3}-\frac{2n}{3} \quad(\tau_R T={\rm cste}),\qquad g_n(\infty)=-\frac{4}{3}-n \quad(\tau_R ={\rm cste}).    
\eeq
These power laws characterize the hydrodynamic fixed point that we shall discuss further later. Note that these fixed point values do not depend on the truncation, in contrast to what happens in the free streaming case where the value of the stable fixed point depends (weakly) on the order of the truncation. The fixed point values in Eq.~(\ref{eq:gninf}) depend only on the time dependence of $\tau_R$, and that corresponding to $g_0$ (-4/3) is universal.         

\subsection{Asymptotic behavior from the kinetic equation}
\label{sec:asymptkinetic}

It is instructive to see how the fixed point behavior emerges from the solution of the kinetic equation. In doing so, we shall also be able to determine the coefficient of the leading power law, $B_n^{(n)}$, as well as that of the subleading contribution,  $B_n^{(n+1)}$. 

Let us then return to the equations of motion for ${\cal L}_n$, that is, Eq.~(\ref{eq:eomL}). 
There is an intriguing feature of this equation whose solution could contain a priori exponentially decaying contributions because of the last term, which  seems to be incompatible with the gradient expansion. Let us however rewrite Eq.~(\ref{eq:eomL}) as follows 
\beq\label{eqforgn}
g_n(\tau)=\tau\del_\tau \ln{\cal L}_n=-a_n-b_n\frac{\L_{n-1}}{\L_n}-c_n\frac{\L_{n+1}}{\L_n}-\frac{\tau}{\tau_R}.
\eeq
At large time, we can ignore the constant term $a_n$, as well as the ratio ${\L_{n+1}}/{\L_n}$, which is of order $1/\tau$. Then, in order to avoid the appearance of exponential terms, it is sufficient that the remaining two terms cancel, that is
\beq\label{recursionBnn}
-b_n \frac{\L_{n-1}}{\L_n}-\frac{\tau}{\tau_R}=0,\qquad \L_n=-b_n\frac{\tau_R}{\tau}\L_{n-1}.
\eeq
This indeed fixes the leading order in the gradient expansion in agreement with what was obtained before. Eq.~(\ref{recursionBnn}) provides a simple recursion relation from which one can deduce 
\beq
\alpha_n^{(n)}=\varepsilon \tau_R^n\prod_{i=1}^n\left(-b_i  \right)=\varepsilon \tau_R^n \,B_n^{(n)}.
\eeq
In particular, $\alpha_0^{(0)}=\varepsilon=\L_0$, and $\alpha_1^{(1)}=-b_1\varepsilon\tau_R$, from which the expression of the shear viscosity follows, $\eta=(b_1/2)\varepsilon\tau_R$ (see Eq.~(\ref{eq:pressA_hydro})).     
Note the factorial      
growth of the coefficient ($b_n\sim n$ at large $n$), at the origin of the divergence of the gradient expansion \cite{Heller:2013fn,Heller:2015dha,Basar:2015ava,Aniceto:2015mto}.

We can push the analysis to the next-to-leading order. The cancellation of the large ($\propto \tau$) terms leading to Eq.~(\ref{recursionBnn}) left aside a possible constant contribution that we can determine. We then return to Eq.~(\ref{eqforgn}) and keep the leading order terms at large $\tau$, i.e., 
\beq
g_n(\tau\to\infty)=-a_n-b_n\frac{\L_{n-1}}{\L_n}-\frac{\tau}{\tau_R}+{\cal O}\left(\frac{1}{\tau}\right).
\eeq    
By using the expansion of the moments to the next to leading order, \beq
\L_n=\frac{\alpha_n^{(n)}}{\tau^n}+\frac{\alpha_n^{(n+1)}}{\tau^{n+1}}.
\eeq
we can then obtain for the coefficients $\alpha_n^{(n+1)}$ the following recursion relation
\beq
\frac{\alpha_n^{(n+1)}}{\alpha_n^{(n)}}=-\tau_R\left[ g_n(\infty)+a_n   +b_n \frac{\alpha_{n-1}^{(n)}}{\alpha_n^{(n)}} \right],
\eeq
with $g_n(\infty)$ given by Eq.~(\ref{eq:gninf}) (and for $n=1$ the last term vanishes, i.e. $\alpha_0^{(n)}=0$). One then gets  
\beq\label{B1B2a}
B_n^{(n)}=\prod_{i=1}^n\left(-b_i  \right),\qquad \frac{B_n^{(n+1)}}{B_n^{n}}=-\sum_{i=1}^n\left[g_i(\infty)+a_i\right].
\eeq
  One can deduce in particular from the relations above the value of the second order transport coefficient in Eq.~(\ref{eq:pressA_hydro}). We have indeed,   
  \beq
  \lambda_1-\eta\tau_\pi=\frac{3}{4}\alpha_1^{(2)}=-\frac{3}{4}\tau_R^2\, \varepsilon\, b_1(2-a_1),
  \eeq
   where the last expression holds for the conformal case.          

The same results can be obtained by solving directly the coupled equations for the moments, searching  a solution in the form of a gradient expansion. As an illustration, consider the two-moment truncation, i.e., Eqs.~(\ref{eq:l0l1}), assuming here that $\tau_R$ is constant for simplicity. Using the Ansatz 
\beq\label{ansatzL1}
\L_1=\L_0\left[B_1^{(1)}\,\frac{\tau_R}{\tau}+B_1^{(2)}\left(\frac{\tau_R}{\tau}   \right)^2+\cdots\right]
\eeq  
for $\L_1$, we obtain, after a simple calculation, the following solution for the energy density $\L_0$     
\beq\label{L0B}
\L_0(\tau)&\approx& \tau^{-a_0}\exp\left\{c_0B_1^{(1)}\frac{\tau_R}{\tau}+\frac{c_0}{2}B_1^{(2)}\left(\frac{\tau_R}{\tau}   \right)^2\right\}\nn
&\approx& \tau^{-a_0}\left( 1+  c_0B_1^{(1)}\,\frac{\tau_R}{\tau} +\left[\frac{c_0^2\left(B_1^{(1)}\right)^2}{2}+\frac{c_0}{2}B_1^{(2)}\right]\left(\frac{\tau_R}{\tau}   \right)^2   \right),
\eeq
together with the explicit values of the coefficients  
\beq\label{B1B2}
B_1^{(1)}=-b_1,\qquad B_1^{(2)}=-b_1[1+a_0-a_1].
\eeq
These are identical to those obtained earlier, Eq.~(\ref{B1B2a}), for the case of a constant $\tau_R$. Note that these first two coefficients $B_1^{(1)}$ and $B_1^{(2)}$ are given exactly by the two-moment truncation (this would not be the case for the coefficient $B_2^{(2)}$ which involves $b_2$, hence $\L_2$).\footnote{The values of the coefficients $B_1^{(1)}$ and $B_1^{(2)}$ obtained in this section  agree with those given in  \cite{Heller:2018qvh} for both constant and conformal $\tau_R$.}  

This example  illustrates a subtle aspect of the gradient expansion. We have argued earlier that the coefficients $\alpha_0^{(n)}$ vanish, that is, there is no genuine gradient expansion for the energy density, in the sense of a constitutive equation analogous to Eq.~(\ref{ansatzL1}) for $\L_1$. However, such a gradient expansion is generated dynamically, through the coupling of $\L_0$ to higher moments, as demonstrated in Eq.~(\ref{L0B}). It can be seen in particular that the gradient terms in Eq.~(\ref{L0B}) are all multiplied by $c_0$, the coefficient that couples $\L_0$ to $\L_1$. Such a gradient expansion of $\L_0$ needs to be properly identified when extracting the values of the coefficients $B_n^{(m)}$ from the solution of the moment equations. 
         
\subsection{The attractor solution}
\label{sec:attractor}

\begin{figure}
\begin{center}
\includegraphics[width=.450\textwidth] {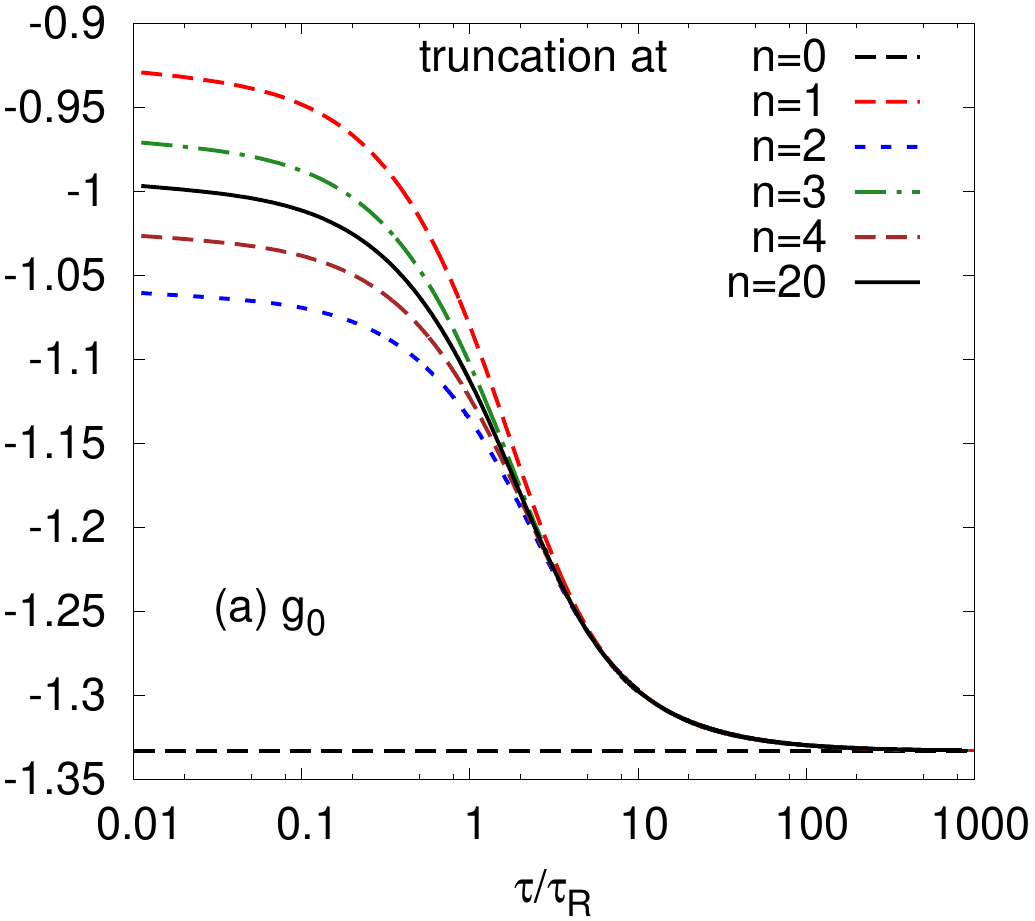}
\includegraphics[width=.450\textwidth] {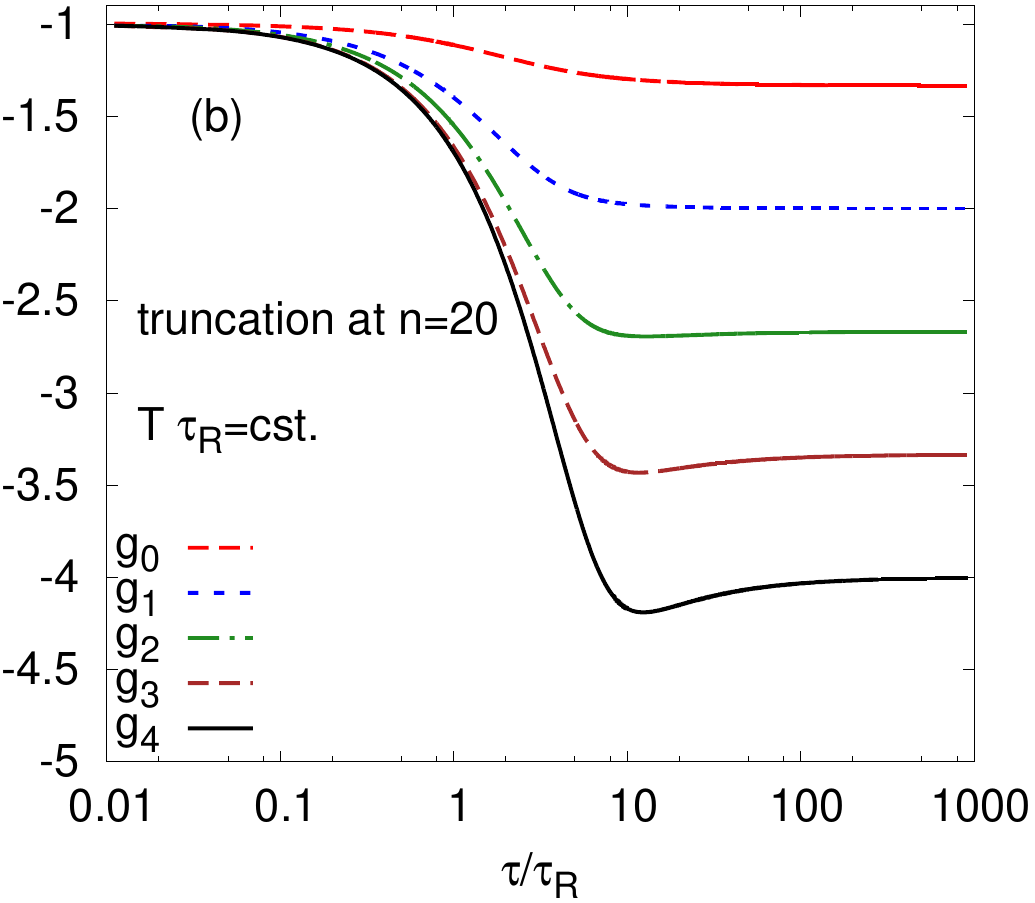}
\caption{ Attractor solution of the moment equations (conformal case). Left panel: attractor solution for $g_0$, for various truncations (n=1 is the two-moment truncation). Right panel: the attractor solutions for the moments $g_0,\cdots, g_4$, calculated with the ``exact'' solution (n=20).    \label{fig:attractor}
}
\end{center}
\end{figure}

We have now a more complete picture of the general solution of the kinetic equation. At small times, i.e. $\tau\ll\tR$, the collision rate is small compared to the expansion rate and 
the collisions play little role:  the evolution  is then dominated by free streaming. On the contrary, at later times when the collision rate exceeds the expansion rate, $\tau\gg \tau_R$, the collisions dominate and the evolution is controlled by  the hydrodynamic fixed point. It is interesting to consider the particular solution of the kinetic equation that starts at the free streaming fixed point, that is with a flat distribution and no longitudinal pressure, and follow its evolution to the hydrodynammic fixed point.  We call this particular solution the ``attractor solution'' since the solutions corresponding to different initial conditions will eventually converge to this attractor solution at late time. Thus defined, the attractor solution joins smoothly the two (stable) fixed points that we have identified. Note that, as we shall discuss further in Sec.~\ref{attractorsolution}, the attractor depends on the value of the initial time $\tau_0$. We assume here that $\tau_0\ll \tau_R$, so that there is a sizeable  region ($\tau_0<\tau<\tau_R$) of the attractor that is sensitive to the free streaming fixed point. 

\Fig{fig:attractor} depicts the attractor solution obtained from various truncations of  the coupled moment equations. The transition from the free streaming regime to the hydrodynamic regime  is clearly visible. It occurs, as expected, when $\tau\sim \tau_R$. The dispersion of the curves at small time on the left panel reflects the slow convergence of the truncation towards the free-streaming fixed point, as discussed in Sec.~\ref{sec:freestreaming}. Note that this (weak) sensitivity to the initial conditions is quickly washed out as soon as the collision rate becomes comparable to the expansion rate.   

Also shown in \Fig{fig:attractor} (right panel) are the attractor solutions for the logarithmic derivatives of first few moments, $g_0,\ldots, g_4$,   obtained by solving the coupled moment equations with truncation at $n=20$ (i.e., keeping 21 moments), which coincides numerically with the exact solution. In this case, all moments start at their free-streaming fixed point value, -1, and then evolve towards their hydrodynamical fixed point values, $g_n(\infty)$ given in  \Eq{eq:gninf}. The decrease with $n$ of theses fixed point values reflects the fact that higher moments are more efficiently damped at late times. As already emphasized these hydrodynamic fixed point values are independent on the truncation. In particular the fixed point values of $g_0$ and $g_1$ are perfectly captured by the two-moment truncation.

\begin{figure}
\begin{center}
\includegraphics[width=.40\textwidth] {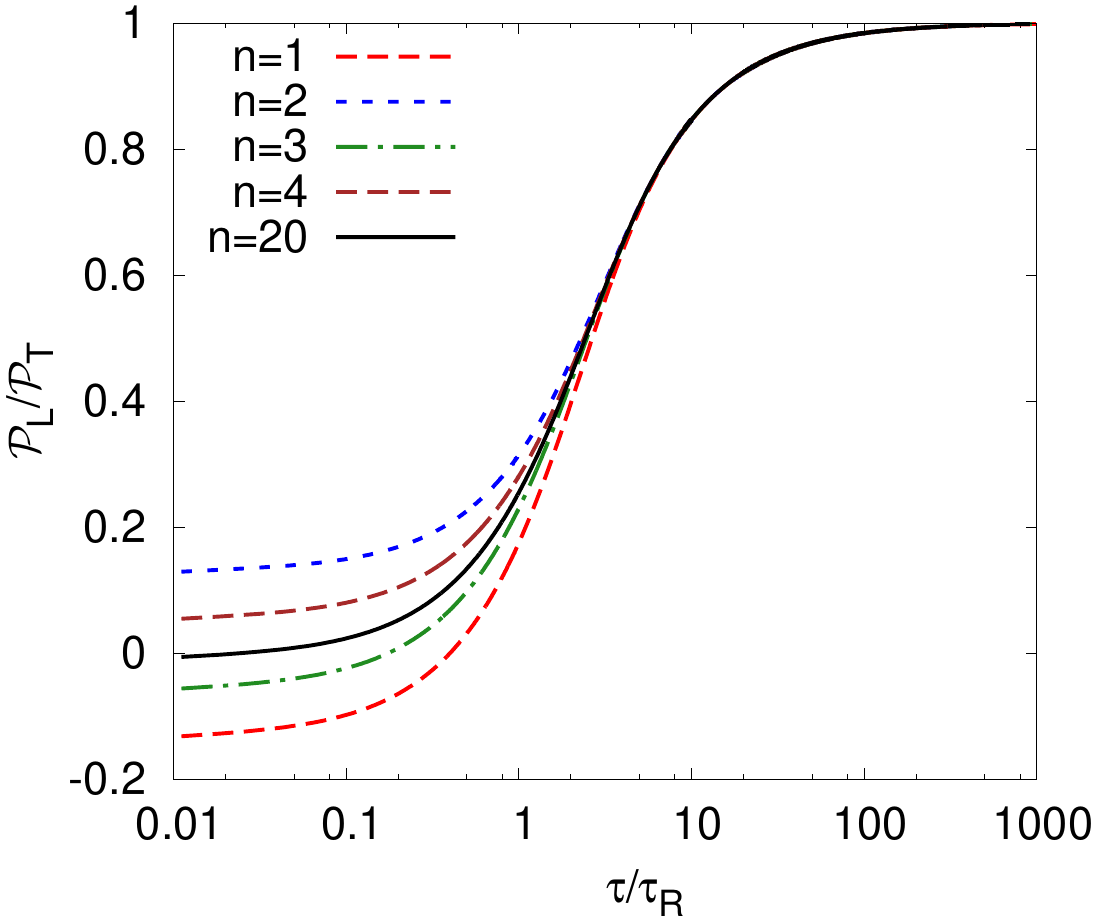}
\includegraphics[width=.40\textwidth] {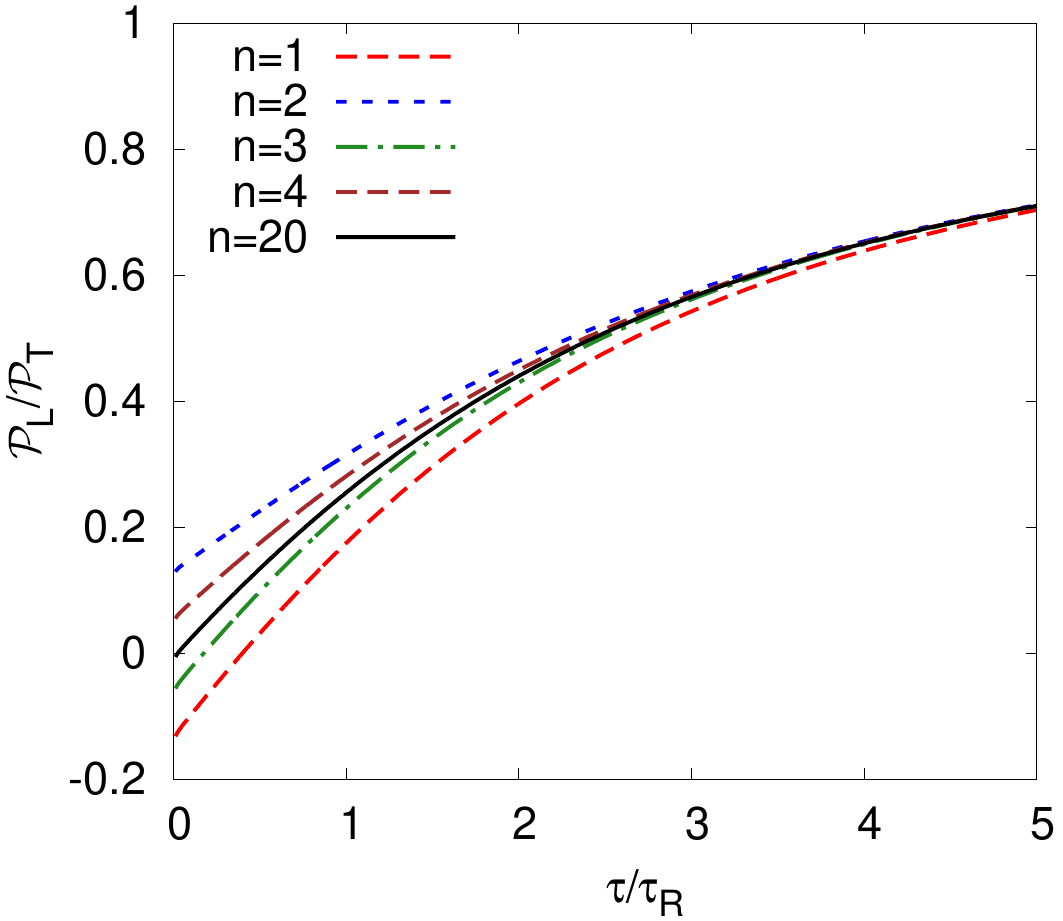}
\caption{ Attractor solution for the pressure ratio (conformal case) as a function of $\tau/\tau_R$ in  a logarithmic scale (left) and a linear scale (right).    \label{fig:attractor2}
}
\end{center}
\end{figure}
The existence of attractor solutions for the  $g_n$'s translates into corresponding attractors for other quantities, such as the ratios of moments, in particular the ratio $\L_1/\L_0$ and hence the pressure anisotropy $\P_L/\P_T$ (see Eq.~(\ref{PLoverPT})).  This is indeed the case, as shown in \Fig{fig:attractor2} for the pressure ratio $\P_L/\P_T$. The insensitivity of the attractor at late times to the truncation reflects the  ``universality'' of the hydrodynamic fixed point. A similar feature was pointed out concerning Fig.~\ref{fig:exactI},  where it was observed that the solutions corresponding to the four distinct initial conditions converge to a unique curve, which we now recognize as the hydrodynamic attractor, when the time $\tau$ exceeds a few times $\tau_R$.

\section{ The approach to hydrodynamics within the two-moment truncation}\label{hydrodynamisation}

This section contains a detailed study of the two-moment truncation, i.e. of the coupled system of equations (\ref{eq:l0l1}). These are obtained from the exact equations
\begin{subequations}
\label{eq:l0l1b}
\begin{align}
\partial_\tau \L_0 + \frac{1}{\tau}(a_0 \L_0 + c_0 \L_1) =&\; 0\,,\\
\partial_\tau \L_1 + \frac{1}{\tau}(b_1 \L_0 + a_1 \L_1+c_1\L_2) =&\; -\frac{\L_1}{\tR}
\end{align}
\end{subequations}
by dropping the term proportional to $\L_2$ in the second equation. 
 As we have repeatedly emphasized, this simple truncation captures the main qualitative features of the free streaming, and the damping   of the $\L_1$ moment drives the system towards the hydrodynamical regime at late times.  Because of its simplicity, it allows for a semi-analytical treatment that provides insight into the approach to the hydrodynamic regime. At the end of this section we shall discuss the role of the moment $\L_2$, and through it of that of the higher moments: as we shall see, the effect of these moments can be accommodated at late time by a simple renormalization of the dynamics captured by the two-moment truncation.

\subsection{Perturbative corrections to free streaming at small time}\label{sec:perturbation}

We can rewrite the system of equations (\ref{eq:l0l1})   in a matrix form
\beq\label{FSb}
\frac{\partial}{\partial t}
\left(
\begin{array}{c}
 {\cal L}_0 \\
  {\cal L}_1
\end{array}
\right)
=
-\left(
\begin{array}{cc}
 a_0 & c_0 \\
b_1 &  a_1 +r_0\rme^t \end{array}
\right)
\left(
\begin{array}{c}
 {\cal L}_0 \\
{\cal L}_1
\end{array}
\right)= -M(t)\left(
\begin{array}{c}
 {\cal L}_0(t) \\
  {\cal L}_1(t)
\end{array}
\right).
\eeq
where we have set $\tau=\tau_0\rme^t$, and $r_0=\tau_0/\tau_R$ (we suppose $\tau_R={\rm Cste}$ in this subsection).
This is a linear, homogeneous, system of equations with time dependent coefficients. 
At early times, i.e. when $\tau/\tau_R\lesssim 1$, one can treat the effect of collisions by using perturbation theory, that is we set 
\beq
M=H_0+V(t),\qquad H_0=\left(
\begin{array}{cc}
 a_0 & c_0 \\
b_1 &  a_1 \end{array}
\right),\qquad 
V(t)=\left(
\begin{array}{cc}
 0 & 0 \\
0 &  r_0\rme^t \end{array}
\right),
\eeq
where $H_0$ represents free streaming and $V(t)$ is the perturbation caused by the collisions. By applying the standard  techniques of time-dependent perturbation theory, we can then write the solution in the form of a time-ordered exponential
\beq\label{FSb}
\left(
\begin{array}{c}
 {\cal L}_0(t) \\
  {\cal L}_1(t)
\end{array}
\right)
= \rme^{-H_0 t}\,{\rm T}\exp\left\{-\int_{0}^t \rmd t'\, V_I(t')   \right\}\left(
\begin{array}{c}
 {\cal L}_0(0) \\
  {\cal L}_1(0)
\end{array}    
\right),     
\eeq
where    
\beq
V_I(t)=\rme^{H_0t}V(t) \rme^{-H_0t}.    
\eeq  
Loosely speaking, the expansion of the time-ordered exponential in powers of $V$ corresponds to an expansion in the number of collisions, the linear term corresponding to one collision, the second order term to two collisions, and so on. 
  
\begin{figure}
\begin{center}
\includegraphics[width=0.45\textwidth] {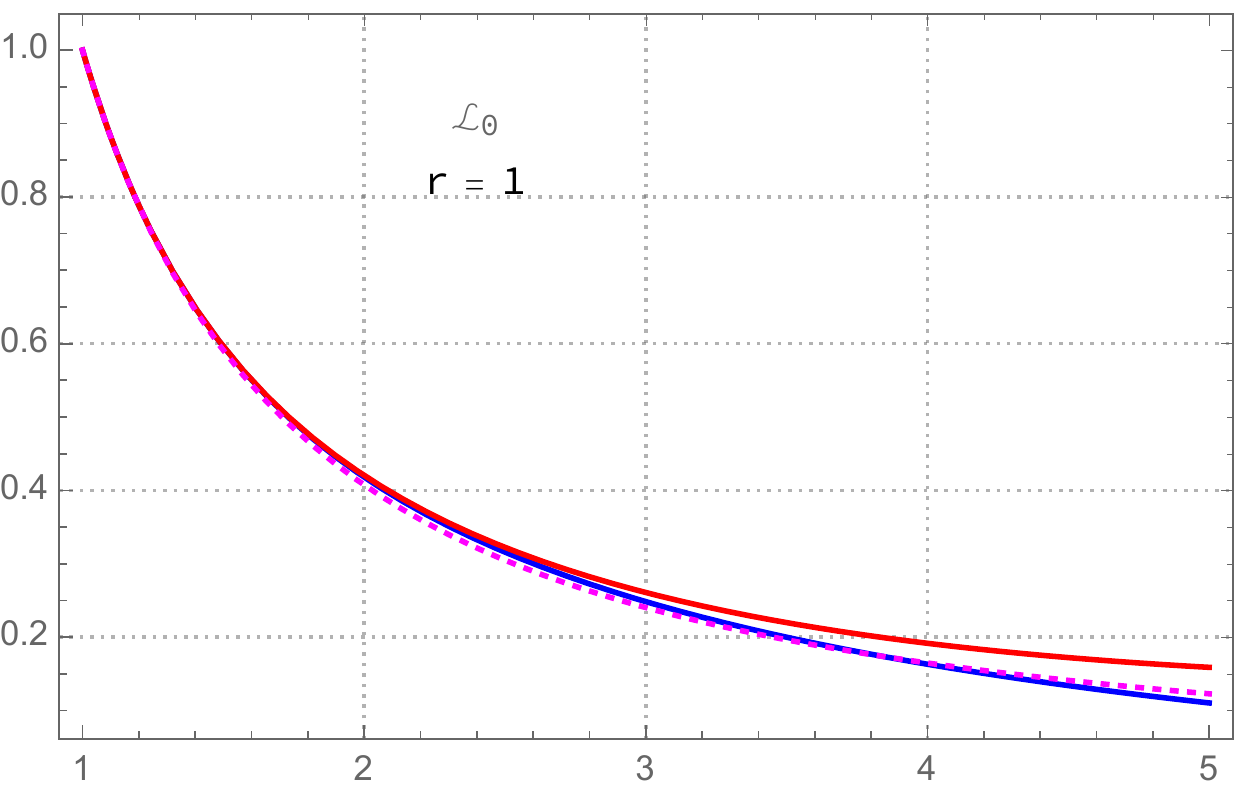}\includegraphics[width=0.45\textwidth] {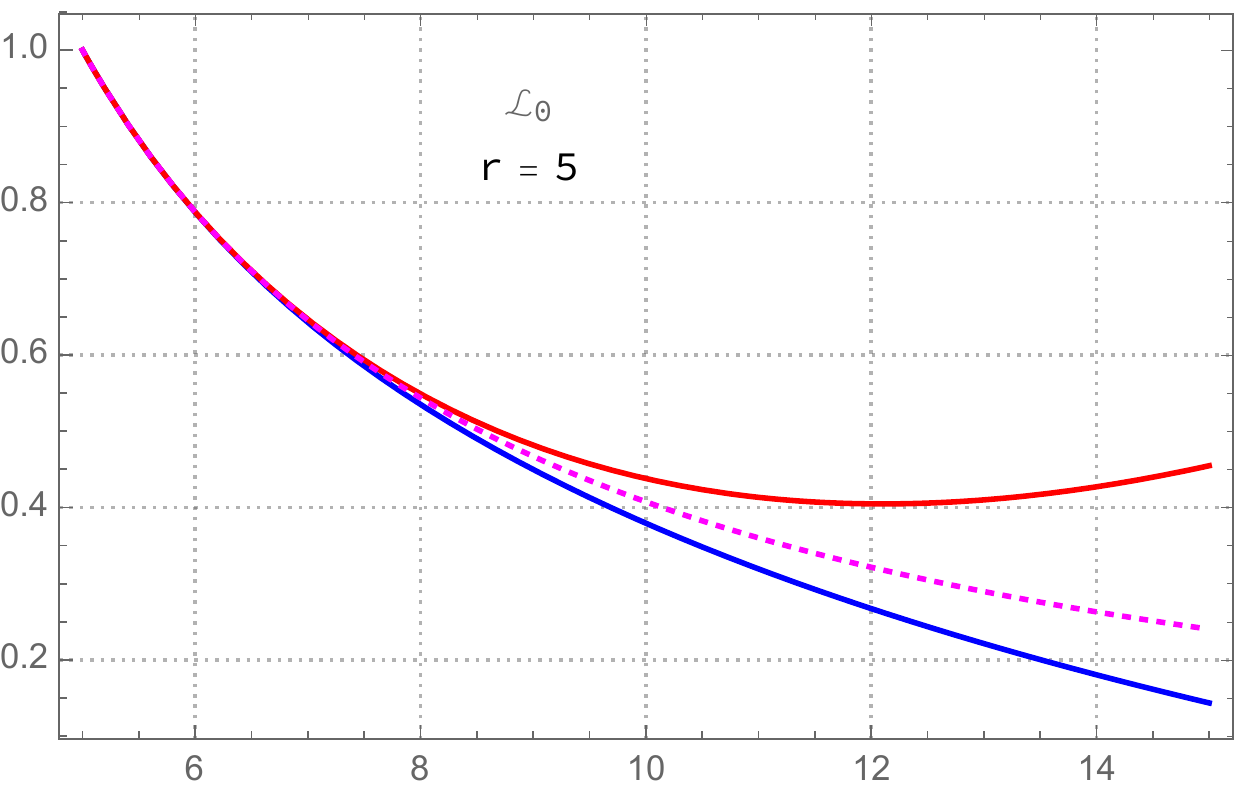}\\
\includegraphics[width=0.45\textwidth] {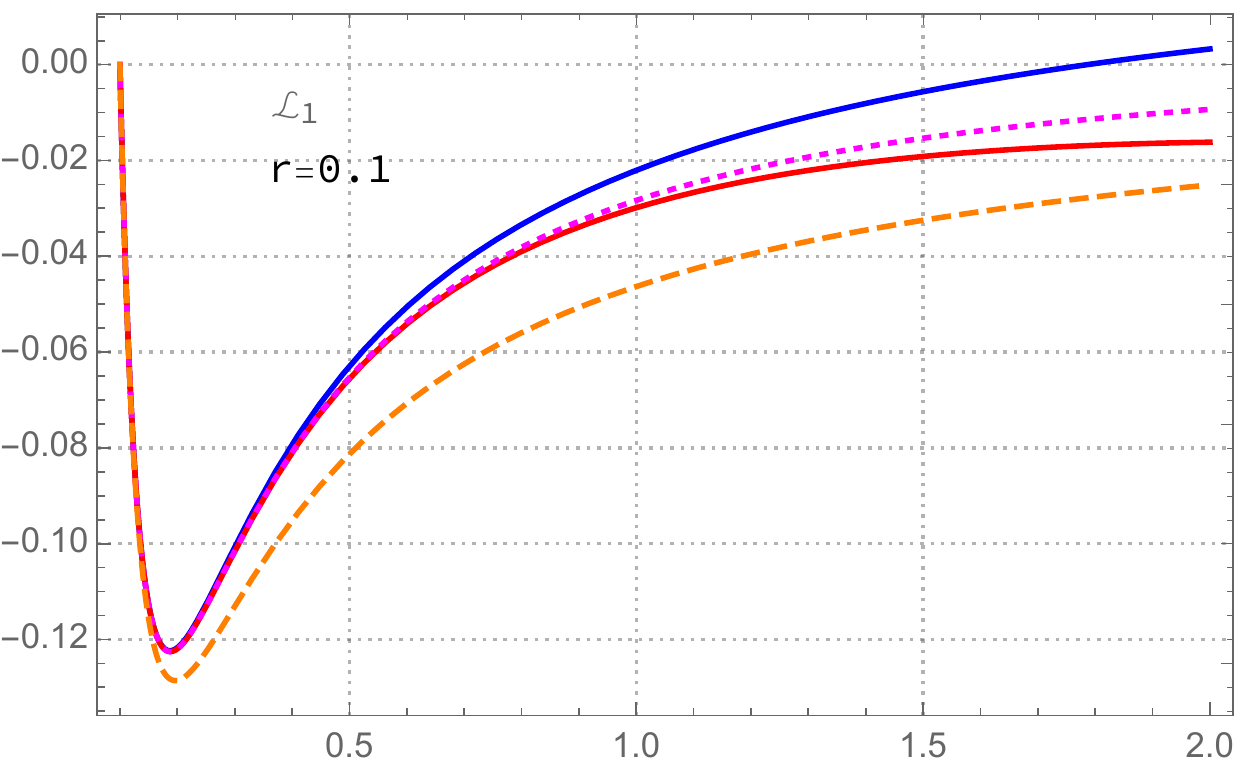}\includegraphics[width=0.45\textwidth] {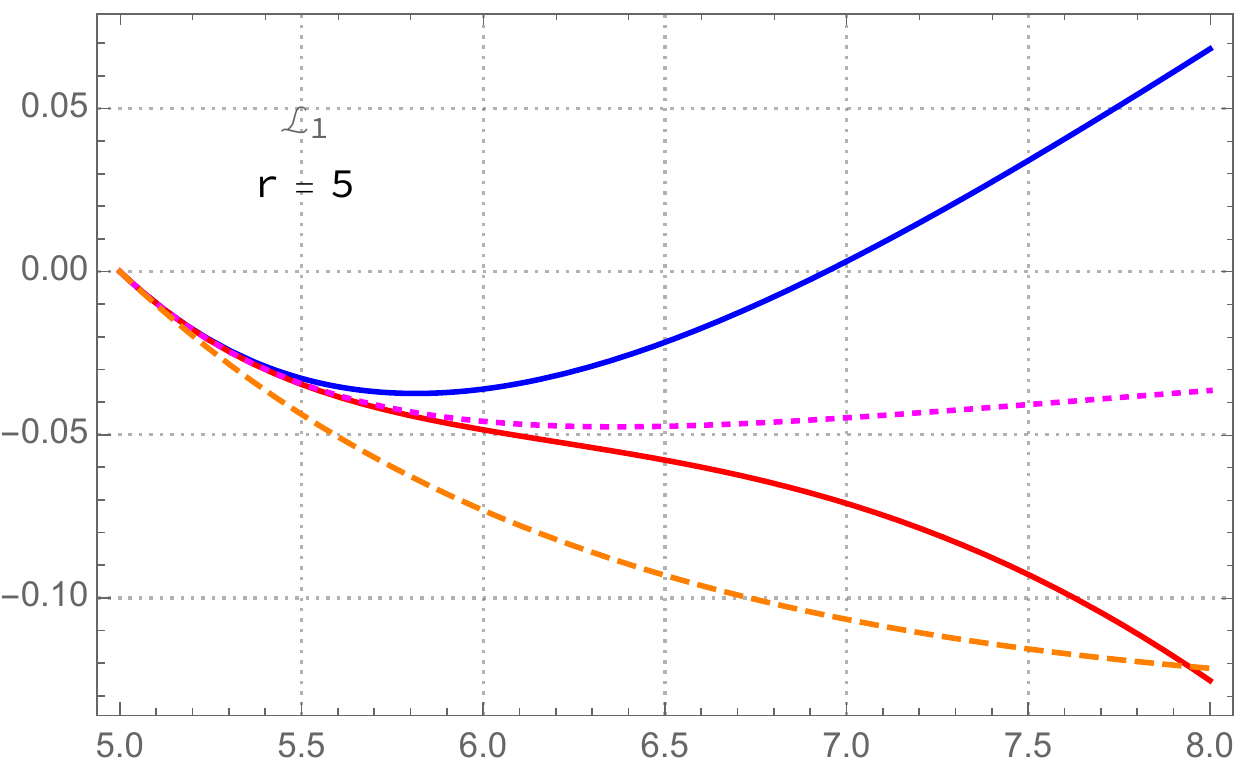}\\
\includegraphics[width=0.45\textwidth] {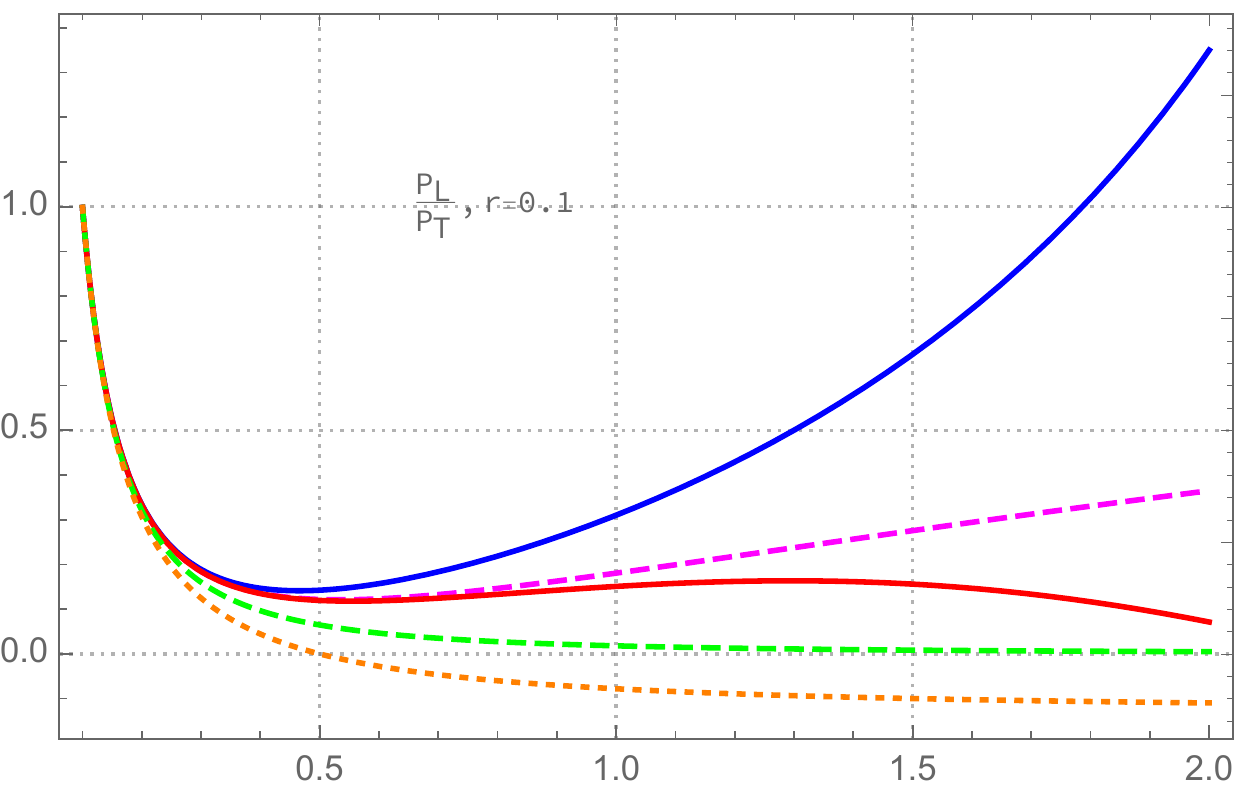}\includegraphics[width=0.45\textwidth] {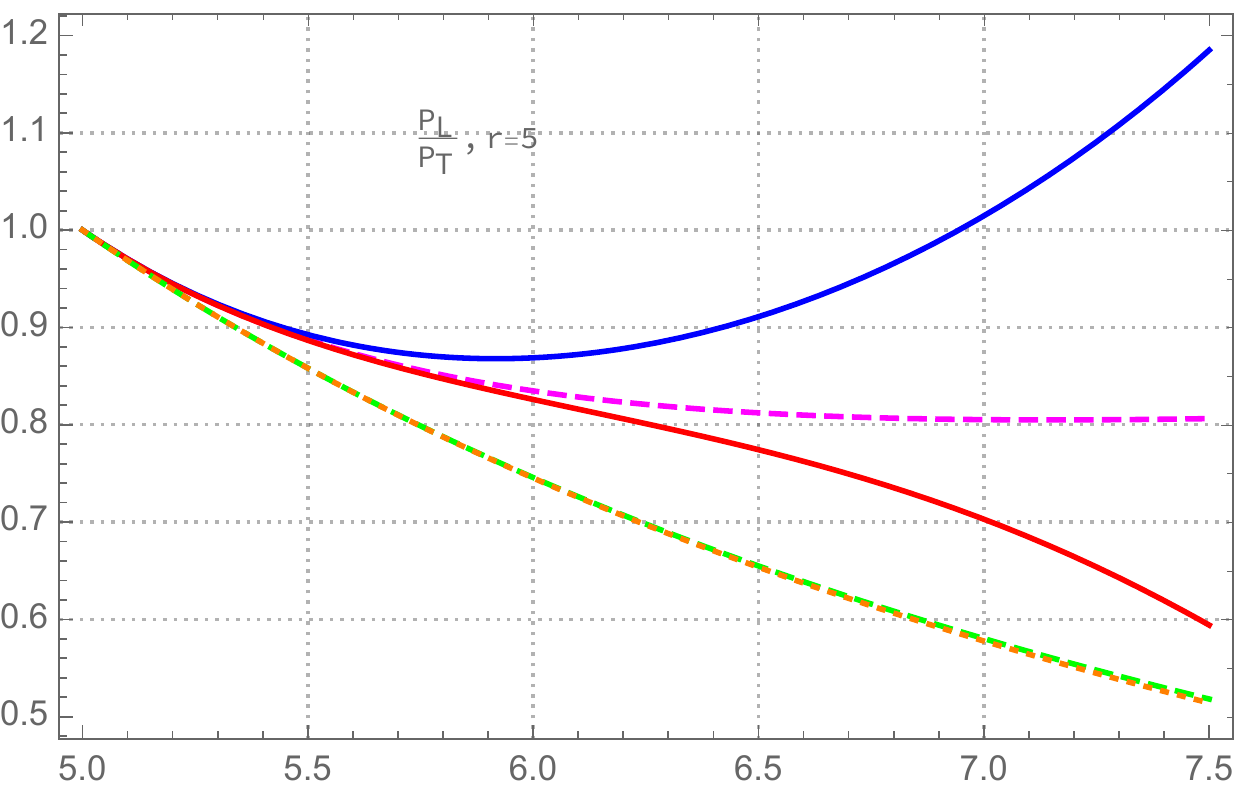}\\\includegraphics[width=0.45\textwidth] {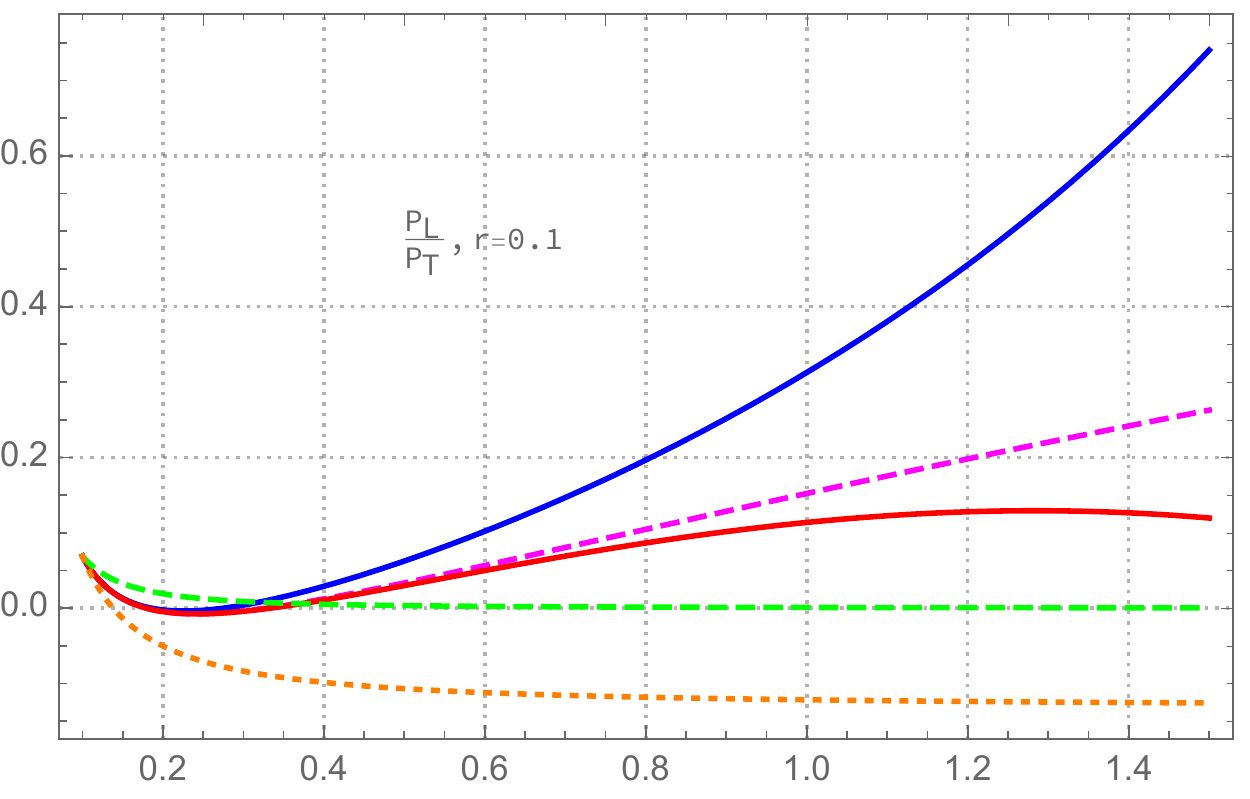}\includegraphics[width=0.45\textwidth] {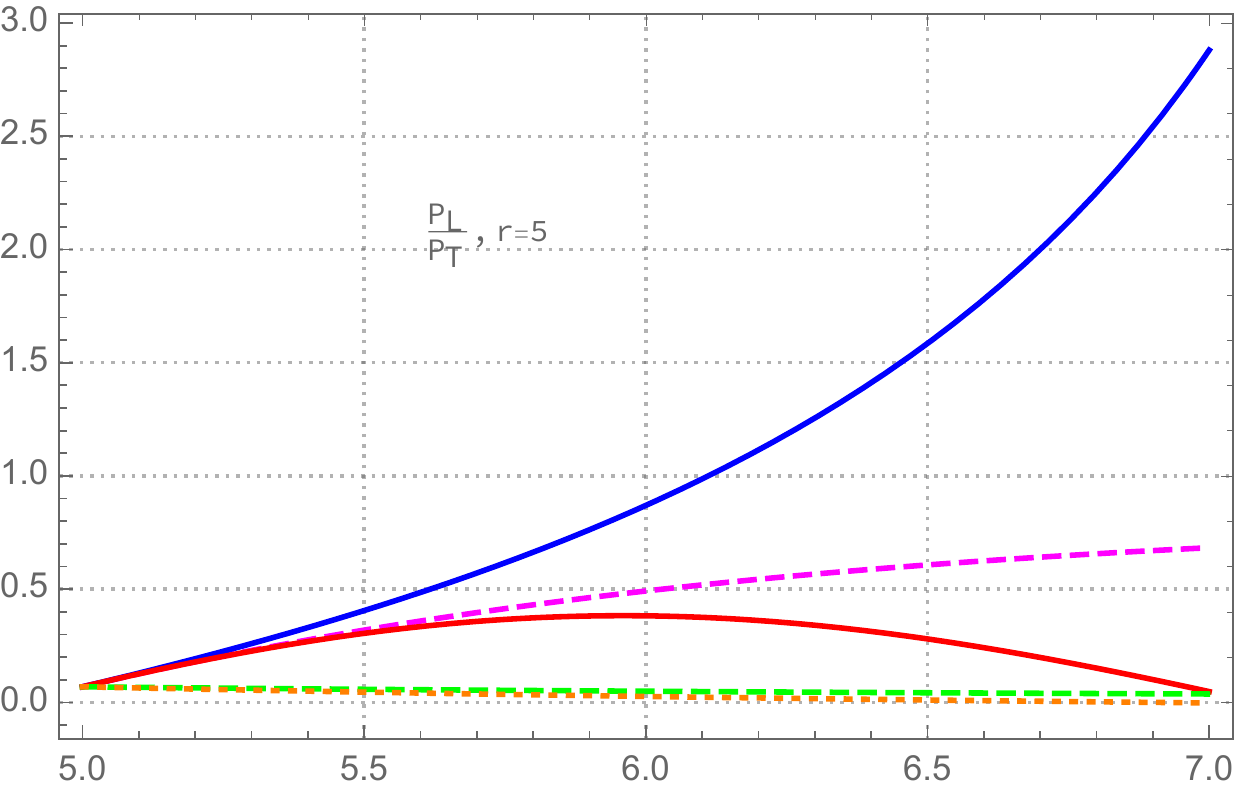}
\caption{(Color online.) From top to bottom, the moments $\L_0$, $\L_1$, and the ratio $\P_L/\P_T$ for isotropic initial conditions ($\Lambda_0=0$), and  for flat initial conditions ($\Lambda_0=-0.45$). Full lines: first order (blue) and second order (red) perturbation theory, as a function of $\tau/\tau_R$, for $r_0=0.1$ (left) and $r_0=5$ (right) (for $\L_0$ the left curve corresponds to $r_0=1$, as the curves corresponding to $r_0=0.1$ would be indistinguishable from the complete solution).  Dashed (magenta) lines: exact solution of the two-moment truncation.  Orange dotted lines: free-streaming obtained from the two moment equations. Dashed green lines: exact free streaming. \label{fig:pert_small_l0}}     
\end{center}    
\end{figure}

The moments obtained up to second order are displayed in Fig.~\ref{fig:pert_small_l0}, for a small and a large value of $r_0$. Note that all curves start at $\tau_0/\tau_R=r_0$. In both cases, perturbation theory accounts very well for the time variations of the moments  and their deviations from free streaming  at early time, i.e., when $\tau-\tau_0\lesssim \tau_R$. Note that the deviation from free streaming can be sizeable before $\tau-\tau_0\sim \tau_R$. This depends somewhat on the quantity one looks at, and on the initial condition. In the case of the ratio $\P_L/\P_T$, the short time behavior is dominated by free streaming when the initial conditions are isotropic. This is not so for flat initial conditions, where collisions produce a deviation from free streaming already at early time. We shall return to this aspect shortly. Note finally the artefact of the two-moment truncation that we have already emphasized: the free streaming tends to drive the longitudinal pressure to negative values. This unphysical feature is absent in the exact free streaming calculation, and is also much attenuated when the collision rate is sufficiently high, as can be see in the right panels of Fig.~\ref{fig:pert_small_l0}.
                                                                                    
Although we could increase the range of validity of perturbation theory by including higher order terms, the plots in Fig.~\ref{fig:pert_small_l0} already suggest that this will not give a consistent account of the late time behaviour, as deviations from one order to the next in perturbation theory grows rapidly with time (this is most clearly visible in the right panels of Fig.~\ref{fig:pert_small_l0}, corresponding to the large collision rate $r_0=5$). In fact, at late time, one expects hydrodynamics to set in, and this regime is not expected to be reached by perturbing free streaming to any order in perturbation theory. Indeed, as we have seen in the previous section, it is controlled by a different fixed point than the free streaming one around which one is expanding. 
                    
 In order to study the large time behaviour of the solution it is then necessary to go beyond perturbation theory. To do so, it is more convenient to transform the linear system into a single differential equation.

\subsection{Reducing the linear system to a second order differential equation}\label{equadiffL0L1}

 Such an equation is easily obtained by taking a time derivative of the first equation (\ref{eq:l0l1}) and using the second equation in order to eliminate $\L_1$. One gets then  
 a second order linear differential equation for $\L_0$
\beq\label{eqforL0}
\tau \ddot\L_0+\left(1+a_0+a_1+\frac{\tau}{\tR}\right)\dot\L_0+\frac{1}{\tau}\left(a_1a_0-c_0b_1+\frac{a_0\tau}{\tau_R}\right) \L_0 =0.
\eeq
 Note that this equation is valid for an arbitrary (e.g. time dependent or time independent) relaxation time $\tau_R$.  This single equation for $\L_0$ can be viewed as an approximation to the exact equation  for the energy density (Eq.~(\ref{eq:rta_0a}), for $n=0$). In the variable $t=\ln(\tau/\tau_0)$,  this equation reads
 \beq\label{eqforL0t}
 \frac{\del \L_0}{\del t^2}+\left(a_0+a_1+r_0\rme^t\right)\frac{\del \L_0}{\del t}+\left( a_1a_0-c_0b_1+a_0r_0\rme^t\right)\L_0=0.
 \eeq
 Once $\L_0$ is known, $\L_1$ can be determined from the first of Eqs.~(\ref{eq:l0l1b}). Alternatively, one may obtain $\L_1$ by solving an equation analogous to Eq.~(\ref{eqforL0t}), namely
  \beq\label{eqforL1t}
 \frac{\del \L_1}{\del t^2}+\left(1+a_0+a_1+r_0\rme^t\right)\frac{\del \L_1}{\del t}+\left( a_1a_0-c_0b_1+(1+a_0)r_0\rme^t\right)\L_1=0.
 \eeq
 In contrast to Eqs.~(\ref{eqforL0}) or (\ref{eqforL0t}) this equation is valid only for constant $\tau_R$ (a derivative of the second equation (\ref{eq:l0l1}) is involved in its derivation). We shall see later how to treat the conformal case, and focus for the time being on the case of constant $\tau_R$.

Since the equations (\ref{eqforL0t}) and (\ref{eqforL1t}) are  of second order, we need to specify the values of the functions $\L_0$ and $\L_1$ and their time derivatives at $t=0$. These are easily obtained from the equivalent linear problem (\ref{eq:l0l1}):
  \beq\label{FSb2b}
\left(
\begin{array}{c}
 \dot{\cal L}_0 \\
  \dot{\cal L}_1
\end{array}
\right)
=
-\left(
\begin{array}{cc}
 a_0 & c_0 \\
b_1 &  a_1 +r_0 \end{array}
\right)
\left(
\begin{array}{c}
 1 \\
\Lambda_0
\end{array}
\right),
\eeq
where we have set $\L_0(t=0)=1$, and $\L_1(t=0)=\Lambda_0$. 
Note that since the equation is linear, the solution $\L_0$ is defined to within a multiplicative constant. Measuring all moments in units of the initial energy density, we fix all initial conditions so that $\L_0=1$ initially, leaving $\Lambda_0=\L_1/\L_0$ at the initial time as the only parameter. Recall that physically acceptable values of $\Lambda_0$ range from $\Lambda_0=-0.5$ corresponding to a flat distribution, to $\Lambda_0=1$ corresponding to an isotropic distribution.

The equation $\dot\L_0=-a_0-c_0 \Lambda_0$ is exact. It involves only $\L_0$ and $\L_1$ and no other moment, and it is independant of the collision rate.  
Since $a_0$ and $c_0$ are both positive, $\dot \L_0<0$ 
 for all physical values of $\Lambda_0$, and $\dot\L_0$ goes from $-4/3$ for an isotropic distribution to $-1$ for a flat distribution.  
 
 The equation $\dot\L_1=-b_1-(a_1+r_0)\Lambda_0$  
 indicates that the sign of $\dot\L_1$, may vary depending on the values of $\Lambda_0$ and $r_0$. For all positive values of $r_0$, $\dot\L_1$ is a decreasing function of $\Lambda_0$. For $\Lambda_0=0$, corresponding to isotropic initial conditions, $\dot\L_1=-b_1=-8/15$ and is independent of $r_0$. (Note that the value $\dot\L_1=-b_1$ coincides with the slope of the exact free streaming solution.)   As $\Lambda_0$ decreases 
 $\dot\L_1$ increases and vanishes for $\Lambda_0=\bar\Lambda_0=-56/(190+105 r_0)$. Note that $\bar\Lambda_0\simeq -0.295$ for $r_0=0$, and $\bar\Lambda_0$ remains negative as $r_0\to\infty$. Thus for a flat initial condition, i.e. $\Lambda\gtrsim -0.5$, $\dot\L_1>0$. This behavior is consistent with that of the exact free streaming solution displayed in Fig.~\ref{fig:FSl1a0105}.
 
 The expression of $\dot\L_1$ obtained from the two-moment equations is only approximate.   The  exact equation, which can be obtained from Eq.~(\ref{eq:l0l1b}), involves not only $\L_0$ and $\L_1$, but also $\L_2$. It reads 
 \beq \label{dotL1}
 \dot\L_1=-b_1-\left(a_1+r_0\right) \Lambda_0 -c_1 \frac{\F_2(1/\xi_0)}{\F_0(1/\xi_0)}, 
  \eeq 
 where $\xi_0$ is such that  $\F_1(1/\xi_0)/\F_0(1/\xi_0)=\Lambda_0$. One can estimate the effect of the correction due to $\L_2$ in the two limiting cases of isotropic and flat initial distributions.      
 For isotropic initial conditions, $\xi_0=1$ and $\F_2=0$, so the correction due to $\L_2$ vanish, and $\dot\L_1=-b_1$, as we have already observed. 
 Near the flat distribution, ${\L_2(1/\xi_0)}/{\L_0(1/\xi_0)}\simeq A_2=3/8$, and $\Lambda_0\simeq -0.5$. We have therefore, in the exact case
\beq
\dot\L_1=\frac{1}{2}+\frac{r_0}{2},
\eeq
while, neglecting the contribution from $\L_2$, we get instead
\beq
\dot\L_1=\frac{13}{35}+\frac{r_0}{2}.
\eeq
 Such a correction has an impact on the behavior of the ratio $\P_L/\P_T$ at small $\tau$. We have indeed 
  \beq
\left.\frac{\rmd }{\rmd t}\frac{\P_L}{\P_T}\right|_{t=0}=\frac{3\dot\L_1-3\Lambda_0\dot\L_0}{(1-\Lambda_0)^2}=\tau_0\left.\frac{\rmd }{\rmd \tau}\frac{\P_L}{\P_T}\right|_{\tau_0}.
\eeq
This relation holds for any $\tau_R$, constant or not. A simple calculation yields for the slope at the origin 
 (expressed in terms of $\tau/\tau_R$ for constant $\tau_R$)
 \beq
 \tau_R\left.\frac{\rmd }{\rmd \tau}\frac{\P_L}{\P_T}\right|_{\tau_0}=\frac{2}{3}  \quad {\rm (exact)},\qquad \tau_R\left.\frac{\rmd }{\rmd \tau}\frac{\P_L}{\P_T}\right|_{\tau_0}=-\frac{6}{35 r_0}+\frac{2}{3}   \quad   {\rm (without}\;\L_2) .
 \eeq
 For a non constant $\tau_R$, we use the variable $w=\tau/\tau_R$ (see next subsection) and express ${\P_L}/{\P_T}$ in terms of $w$. That is, 
 \beq
 \left.\frac{\rmd }{\rmd w}\frac{\P_L}{\P_T}\right|_{w_0}=\left(\frac{\rmd }{\rmd \tau}\frac{\P_L}{\P_T}\right)   \frac{\rmd\tau}{\rmd w}=\left(\tau_0\frac{\rmd }{\rmd \tau}\frac{\P_L}{\P_T}\right)\frac{\tau_R/\tau_0}{1+g_0/4}=\frac{2r_0}{3}\frac{1}{r_0}\frac{1}{3/4}=\frac{8}{9},
 \eeq
 where we have used $g_0=-1$.   
Thus, for the exact case the slope at the origin is always positive, which is physically expected: for a flat distribution the longitudinal pressure vanishes, and cannot therefore decrease. However, if one ignore the contribution of $\L_2$, as we do in the two-moment truncation, there is a value of $r_0$, $r_0=9/35$, below which the slope is negative. This is the situation illustrated in \Fig{fig:exactI} for initial conditions I. 

\begin{figure}
\begin{center}
\includegraphics[width=0.45\textwidth] {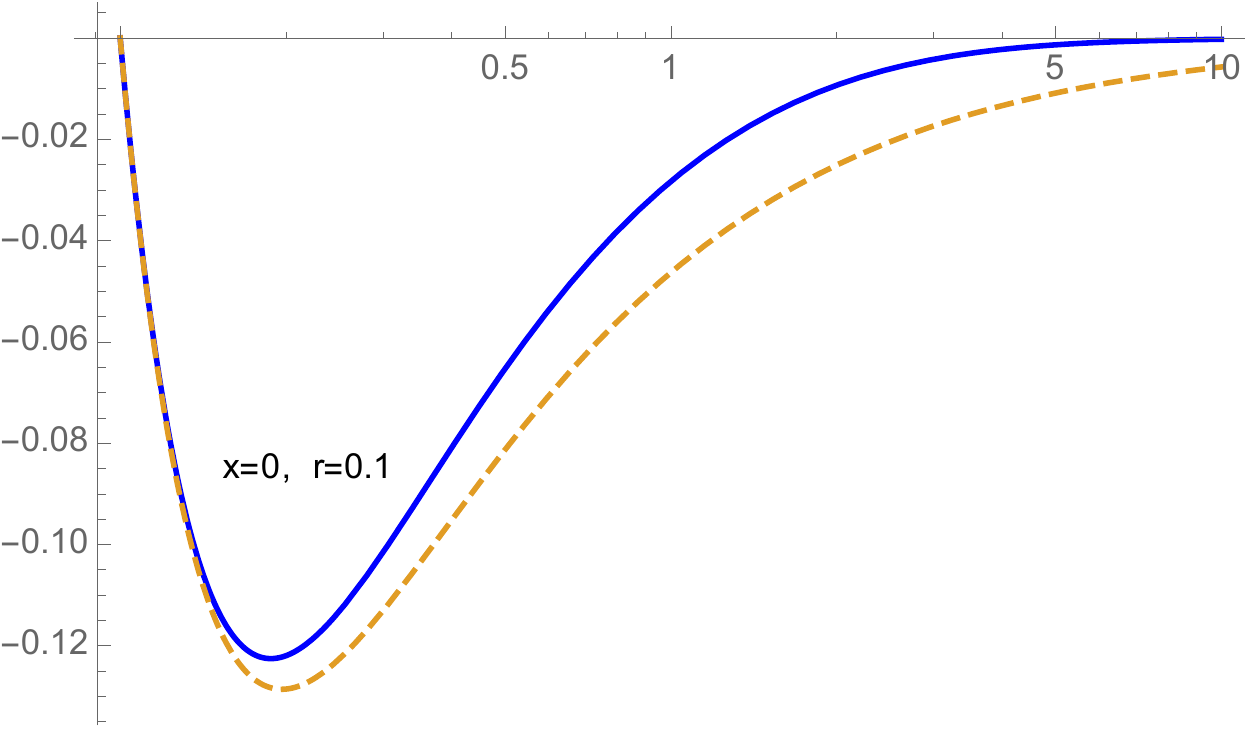}\includegraphics[width=0.45\textwidth] {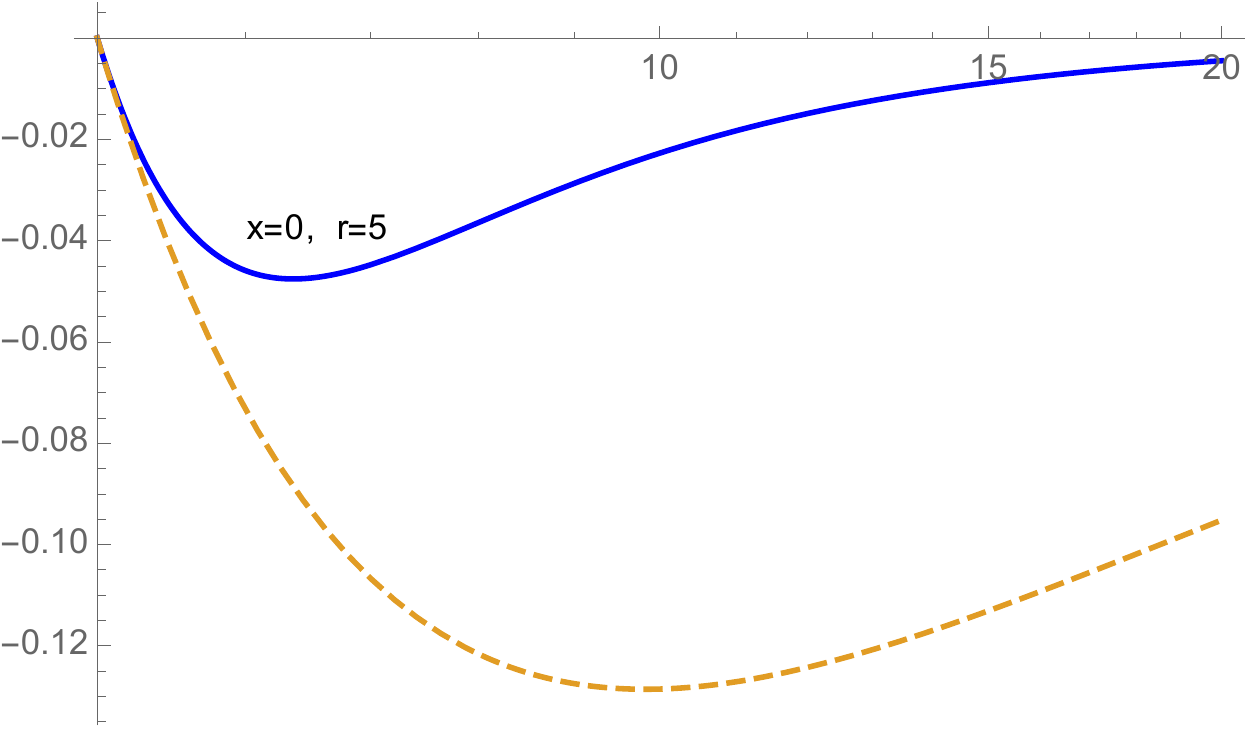}
\caption{(Color online.) The moment  $\L_1$  for isotropic initial conditions, as a function of $\tau/\tau_R$ on a logarithmic scale, for a small collision rate, $r_0=0.1$ and a large collision rate $r_0=5$. The full (blue) line is the complete solution with collisions, while the dashed line is the free streaming solution.  \label{fig:L0x0r}
}
\end{center}
\end{figure}

We now consider the effects of the collisions beyond those just described, and  that concerned the short time behaviour. We consider only isotropic initial conditions, that is $\Lambda_0=0$, and study how the time dependence of $\L_0$ and $\L_1$ is affected by the change of $r_0$. This is illustrated in Fig.~\ref{fig:L0x0r}. The effect of the collisions in accelerating the damping of the moment $\L_1$ is clearly visible. We note however that the short time behaviour is not modified: this is in line with the fact that the initial conditions do not depend on $r_0$ in the isotropic case. Note that the free streaming curves are the same for the two values of $r_0$. The apparent difference is due to the change in the relative expansion rate versus the collision rate (in other words the two curves would be the same if plotted as a function of $\tau/\tau_0$ instead of $\tau/\tau_R=\tau/(r_0\tau_0)$ as done here).  
      
\subsection{Gradient expansions at late time}\label{sec:gradexpansion}
     
As explained in the previous section, one expects  the solution     at late time to be well represented by an expansion in powers of $1/\tau$, i.e., by a gradient expansion. 
In order to study such an expansion in a systematic fashion, it is convenient to write Eq.~(\ref{eqforL0}) in terms of the function $g_0(\tau)=\rmd\ln \L_0(\tau)/\rmd\ln\tau$ (see Eq.~(\ref{logderivative})). A simple calculation yields 
\beq\label{eqforbg0} 
\tau\frac{\rmd g_0}{\rmd \tau}+g_0^2+\left(a_0+a_1+\frac{\tau}{\tR}\right)g_0+a_1a_0-c_0b_1+\frac{a_0\tau}{\tau_R} =0.
\eeq
This is a first order, non linear equation for $g_0$. It is furthermore useful to perform a change of variables, setting 
\beq
w=\frac{\tau}{\tau_R}, 
\eeq
and 
 assuming  the mapping between $w$ and $\tau $ to be invertible, which is the case in practice. We may then consider $\L_0$ as a function of $w$, i.e. $\L_0(\tau(w))$,  so that (with a slight abuse of notation)
\beq
\frac{\rmd\ln \L_0(w)}{\rmd  \ln w}=g_0 \frac{\rmd \ln\tau}{\rmd\ln w}.
\eeq
If $\tau_R$ is a constant,   ${\rmd \ln\tau}/{\rmd\ln w}=1$. If $\tau_R T$ is a constant, then ${\rmd\ln w}/{\rmd \ln\tau}= (1+g_0/4)$, where we have used  $\L_0(\tau)\propto T^4(\tau)$ so that ${\rmd \ln T}/{\rmd \ln\tau}={g_0}/{4}$.
It is then straightforward to transform Eq.~(\ref{eqforbg0}) into
\beq\label{eqforg0}
w g_0' + g_0^2 +(a_0+a_1+w) g_0 +wa_0+a_0a_1-b_1c_0 =0,\qquad g_0'\equiv\frac{\rmd g_0}{\rmd w}\label{eqfor g0},
\eeq
which is valid for the case of constant $\tau_R$. A similar equation holds in the conformal case, with  $wg_0'$  replaced by $wg_0'\left(1+\frac{g_0}{4}\right)$.\footnote{This equation, for the conformal case, can also be written  in terms of the function $f=\rmd \ln w/\rmd\ln\tau=1+g_0/4$. It then takes the form ($f'=\rmd f/\rmd w$)
$$
4w f' f+ 16  f^2+\left( -32 +4\left(a_0+a_1+w    \right)\right)f + 16-4(a_0+a_1)+a_0a_1-c_0b_1+(a_0-4)w    =0.$$
This equation (to within inessential details) is the equation whose asymptotic solution is analyzed thoroughly in Refs.~\cite{Heller:2015dha} and \cite{Basar:2015ava} (see also Appendix~\ref{stabilityanalysis}).}                  

We look now for a  solution of Eq.~(\ref{eqforg0}) at large time of the form
\beq\label{gradexpg0w}
g_0(w)=\sum_{n=0}\frac{\gamma_n}{w^n},
\eeq
where the coefficients $\gamma_n$ are determined by solving the equation order by order. The first two coefficients are independent of the choice of $\tau_R$. They read 
\beq\label{gradientcoef0and1}
   \gamma_0=-a_0=-\frac{4}{3},\quad \gamma_1=b_1 c_0=\frac{16}{45}.
   \eeq
The higher order coefficient depend on the choice of $\tau_R$. For instance
\beq\label{gradientcoefgamma2}
 \gamma_2=\frac{b_1 c_0}{4} \left(3 a_0-4 a_1+4\right)\;\; (\tau_R T={\rm Cste}\;),\quad \gamma_2=b_1 c_0\left(1+a_0-a_1\right)\;\;(\tau_R ={\rm Cste}).\nn
   \eeq
   Note that $\gamma_1$ and $\gamma_2$ are proportional to $c_0$, that is to the coupling of $\L_0$ to $\L_1$: it is indeed via this coupling that the gradient expansion of $\L_0$ emerges dynamically, as already explained. 
Further details are given in Appendix~\ref{gradientg0}.

\subsection{Fixed point analysis and attractor solution}\label{attractorsolution}

Equation (\ref{eqforg0}) also lends itself to a simple analysis in terms of fixed points. It is convenient to rewrite this equation as follows\footnote{In the conformal setting $wg_0'$ is to be multiplied by $(1+g_0/4)$.}
\beq\label{betag0w}
w\frac{\rmd g_0}{\rmd w}=\beta(w,g_0),
\qquad 
\beta(w,g_0)=\beta(g_0)-w(g_0+a_0),
\eeq
where $\beta(g_0)$ is the function introduced in Eq.~(\ref{eqg0FSfp}) for the free streaming case, and which we rewrite here for convenience
\beq\label{beta0}
\beta(g_0)=-g_0^2-(a_0+a_1) g_0-a_0a_1+c_0b_1.
\eeq 
 The function $\beta(w,g_0)$ plays a role similar to that of the function $\beta(g_0)$ in the free streaming case. However, since it depends on $w$, we do not have true fixed points as in the free streaming case. Nevertheless we shall see that the function $\beta(w,g_0)$ is helpful to understand the main features of the solution.   

  \begin{figure}
\begin{center}
\includegraphics[width=0.5\textwidth] {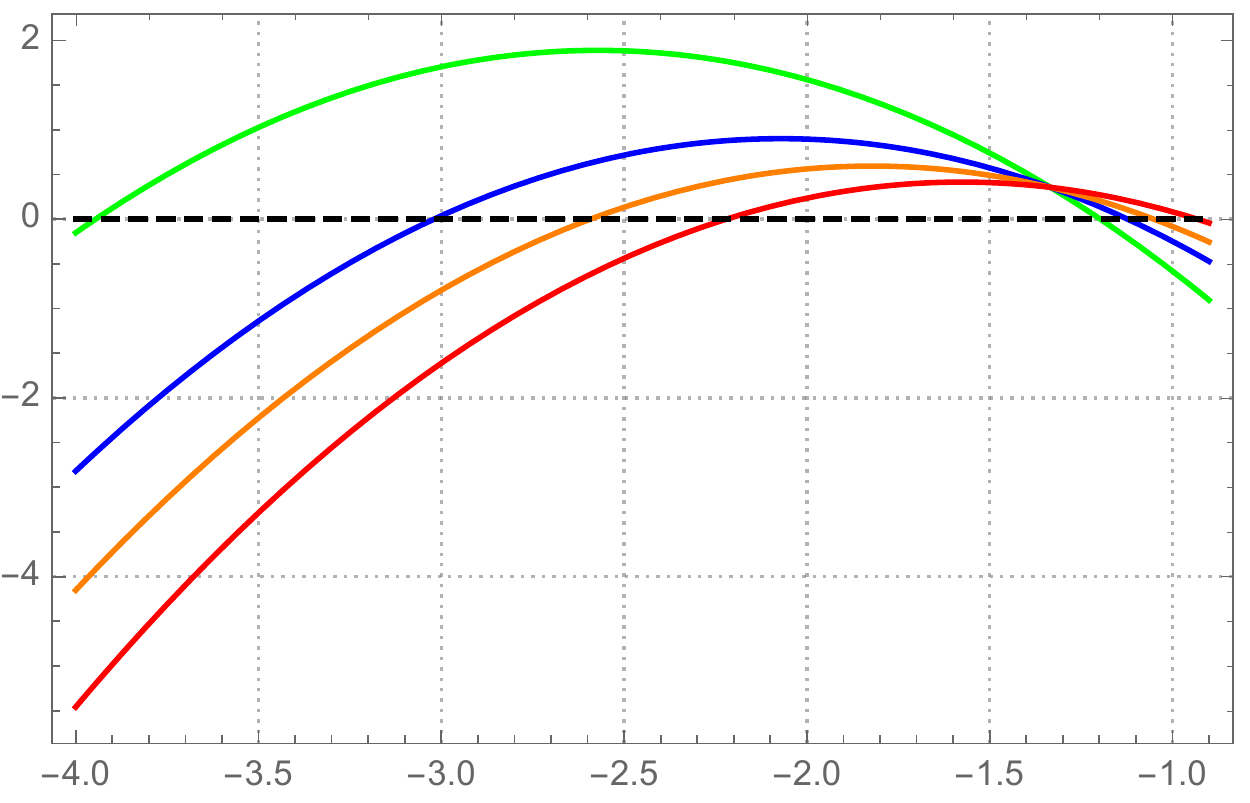}\includegraphics[width=0.5\textwidth] {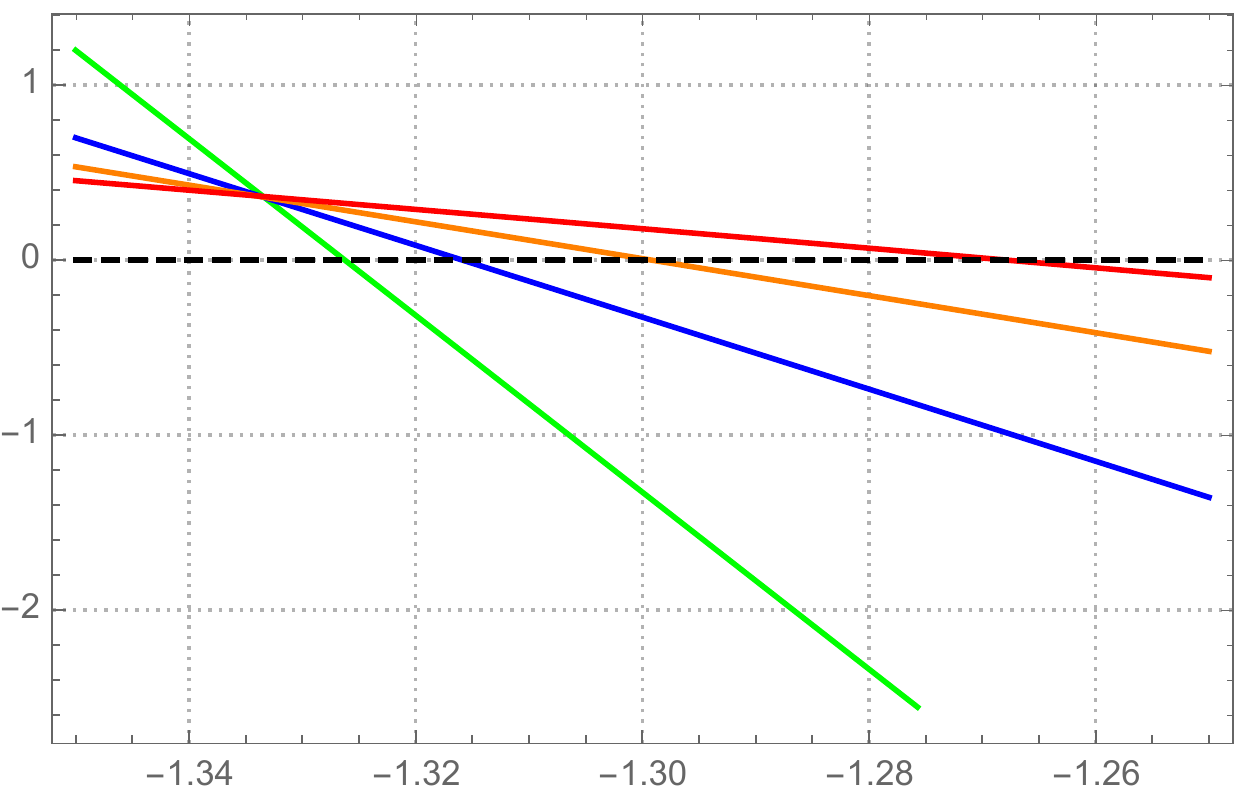}
\caption{(Color online.) The function $\beta(g_0,w)$ as a function of $g_0$ for different values of $w $. Left, from bottom to top: $w=0.01$ (red), $w=0.5$ (orange), $w=1$ (blue), $w=2$ (green). The pseudo fixed point are located at the intersection of these curves with the horizontal dashed line. The attractive fixed point is on the right, the repulsive one on the left. Note that all curves cross for $g_0=-4/3$. The right panel shows the approach of the hydrodynamic fixed point $g_0=-4/3$ as $w$ tends to infinity: $w=5$ (red), $w=10$ (blue), $w=20$ (orange), $w=50$ (green). Recall that for very small values of $w$, the stable (free streaming) fixed point sits at $g_0=-0.929$ (the approximation to $-1$ in the two-moment truncation). As $w$ increases, this fixed point moves continuously towards the hydrodynamical fixed point $g_0=-4/3$.   
\label{fig:fixedpoint1}
}
\end{center}
\end{figure}

A plot of the function $\beta(g_0,w)$ as a function of $g_0$ for different values of $w $ is given in Fig.~\ref{fig:fixedpoint1}. The difference between  $\beta(g_0,w)$ and $\beta(g_0)$  is given by the quantity linear in $g_0$ and $w$, $-w(g_0+a_0)$. This term is small near the free streaming fixed point, i.e.,  for small $w$. It vanishes at the hydrodynamical fixe point, where $g_0=-a_0$. Note that  all curves cross at this particular point, since there the dependence on $w$ disappears. As $w$ increases, the slope of $\beta(g_0,w)$ viewed as a function of $g_0$ is simply $-w$, and it becomes infinite as $w$ becomes infinite. 

When $w\simeq 0$, the function $\beta(g_0,w)$ has two zeroes in the vicinity of the two free streaming stable and unstable fixed points. We shall refer to these zeroes as pseudo fixed points.  The motion of these pseudo fixed points as $w$ increases is clearly visible on the left panel of Fig.~\ref{fig:fixedpoint1}. As $w$ becomes large the unstable pseudo fixed point is pushed to large (eventually infinite) negative values of $g_0$, while the original stable pseudo fixed point approaches the hydrodynamic fixed point located at $g_0=-a_0$.  The expansion of the location of the stable pseudo fixed point at large $w$ reads
\beq
g_{\rm fp}(w)=-\frac{4}{3}+\frac{16}{45 w}-\frac{32}{189 w^2}-\frac{4544}{99225
   w^3}+O\left(\frac{1}{w^4}\right)
   \eeq 
    Note that the first two terms in this expansion coincide with the first two terms in the gradient expansion of $g_0(w)$. This is no accident as we shall see shortly. 
    
 A stability analysis can be carried out, as we did earlier for the free streaming fixed point. To do so, we start from Eq.~(\ref{betag0w}), or the equivalent equation for the conformal case. By expanding a generic solution about the fixed point $g_0+a_0\simeq b_1c_0/w$, and linearizing, one finds 
 that $\delta g(w) \propto \rme^{-Sw} w^{\beta+CS^2/4}$, with $\delta g$ denoting the deviation from the fixed point solution and $S=1, \beta=a_0-a_1, C=b_1c_0$ for the case of constant $\tau_R$, and $S=3/2, \beta=3(a_0-a_1)/2, C=b_1 c_0$ for the conformal case (see Appendix~ \ref{stabilityanalysis} for more details). In either case a generic solution relaxes exponentially fast to the fixed point solution (i.e. towards hydrodynamics).

Consider now the attractor solution, that is the solution that starts at time $\tau_0=r_0\, \tau_R$ in the vicinity of the stable free streaming fixed point, i.e. $g_0=-0.93$. As $w$ increases, $g_0$ decreases, and eventually reaches the hydrodynamical regime. The approach to the hydrodynamical fixed point is subtle however. Note that $\beta(g_0=-a_0,w)=b_1c_0$, that is, this is the result one obtains if one sets $g_0=-a_0$ in Eq.~(\ref{betag0w}).  However, as $w\to\infty$, $\beta(g_0,w)$ is linear in the vicinity of the point where all curves cross, and we have there $\beta(g_0,w)\sim b_1c_0 -w(g_0+a_0)$. The pseudo fixed point, determined by the condition $\beta(g_0,w)= 0$, is located at $g_0\sim -a_0+b_1c_0/w$. This moves smoothly toward $-a_0$ as $w$ increases, keeping $\beta(g_0,w)=0$, that is, the pseudo fixed point moves smoothly toward the hydrodynamic fixed point.  This is clearly seen in the right panel of Fig.~\ref{fig:fixedpoint1}.

 \begin{figure}
\begin{center}
\includegraphics[width=0.75\textwidth] {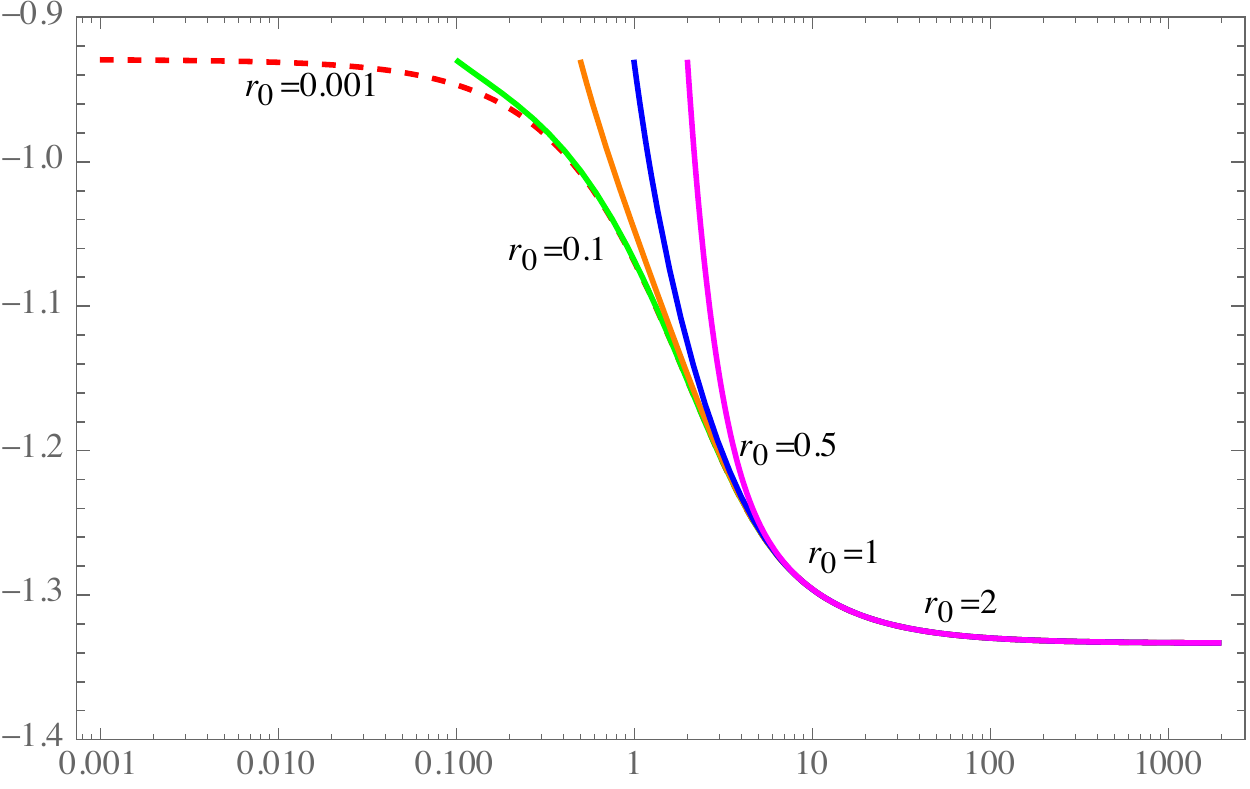}
\caption{(Color online) The attractor solution of the two-moment truncations for various values of $r_0$ (from $0.001$, $0.1$, $0.5$, $1.0$ to $2.0$, corresponding to
red-dashed, green, brown, blue and magenta lines respectively), as a function of $\tau/\tau_R$ in logarithmic scale (constant $\tau_R$). 
\label{fig:fixedpoint2}
}
\end{center}
\end{figure}
      
The full attractor solution has already been discussed in Sect.~\ref{sec:attractor}. We complete here this analysis by examining the effect of the initial time $\tau_0$, or more properly, the effect of changing the ratio $r_0$ between the collision rate and the expansion rate at the initial time. This is illustrated in Fig.~\ref{fig:fixedpoint2}. If $r_0$ is sufficiently small, there is a time regime dominated by free streaming, that is a regime where the attractor remains in the vicinity of the free streaming fixed point. As $r_0$ increases, this regime gradually disappears, and for $r_0\gtrsim 1$, the initial phase of the attractor is dominated by the effect of collisions, exhibiting a rapid transition  toward the hydrodynamic fixed point. 

\begin{figure}
\begin{center}
\includegraphics[width=0.75\textwidth] {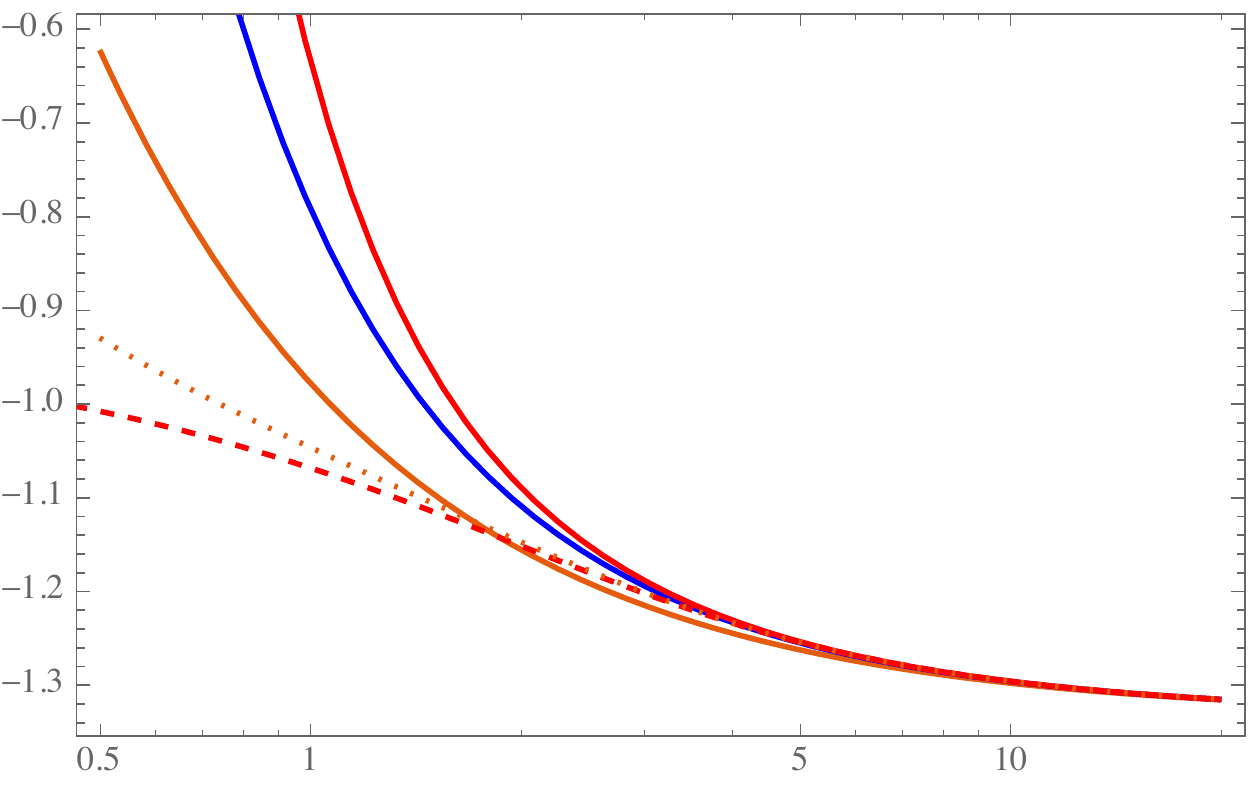}
\caption{(Color online. The attractor solution ot the two-moment truncation for $r_0=0.01$ (dashed line), and $r_0=0.5$ (dotted line),  compared with the gradient expansion of $g_0(w)$ to order 1 (orange), 2 (blue) and 3 (red), as a function of $w=\tau/\tau_R$ in logarithmic scale (constant $\tau_R$). 
\label{fig:attractorb2}
}
\end{center}
\end{figure}

Note that the attractor near the hydrodynamic fixed point becomes insensitive to the starting point after a few collisions. When this happens, the attractor is well described by the first few terms in the gradient expansion, as illustrated in Fig.~\ref{fig:attractorb2}. However, the deviations become significant as soon as $\tau/\tau_R\lesssim 1$: in this region the attractor starts to feel the effect of the free streaming fixed point (assuming that $r_0$ is small enough), an effect which, of course, cannot be captured by the gradient expansion. 

\subsection{The role of higher moments}   

So far in this section, we have focussed on the two-moment truncation. In this last subsection, we examine the effects of the higher moments, and examine in particular how they can be accounted for by a simple renormalization 
of the equations of the two-moment truncation.

\subsubsection{Corrections to gradient expansion from higher moments}    
The generalization of Eq.~(\ref{eqforbg0}) obtained by keeping the contribution of $\L_2$ reads
\beq
\tau\frac{\rmd g_0}{\rmd \tau}=- \left[g_0^2+(a_0+a_1+w) g_0+a_0a_1-c_0b_1+a_0 w\right] +c_0c_1 \frac{\L_2}{\L_0}.
\eeq
In this form this equation generalizes Eq.~(\ref{eqg0FSfp}) obtained in the free streaming case. However the role of the last term, proportional to $\L_2/\L_0$ is here more subtle. Indeed, while in the free streaming case the ratio $\L_2/\L_0$ is a constant near the fixed points, here, near the hydrodynamic fixed point, $\L_2/\L_0\sim -b_1 b_2/w^2$, and this affects the gradient expansion. We shall verify, however, that as $\L_2/\L_0$ vanishes as $1/w^2$ in the vicinity of the hydrodynamic  fixed point,  the contribution of $\L_2$  does not affect the value of $g_0$ at the fixed point, nor its first order correction. 

To proceed, we start by rewriting the equations (\ref{eq:l0l1b}) in terms of the logarithmic derivatives (\ref{logderivative}). We get
\beq\label{eqsforg0g1}
&&g_0+a_0+c_0\frac{\L_1}{\L_0}=0\nn
&&g_1+a_1+w+b_1\frac{\L_0}{\L_1}+c_1\frac{\L_2}{\L_1}=0.
\eeq
Ignoring temporarily the contribution proportional to $\L_2$ in the second equation, one can eliminate the ratio $\L_1/\L_0$ between the two equations, and  obtain 
 \beq\label{relg0g1}
g_0 = -a_0 + \frac{b_1 c_0}{a_1+g_1+w}.
\eeq
This relation, which is an exact relation in the two-moment truncation, 
allows us to recover the first and second order terms of the gradient expansion of $g_0$:
\beq
g_0 \simeq -a_0+\frac{b_1 c_0}{w}\left(1-\frac{a_1+g_1}{w}+\cdots\right),
\eeq
where the first two terms are independent of the choice of  $\tR$. In the second order term ($\sim 1/w^2$), we can replace $g_1$ by $g_1(\infty)$, and get for the coefficient of $1/w^2$, $-(a_1+g_1(\infty)) b_1 c_0=\gamma_2$, where $\gamma_2$ is the second order coefficient obtained by other means in Eqs.~\ref{gradientcoefconformal} and \ref{gradientcoeffconstanttau}. 

In order to estimate the effect of $\L_2$ on the gradient expansion, we start from the equation for $\L_2$ 
\beq\label{eqforL2}   
-\tau\frac{\del \L_2}{\del \tau}=a_2\L_2+b_2\L_1+c_2\L_3+w\L_2, 
\eeq
from which, dividing by $\L_2$ and neglecting the contribution proportional to $\L_3$, we get
\beq\label{estimateL2L10}
-g_2=a_2+b_2\frac{\L_1}{\L_2}+w,\qquad \frac{\L_1}{\L_2}=-\frac{1}{b_2}(g_2+a_2+w).
\eeq
We can then use this result in Eqs.~(\ref{eqsforg0g1}), and get an ``improved'' expression for  $g_0$,
\beq    
 g_0 =-a_0 +\cfrac{b_1c_0}{g_1+a_1+w - \cfrac{b_2 c_1}{g_2 + a_2+w}}.
\eeq
This equation allows us to obtain the gradient expansion of $g_0$ up to order $1/w^3$. By expanding to the required order, we get
\beq
g_0=-a_0+\frac{b_1c_0}{w}\left[1-\frac{g_1+a_1}{w}+\left(\frac{g_1+a_1}{w}\right)^2+\frac{b_2c_1}{w^2}+\ldots\right].
\eeq
To obtain this result, we have used the leading order behavior  $g_2+a_2\to g_2(\infty)+a_2$, and ignored this constant term in replacing $b_2c_1/(g_2+a_2+w)\to b_2c_1/w$.

The coefficient of the second order term involves $g_1(\infty)$ and equals $\gamma_2$, as we have just mentioned. The coefficient of the third order term requires the expansion of $g_1+a_1$ to order $1/w$, which can be obtained in the two-moment truncation and is given explicitly in Eq.~\ref{gradg1}. We then get, for the full contribution of order $w^{-3}$ to $g_0$, 
\beq
b_1c_0\left[\frac{(\gamma_1\gamma_3-\gamma^2_2)b_1c_0}{\gamma_1^3}+\left(\frac{\gamma_2b_1c_0}{\gamma_1^2}\right)^2+b_2c_1\right]=\gamma _3+b_1c_0b_2c_1,
\eeq
revealing, in addition to the term $\gamma_3$ coming from the expansion of $g_1+a_1$, the additional contribution coming from the moment $\L_2$.\footnote{It can be verified that this complete third order contribution agrees with that given in Ref.~\cite{Heller:2018qvh}.} The latter, proportional to $b_1c_0 b_2c_1$, that is to the coefficients that couple $\L_0$ to $\L_1$ and $\L_1$ to $\L_2$, reflects the indirect, dynamical, origin of this contribution. We shall return to these dynamical corrections in the next section, in the broader context of viscous hydrodynamics. At this point, we shall examine how they can be handled as a simple renormalization of the two-moment truncation.

\subsubsection{Renormalized relaxation time from higher moments}
Let us then return to the equation for $\L_1$, which we write as follows
\beq\label{renormalizedtwomoments}
\del_\tau \L_1=-\frac{1}{\tau}\left(a_1 \L_1+b_0\L_0\right) -\left[ 1+\frac{c_1}{w}\frac{\L_2}{\L_1}  \right]\frac{\L_1}{\tau_R}.
\eeq
This writing suggests interpreting the effect of $\L_2$ as a correction of the relaxation time $\tau_R$ (or equivalently of the viscosity $\eta/s=\tau_R T/5$), viz., $ \tau_R\to Z_{\eta/s}  \tau_R$, with \cite{Blaizot:2017ucy}
\beq
\label{eq:zetas}
Z_{\eta/s} \equiv \left[1+\frac{c_1}{w}\frac{\L_2}{\L_1}\right]^{-1} \,.
\eeq
Note that since both $c_1= -12/35$ and $\L_2/\L_1$ are negative, $Z_{\eta/s}<1$ corresponds to a lowering of the effective viscosity.  As $w\to\infty$, i.e., in the vicinity of the hydrodynamic fixed point, $Z_{\eta/s}\to 1$, indicating that there the two-moment truncation is accurate.
 \begin{figure}[h]
\begin{center}
\includegraphics[width=0.45\textwidth] {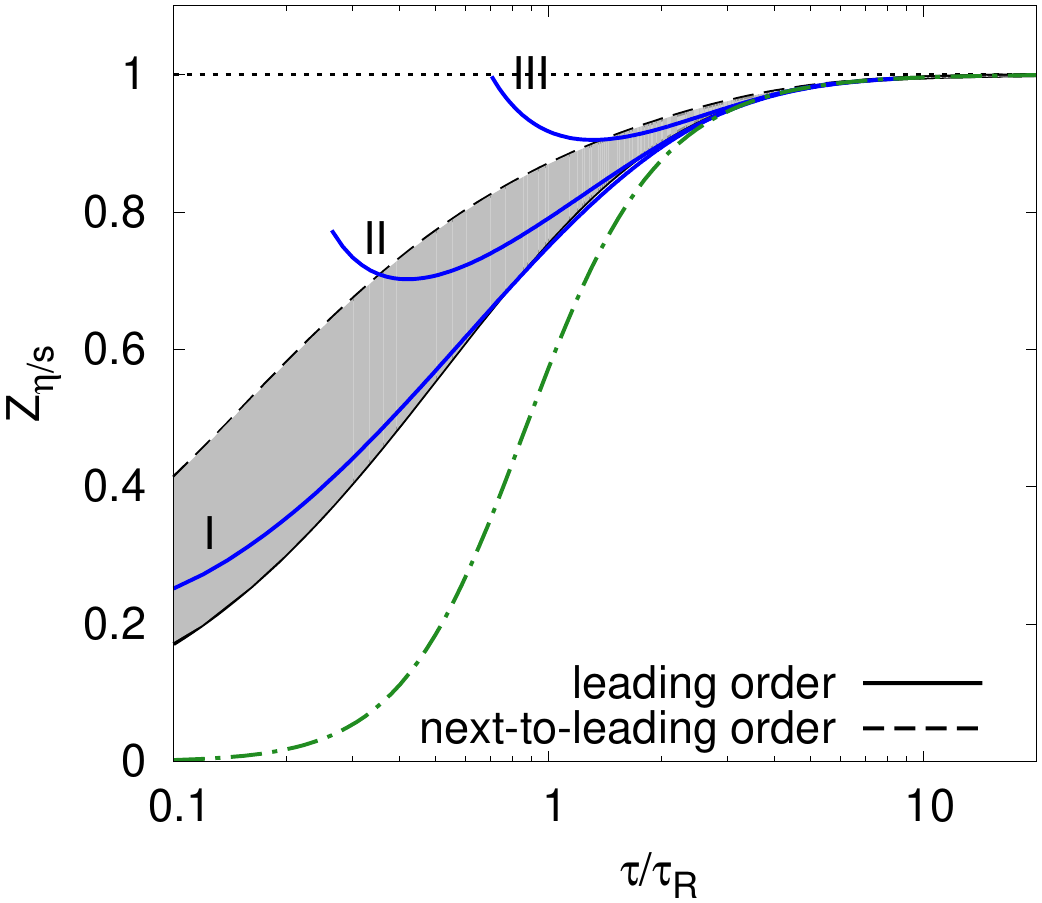}
\includegraphics[width=0.45\textwidth] {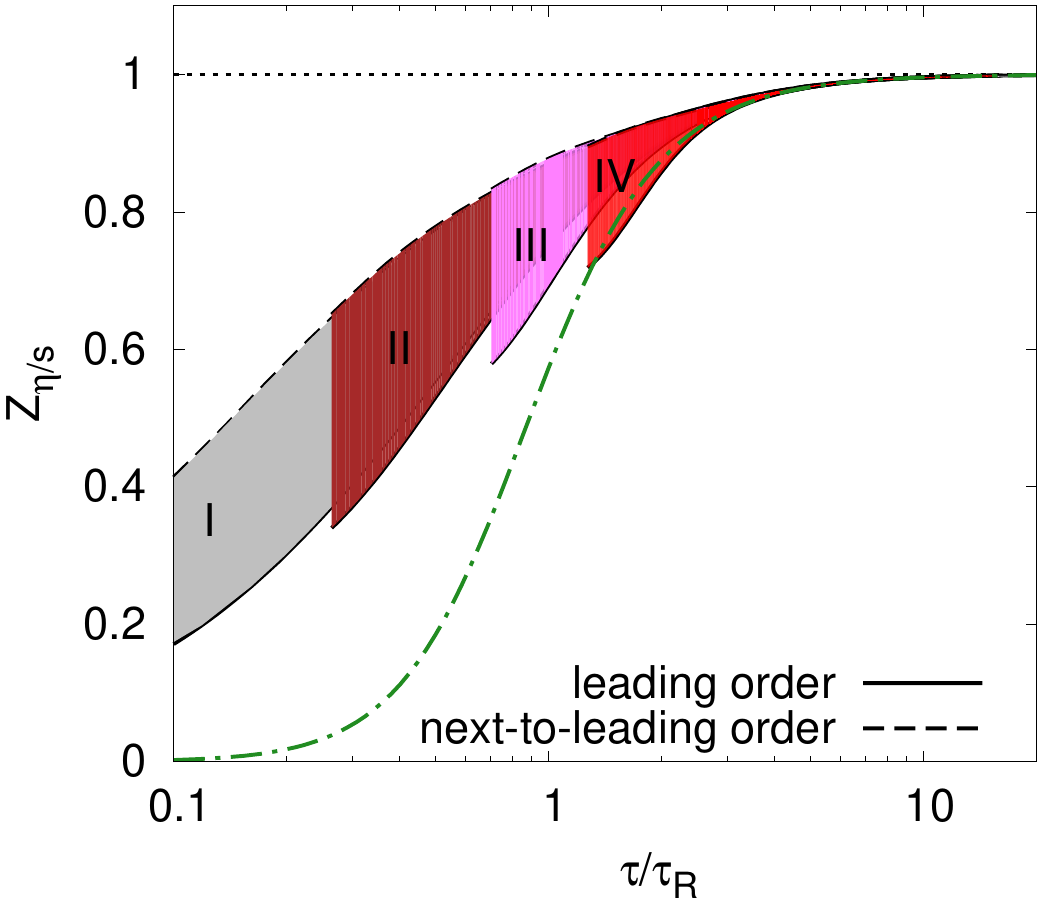}
\caption{(Color online.)   The renormalization factor $Z_{\eta/s}$, \Eq{eq:zetas}, calculated 
according to different approximations for $\L_2/\L_1$. The green dash-dotted
line is obtained from the first two terms in the gradient expansion. The bands correspond to results from the attractor
solution of $g_2$ (leading order) and with both $g_2$ and $g_3$ (next-to-leading order), for  different values of the ratio of $\tau_0/\tR$.  Blues lines are from solving \Eq{eqforL2} as discussed in the text. 
\label{fig:exact3}   
}    
\end{center}
\end{figure}

A plot of the function $Z_{\eta/s}$ as a function of $\tau/\tau_R$ is given in Fig.~\ref{fig:exact3}, for various determinations of $\L_2/\L_1$. The dashed-dotted (green) line corresponds to the   first two terms in the gradient expansion, which can be obtained from the general formulae in Sec.~\ref{sec:asymptkinetic}, or directly from Eq.~(\ref{estimateL2L10}),  
\beq
\frac{\L_2}{\L_1}=-\frac{b_2}{w+g_2+a_2}\simeq -\frac{b_2}{w}\left( 1-\frac{a_2+g_2(\infty)}{w} \right).
\eeq    
         \begin{figure}[h]
\begin{center}
\includegraphics[width=0.5\textwidth] {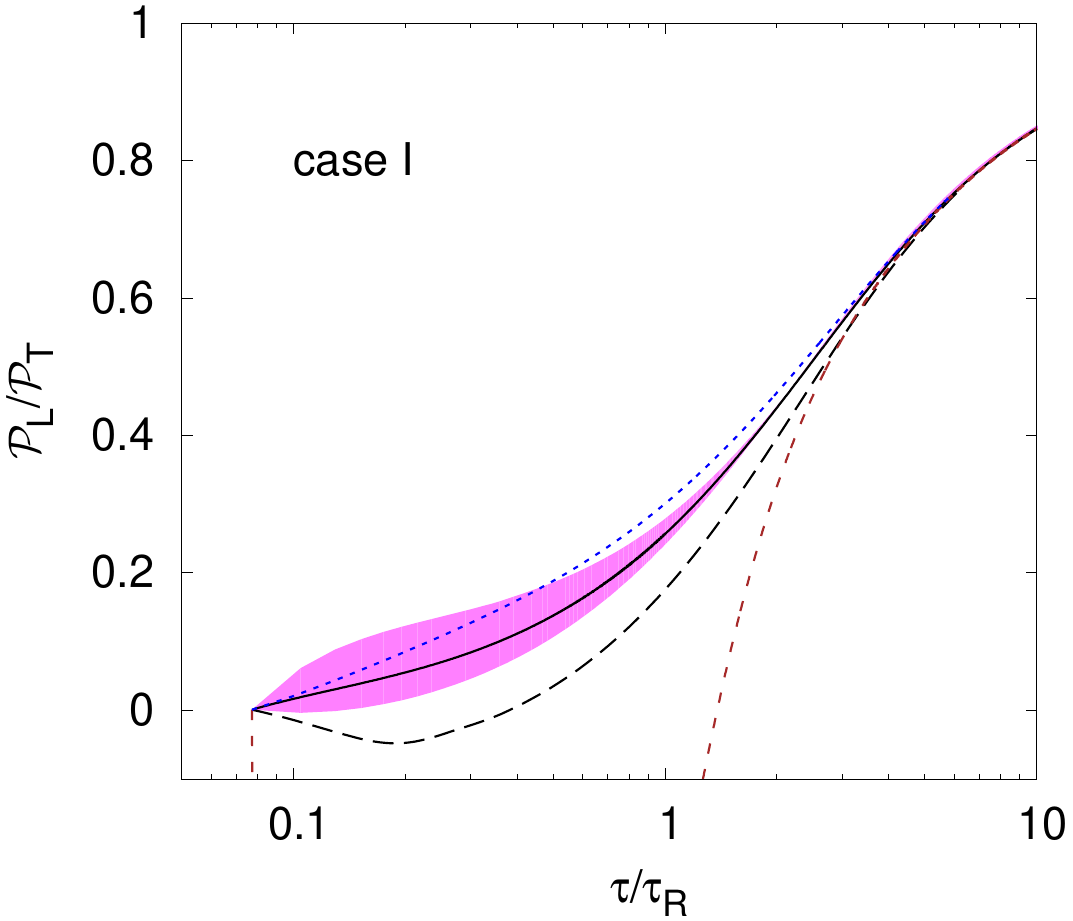}\includegraphics[width=0.5\textwidth] {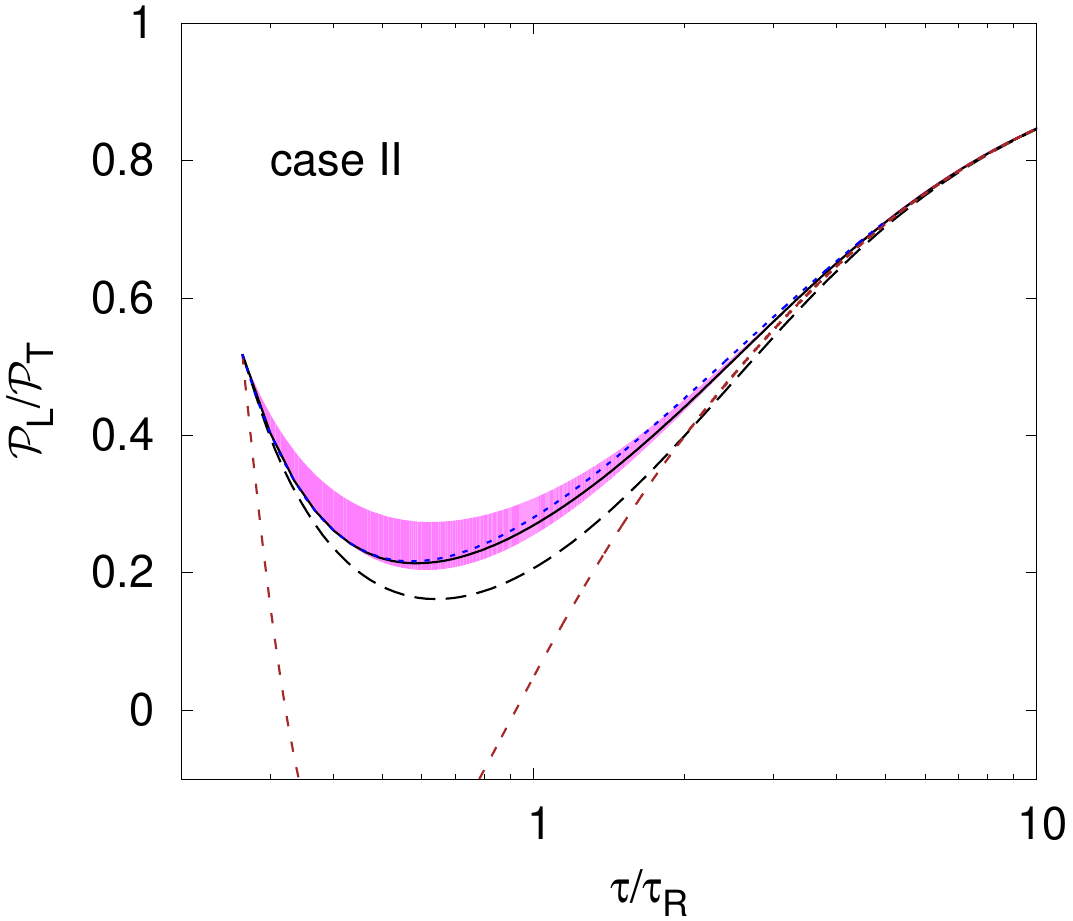}\\
\includegraphics[width=0.5\textwidth] {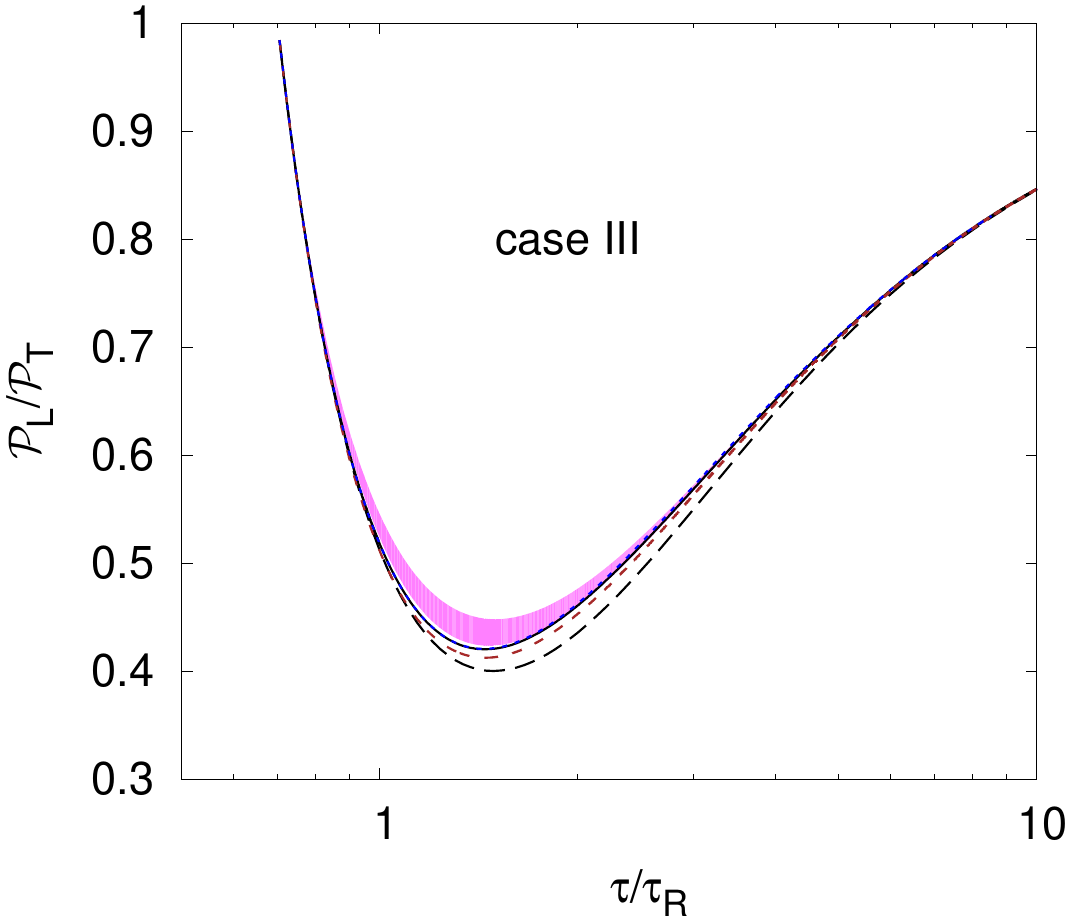}\includegraphics[width=0.5\textwidth] {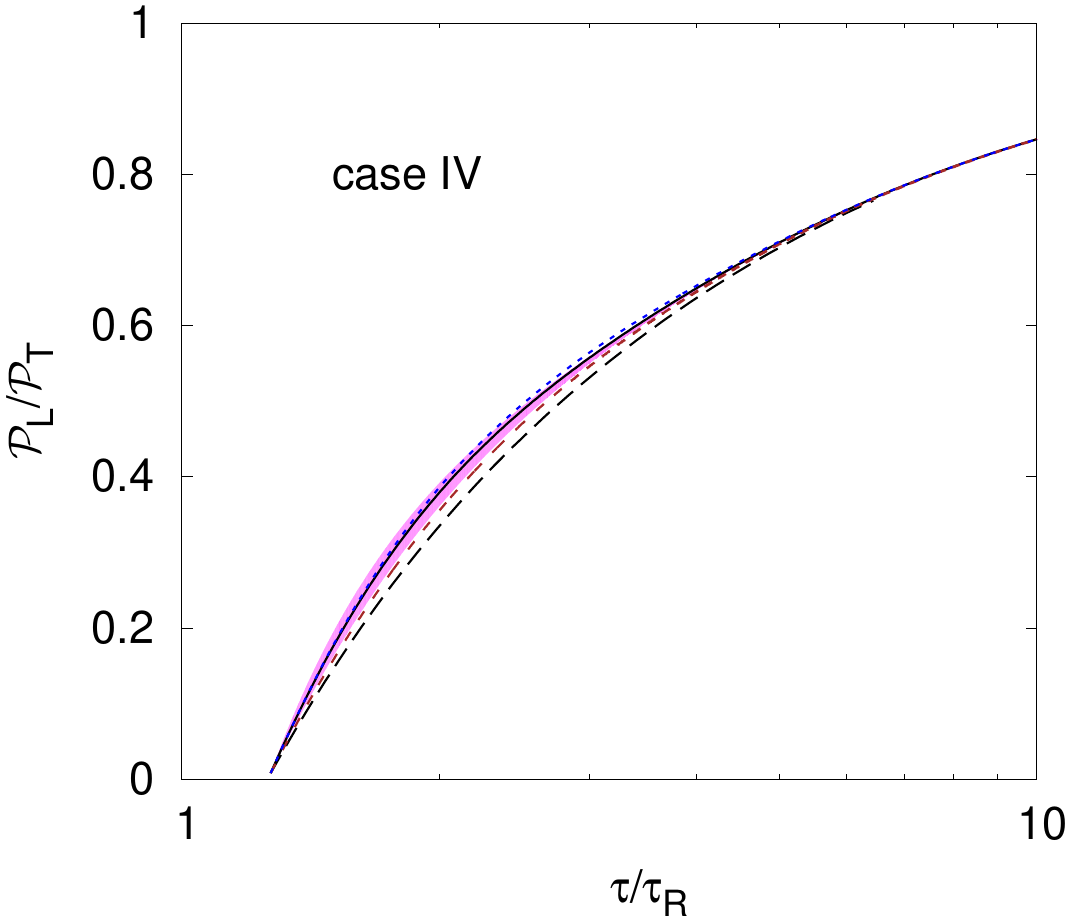}
\caption{(Color online.) Comparison between the exact solution of the kinetic equation (solid 
black lines),
of the two-moment truncation (dashed black lines), and two-moment truncation 
with a renormalization
of $\L_2/\L_1$ obtained via different schemes: by using two-terms in the gradient 
expansion (red dashed
lines),  by using the attractor solutions (pink bands) and solving 
effectively \Eq{eqforL2} without the $\L_3$ term
(blue dotted lines).
\label{fig:exact2}       
}    
\end{center}
\end{figure}  
Of course such an estimate makes sense only when $w\gtrsim 1$, i.e., in the vicinity of the hydrodynamic fixed point. As one moves away from this fixed point, i.e. reaching values $w\lesssim 1$, 
$\L_2/\L_1$ becomes sizeable, the gradient expansion breaks down and does not represent accurately the solution of the kinetic equation: this is indeed the region where the influence of the free-streaming fixed point starts to be felt, and correlatively higher moments begin to play a role. A possible way to encode information about this transition region is to express $\L_2/\L_1$ in term of $g_2 $ (see Eq.~(\ref{estimateL2L10})), 
\be
 \frac{\L_2}{\L_1} = -\cfrac{b_2}{a_2+g_2(w)+w}\,,
\ee
and use for $g_2(w)$ the attractor solution (in place of its gradient expansion). We may also improve on this determination by also including the correction coming from $g_3$, viz.
\be
 \frac{\L_2}{\L_1} = -\cfrac{b_2}{a_2+g_2(w)+w - \cfrac{b_3 c_2}{a_3 + g_3(w) + w}}
\ee
and use for both functions $g_2(w) $ and  $g_3(w)$ the attractor solutions. We refer to these two determinations of $Z_{\eta/s}$ as to leading order and next-to-leading order, respectively \cite{Blaizot:2017ucy}. The results obtained in this way correspond to the grey band in Fig.~\ref{fig:exact3} which shows that for values $w\lesssim 1$, the effective viscosity is substantially reduced by the non equilibrium dynamics \cite{Lublinsky:2007mm}.
 As we have seen in Sec.~\ref{attractorsolution}, the attractor solution depends on the initial time $\tau_0$. This is reflected in the right panel of Fig.~\ref{fig:exact3} where the various areas correspond to the  initial conditions considered in Fig.~\ref{fig:exactI}.    

In practical applications the attractor solutions of $g_2$ may not be available, but a good approximation to $\L_2$ can be obtained by solving Eq.~(\ref{eqforL2}), 
 dropping there the contribution from $\L_3$, and using for $\L_1$ the solution of the two-moment truncation.  By doing so, we are assuming  that  moments of high order ($\L_n$ with $n\ge2$) are mostly determined by the  lowest order ones, while corrections from higher ones are minor. We then solve Eq.~(\ref{eqforL2}) as we just indicated, for the various initial conditions of Fig.~\ref{fig:exactI}.  This yields the blue curves in Fig.~\ref{fig:exact3}. The solution depends on the initial conditions, although this dependence quickly disappears when $w\gtrsim 1$.

       To appreciate the impact of the correction factor, we have solved the corresponding ``improved'' equations of the two-moment truncation, that is, injecting into Eq.~(\ref{renormalizedtwomoments}) the value of the factor $Z_{\eta/s}$ determined by the methods indicated above. The results are displayed in Fig.~\ref{fig:exact2}. One sees that, except for the determination based on the first two terms in the gradient expansion,  the exact solution is accurately reproduced with all other methods, and  the corrections  represent in most cases an improvement of the two-moment truncation.  One should emphasize however that the interpretation of the present correction to Eq.~(\ref{renormalizedtwomoments}) in terms of a renormalized viscosity truly makes sense as long as  $Z_{\eta/s}$  is not too small. 
       
       In summary of this section, we have seen that the simple two-moment truncation, which involves only the monopole and the quadrupole components of the distribution function, that is the energy density and the pressure difference $\P_L-\P_T$,  describes rather accurately the whole evolution of the expanding system, and this from the pre-equlibirum, early time regime, all the way to the late time hydrodynamic regime. We shall see in the next section that the two-moment truncation contains exactly the second order viscous hydrodynamics, while the corrections coming from the coupling to higher moments correspond to higher order viscous hydrodynamics. As was shown in Fig.~\ref{fig:exactI} the hydrodynamic regime starts when the pressures are not fully isotropic, which does not imply a large value of the moment $\L_1$. Rather, one finds that hydrodynamics begins when the collision rate becomes comparable to the expansion rate. Coupling to higher moments represents small corrections, but these become large when the expansion rate becomes large compared to the collision rate. Then the hydrodynamic description breaks down, but the dynamics remains well captured by the two-moment truncation.\footnote{To some extent, the two-moment truncation bears some similarity with the so-called anisotropic hydrodynamics \cite{Florkowski:2010cf,Martinez:2010sc}. The coupled equations for $\L_0$ and $\L_1$ capture essentially the same physics as the ``background'' of anisotropic hydrodynamics.}   
   
\section{Hydrodynamics}\label{sec:hydro}
          
In this last section, we exploit the simplicity of the moment equations in order to make closer contact between kinetic theory and hydrodynamics. In particular we use the two-moment truncation, and the corrections arising from the second moment, in order to recover in a simple way  known results from second and third order viscous hydrodynamics. In spite of the simplicity of the present setting, in which the hydrodynamic fields are function only on the proper time (and have no explicit dependence on transverse spatial coordinates), the basic structure of viscous hydrodynamics and its variants emerges naturally, and the values of the relevant transport coefficients are obtained painlessly.  

\subsection{General comments}

The  standard formulation of hydrodynamics involves the expansion in gradients of the viscous part of the energy-momentum tensor, $\pi^{\mu\nu}$. The first order
viscous correction 
involves the first order gradients of the flow four-velocity $u^\mu$, and reads
\beq
\label{eq:etamn}
\pi^{\mu\nu}=\eta \sigma^{\mu\nu}=2\eta \bra \nabla^\mu u^\nu \ket\equiv
\eta \left(\nabla^\mu u^\nu + \nabla^\nu u^\mu - \frac{2}{3} \Delta^{\mu\nu} \nabla\cdot u\right)\,,
\eeq
where the tensor $\Delta^{\mu\nu}=g^{\mu\nu}-u^\mu u^\nu$ projects on directions orthogonal to $u^\mu$ ($u_\mu\Delta^{\mu\nu}=0$).
In \Eq{eq:etamn} and in the following, the tensor structure in angular brackets is
defined as symmetric, traceless and transverse to the 
flow four-velocity $u^\mu$. That is for any second-rank tensor $A^{\mu\nu}$ one defines \cite{Baier:2007ix}
\beq
\langle A^{\mu\nu} \rangle=\frac{1}{2} \Delta^{\mu\alpha}\Delta^{\nu\beta}\left(A_{\alpha\beta}+A_{\beta\alpha}  \right)-\frac{1}{3} \Delta^{\mu\nu}\Delta^{\alpha\beta} A_{\alpha\beta} 
\eeq    
 For a conformal system, 
the  second order terms are constrained by symmetry \cite{Baier:2007ix}, and result in five independent transport coefficients. In the present boost invariant setting, these reduce to two independent coefficients, $\lambda_1$ and $\tau_\pi$, that enter the expansion of $\pi^{\mu\nu}$ as follows \cite{Baier:2007ix}
\begin{align}
\label{eq:tmnBRSSS0}
\pi^{\mu\nu} =\; &\eta \sigma^{\mu\nu} 
-\eta \tau_\pi \left[\bra D\sigma^{\mu\nu}\ket +\frac{1}{3} \sigma^{\mu\nu}\nabla\cdot u
\right]+\lambda_1\langle \sigma^\mu_{\,\lambda}\sigma^{\nu\lambda} \rangle
+ O(\nabla^3)\,,
\end{align}
where we have set $D\equiv u^\mu\partial_\mu$ (in the local rest frame, this operator reduces to the time derivative, i.e., $D=\del_\tau$).
At this point it is customary to use the leading order relation, 
$\pi^{\mu\nu}=\eta\sigma^{\mu\nu}$,
in order to replace $\sigma^{\mu\nu}$ by $\pi^{\mu\nu}$ in the second order terms, which is legitimate since the difference is a contribution of higher order in the gradient expansion. In doing so, we need the derivative of the viscosity which is estimated from the leading order equation of motion, again a legitimate operation at the considered order.  That is, one assumes that $\eta\sim s \sim T^3$, and estimates  $D\eta=3\eta D \ln T$ from  the leading order equation of motion
\beq
D\ln T=-\frac{1}{3} \nabla\cdot u, \qquad \del_\tau\ln T=-\frac{1}{3 \tau},
\eeq
where the second equation holds in the local rest frame. 
One then gets
\begin{align}
\label{eq:tmnBRSSS}
\pi^{\mu\nu} =\; &\eta \sigma^{\mu\nu} 
-\tau_\pi \left[\bra D \pi^{\mu\nu}\ket
+\frac{4}{3}\pi^{\mu\nu}\nabla\cdot u\right]+\frac{\lambda_1}{\eta^2}\langle \pi^\mu_{\,\lambda}\pi^{\nu\lambda} \rangle
+ O(\nabla^3)\,.
\end{align}
\Eq{eq:tmnBRSSS} generalizes the M\"uller-Israel-Stewart hydrodynamics in which
the only second order transport coefficient is the relaxation time $\tau_\pi$ (the coefficient of $\del_\tau$ in the local rest frame).
It is useful to recall how this equation has been obtained.
First, the term quadratic in $\pi$ in 
\Eq{eq:tmnBRSSS} results from the substitution of the leading order  constitutive equation $\pi^{\mu\nu}=\eta\sigma^{\mu\nu}$ in the original equation (\ref{eq:tmnBRSSS0}).
Second, \Eq{eq:tmnBRSSS} is written in such a way that the coefficients of $\tau_\pi$ and $\lambda_1$ transform  separately 
homogeneously under Weyl transformations \cite{Baier:2007ix}. As a result, in addition to $\tau_\pi$,
another second order transport coefficients, $\lambda_1$, appears in the description of a conformal fluid \cite{York:2008rr}. These coefficients have been calculated in kinetic theory, and for massless particles, 
one finds\footnote{These values correspond to a momentum independent relaxation time, hence directly comparable to those that we can extract from our equations.} \cite{Teaney:2013gca}      
\beq
\label{eq:2nd_trans}
\tau_\pi = \frac{5\eta}{sT}\,,
\qquad
\lambda_1=\frac{5}{7}\eta\tau_\pi\,.
\eeq
       
      Equation~(\ref{eq:tmnBRSSS}) can be simplifed in the case of Bjorken flow, where the gradients of the flow four-velocity
are proportional to $1/\tau$. For instance, using coordinates $(\tau, \xi)$, with $t=\tau \cosh \xi$, $z=\tau \sinh\xi$, we get   
\beq
\label{eq:bjork_amb}
\sigma^\xi_{\;\;\xi} = -\frac{4}{3\tau}\,,
\qquad\nabla\cdot u = \frac{1}{\tau}\,,
\eeq 
where we have used the fact that the only nonzero component of $\nabla_\mu u_\nu$ is $\nabla_\xi u_\xi=\tau$.
In fact, for Bjorken flow, there is only one 
independent component of the viscous tensor that is allowed by symmetry. Then, defining $\Pi = \pi^\xi_{\;\;\xi} = -2\pi^x_{\;\; x}= -2\pi^y_{\;\;y}$, one can rewrite 
\Eq{eq:tmnBRSSS} in the simpler form
\beq
\label{eq:eomBRSSS}
\Pi = -\frac{4\eta}{3\tau} - \tau_\pi\left[\partial_\tau \Pi + \frac{4}{3}\frac{\Pi}{\tau} \right] 
+\frac{\lambda_1}{2 \eta^2} \Pi^2 .
\eeq
In the hydrodynamic regime, 
$\Pi$ can be identified to the pressure difference and thus to $\L_1$, viz.
\be
\label{eq:Pihydro}
\Pi =  \frac{2}{3}(\P_L-\P_T) = c_0\L_1\,.
\ee

As we have just recalled, Eq.~(\ref{eq:eomBRSSS}) has been obtained after simplifications that involve the use of both the leading order equation of motion, and the leading order relation between $\pi^{\mu\nu}$ and  the shear tensor $\sigma^{\mu\nu}$. In terms of the $\L$-moments, as we shall see more explicitly in the next subsection, these manipulations involve both the direct expansion of $\L_1$ in gradients, the constitutive equation,  and the  mixing, via the equations of motion, of  terms coming form the expansion  of $\L_2$ to the same order. This is manifest in the fact that two independent linear combinations of the  transport coefficients $\lambda_1$ and $\tau_\pi$ appear in the Chapman-Enskog expansions of $\L_1$ and $\L_2$, as shown in \cite{Blaizot:2017lht} (see also Appendix~\ref{Chapman}). More precisely, and in the notation of the present paper, these two linear combinations are 
\beq\label{lambda1eta}
&&\frac{4}{3\tau^2}(\lambda_1-\eta\tau_\pi)=\frac{\alpha_1^{(2)}}{\tau^2},\nn
&&\frac{4}{3\tau^2}(\lambda_1+\eta\tau_\pi)=\frac{\alpha_2^{(2)}}{\tau^2}.
\eeq
where, in the conformal case (see Sec.~\ref{sec:asymptkinetic})
\beq
&&\alpha_1^{(2)}=-\tau_R^2\, \varepsilon\, b_1(2-a_1)=-\frac{32}{315} \tau_R^2\, \varepsilon,\nn
&& \alpha_2^{(2)}=\tau_R^2\, \varepsilon\,b_1b_2=\frac{64}{105} \tau_R^2\, \varepsilon.
\eeq
It follows that $\lambda_1/\tau_R^2\, \varepsilon=4/21$, and $\eta\tau_\pi/\tau_R^2\, \varepsilon=4/15$, so that in particular $\lambda_1/\eta\tau_\pi=5/7$, in agreement with (\ref{eq:2nd_trans}).

Without the constraint
of conformal covariance,
 the form appearing in \Eq{eq:eomBRSSS}  is not unique. 
 For instance,  the following form of hydrodynamic equation of 
motion is advocated in Refs.~\cite{Denicol:2012cn} and \cite{Jaiswal:2013vta}
\beq
\label{eq:eomMIS}
\Pi = -\frac{4\eta}{3\tau} - \tau_\pi \partial_\tau \Pi - \beta_{\pi\pi}\tau_\pi\frac{\Pi}{\tau}
-\frac{\chi\tau_\pi^2}{\eta}\frac{\Pi^2}{\tau} \,,
\eeq
where $\beta_{\pi\pi}$ and
$\chi$ are dimensionless second order and third order transport coefficients, respectively.
For massless Bosons these coefficients can be evaluated in kinetic theory and found to be
~\cite{Jaiswal:2013vta}
\beq
\beta_{\pi\pi}=\frac{38}{21}\,,\qquad
\chi=\frac{72}{245}\,.
\eeq

What we shall do in the rest of this section is to show how these simple forms of viscous hydrodynamic equations emerge from the moment equations, with the appropriate values of the transport coefficients given in terms of the coefficients $a_n, b_n, c_n$. In fact the second order viscous hydrodynamic equations involve only the two-moment truncation, while the moment $\L_2$ enters explicitly the third order equation. 
    
\subsection{Second order viscous hydrodynamics from the two-moment truncation}

We start with the equations of the two-moment truncation, Eqs.~(\ref{eq:l0l1}). By using the relation (\ref{eq:Pihydro}), $\Pi = c_0\L_1$,    
one can rewrite these equations  as follows, using a notation more familiar in the hydrodynamic context  
\begin{subequations}
\label{eq:l0l1_mis}
\begin{align}
\label{eq:l0l1_mis_a}
\partial_\tau\varepsilon =&\;-\frac{4}{3}\frac{\varepsilon}{\tau} - \frac{\Pi}{\tau}\,,\\
\label{eq:l0l1_mis_b}
\Pi =&\; -\frac{\tR}{\tau} b_1 c_0 \epsilon - \tR \partial_\tau\Pi - a_1 \frac{\tR}{\tau} \Pi
\quad
\leftrightarrow
\quad
\Pi=
-\frac{4\eta}{3\tau} - \tau_\pi \partial_\tau \Pi - \beta_{\pi\pi}\tau_\pi\frac{\Pi}{\tau}.
\end{align}
\end{subequations}     
Note that to obtain the second equation, we have used the leading order relation (\ref{recursionBnn}) to eliminate $\L_0$ in the equation for $\L_1$, as well as the expression for the viscosity $\eta=(b_1/2)\varepsilon \tau_R$.    
Except for the third order viscous term in \Eq{eq:eomMIS}, \Eq{eq:l0l1_mis_b} and \Eq{eq:eomMIS} are identical, provided conformal symmetry is realized, so that  
\be 
\epsilon=3\P \quad\mbox{and}\quad \tau_\pi={5\eta\over sT}=\tau_R\,.
\ee
In addition, we notice that the second order transport
coefficient $\beta_{\pi\pi}$ is precisely the coefficient $a_1$.

In fact,  subtle ambiguities arise when relating  \Eq{eq:l0l1_mis_b} to the hydrodynammic equation. These come in particular  from how one relates the factor $1/\tau$, to tensor structures involving gradients  in the Bjorken flow. Indeed we have 
\be
\label{eq:pi_tau}
\frac{1}{\tau}=\nabla\cdot u =-\frac{3}{4}\sigma^\xi_{\;\xi}= -\frac{3}{4}\frac{\Pi}{\eta}+O(1/\tau^2)
\ee
Since the  leading order relation gives $\Pi
\propto 1/\tau$, we may substitute the factor $1/\tau$ in the last term of \Eq{eq:l0l1_mis_b} by either 
$\nabla\cdot u$ or $\sigma=\sigma^\xi_{\;\xi}$, or any linear combination of these, and obtain  equivalent results at order $1/\tau^2$. Additionally,
the substitution between $1/\tau$ and 
$\Pi$ (or $\L_1/c_0$) among second order terms is also allowed, since such substitutions only modify the equation with viscous corrections at the next order. 
As already mentioned, such  
ambiguities are fixed in the BRSSS hydrodynamics \cite{Baier:2007ix} by 
requiring that the stress tensor  be homogeneous under scale transformations, which then amounts to
consider in the equations only two possible second order terms,
\be
\partial_\tau \Pi + 
\frac{4}{3}\frac{\Pi}{\tau}\quad\mbox{and}\quad   
\Pi^2\,.
\ee
Applying this strategy to \Eq{eq:l0l1_mis_b}  one obtains then
\be
\label{eq:l0l1_brsss}
\Pi = -\frac{4}{3}\frac{\eta}{\tau} 
- \tR \left(\partial_\tau\Pi+\frac{4}{3}\frac{\Pi}{\tau}\right) 
+ \frac{3}{4}\left(a_1-\frac{4}{3}\right) \frac{\tR}{\eta} \Pi^2 + O\left(\frac{1}{\tau}\right)^3\,,
\ee
where the last term in \Eq{eq:pi_tau} has been used. \Eq{eq:l0l1_brsss} is nothing but the BRSSS
hydrodynamic  equations of motion, Eq.~(\ref{eq:eomBRSSS}), with the second order transport coefficient identified as
\be\label{lambda1a}
\lambda_1 = \frac{3}{2}\left(a_1-\frac{4}{3}\right) \eta\tau_\pi=\frac{5}{7}\eta\tau_\pi\,,
\ee
in agreement with \Eq{eq:2nd_trans}.

In all the  derivations of this subsecion, only the leading order term of $\L_1$ in 
the $1/\tau$ expansion has been taken into account, as we have emphasized. The role of higher terms and the moment $\L_2$ will be discussed in the next subsection. \\

\subsection{Third order viscous hydrodynamics, and the effect of the moment $\L_2$}

Third order viscous corrections from the moment equations can be found by
considering the equations of the three-moment truncation
\beq\label{threemoment}
\partial_\tau \L_0 + \frac{1}{\tau}(a_0 \L_0 + c_0 \L_1) =&\; 0\,,\nn
\partial_\tau \L_1 + \frac{1}{\tau}(b_1 \L_0 + a_1 \L_1 + c_2 \L_2) =&\; -\frac{\L_1}{\tR}\,,\nn
\partial_\tau \L_2 + \frac{1}{\tau}(b_2 \L_1 + a_2 \L_2) =&\; -\frac{\L_2}{\tR}\,.
\eeq
It is convenient to define a new quantity
\beq
\Sigma = c_0 \L_2,
\eeq
in analogy  to  Eq.~(\ref{eq:Pihydro}) relating $\L_1$ to $\Pi$. 
On can then rewrite the system of equations (\ref{threemoment}) as follows
(here, we identify $\tau_R$ and $\tau_\pi$ a priori), 
\begin{subequations}
\label{eq:3rd_hydro}
\begin{align}
\partial_\tau \varepsilon \;+\;& \frac{4}{3}\frac{\varepsilon}{\tau} = -\frac{\Pi}{\tau}\,,\\
\label{eq:3rd_hydro_b}
\Pi=&\;
-\frac{4\eta}{3\tau} - \tau_\pi \partial_\tau \Pi - a_1\tau_\pi\frac{\Pi}{\tau}
+\frac{c_1 \tau_\pi\Sigma}{\tau}\,,\\
\label{eq:3rd_hydro_c}
\Sigma=&\;-\frac{\tau_\pi b_2\Pi}{\tau}-\tau_\pi\left(\frac{a_2 \Sigma}{\tau}
+\partial_\tau\Sigma\right)\,.
\end{align}
\end{subequations}
When restricted to the late time behavior, these equations present a systematic generalization of the second order viscous 
hydrodynamics, with the third order viscous correction ($\Sigma$)
adding to the second order equations of motion.
From \Eq{eq:3rd_hydro_c}  $\Sigma$ appears naturally as a new dynamical variable (an explicit time derivative is present in Eq.~(\ref{eq:3rd_hydro_c}), whose evolution is 
coupled to $\Pi$ and to the energy density. In fact, for the third order hydrodynamics, one only needs the first term in the right-hand side of Eq.~(\ref{eq:3rd_hydro_c}), that is 
\be
\label{eq:leadSigma}
\Sigma\simeq -\tau_\pi b_2 \frac{\Pi}{\tau} = 
\frac{4 b_2}{3}\frac{\eta\tau_\pi}{\tau^2}
= 
\frac{3b_2\tau_\pi}{4}\frac{\Pi^2}{\eta},
\ee
where Eq.~(\ref{eq:pi_tau}) has been used to get the last two relations. 
The other terms in Eq.~(\ref{eq:3rd_hydro_c}) involve indeed higher order terms in the gradient expansion. Note that the second equality in Eq.~(\ref{eq:leadSigma}) is nothing but  the leading order expansion of $\L_2$, \Eq{gradientL10}. More precisely, using Eq.~(\ref{lambda1eta}, on finds  
\be
\frac{\lambda_1}{\eta\tau_\pi}= \frac{b_2}{c_0}-1=\frac{5}{7}\,.
\ee
in agreement with Eq.~(\ref{lambda1a}). Note however the presence of the coefficient $b_2$ in this expression, whereas no trace of the second moment is present in the expression (\ref{lambda1a}). This  is another aspect of the ambiguity alluded to earlier concerning the writing of second order hydrodynamic equations. Here this ambiguity, associated with the possibility of reshuffling various terms using equations of motion or lower order relations, rests on relations between the coefficients $a_n, b_n, c_n$. The relation involved here is $b_2-c_0=a_1-a_0$.  It is because of such relations that the gradient expansion of the solution eventually exists, but we lack insight on the systematics of such relations.

Finally, by using the last expression of $\Sigma$ in \Eq{eq:leadSigma} for the last term in 
\Eq{eq:3rd_hydro_b} one determines the third order transport coefficient  
\be
\chi=-\frac{3}{4}b_2 c_1 = \frac{72}{245}\,,
\ee
which reproduces \Eq{eq:eomMIS}. Note that this particular transport coefficient involves the two coupling constants that relate $\L_1$ and $\L_2$ in their respective equations of motion.

\section{Conclusions}

We have considered a simple kinetic description of a gluonic plasma undergoing boost invariant longitudinal expansion, solving a Boltzmann equation within the relaxation time approximation. By using a special set of moments of the distribution function, we have replaced the kinetic equation by an infinite hierarchy of equations for these moments. We have found that a simple two-moment truncation, that involves only the monopole and quadrupole components of the distribution function, works extremely well, even in the case of purely free-streaming motion where many moments get populated. We argued that this is because the free streaming solution of the hierarchy of equations is controlled by two fixed points which are already present in the two-moment truncation, and whose locations are only moderately modified by the coupling to higher moments. Collisions produce a damping of all higher moments, and drive the system to the hydrodynamic regime, characterized by a fixed point of a different nature, where all the moments decay according to well specified power laws. An attractor solution can be defined  to which any solution converges rapidly. The two-moment truncation is analysed in detail, providing semi-analytic control on most aspects of the solution. The two-moment and three-moment truncations contain all the information needed to derive second and third order viscous hydrodynamics, and provide a direct and simple estimate of the corresponding transport coefficients.

The interest of the kinetic framework is to provide a complete description of the time evolution of the system, from the initial pre-equilibrium regime all the way to the late time hydrodynamic regime. Hydrodynamics emerges, as expected, when a few collisions have had time to occur, that is for times of the order of a few times the relaxation time. Observe that viscous hydrodynamics starts to become an accurate description when the ratio $\P_L/\P_T$ is still far from unity, typically of order $0.6$. Such a value is often considered as  an indication of a strong deviation from local equilibrium. We have observed however that this corresponds to a ratio of the angular moments $\L_1/\L_0\approx 0.15$. From that point of view the deviation is not so large. A similar ratio $\P_L/\P_T\approx 0.6$ is also observed at the onset of viscous hydrodynamics in other pictures of the initial evolution, such as that provided by  holographic, strong coupling, techniques \cite{Heller:2011ju}. The strong coupling picture has no direct connection with the kinetic description, and the similarity of the values of   $\P_L/\P_T$ at the onset of hydrodyamics calls for an interpretation within hydrodynamics itself, that is connect the onset with the point where the gradient expansion starts to diverge strongly. 

Finally, we note that although the present paper has focused on a simple kinetic equation, with a relaxation time approximation for the collision kernel, we believe that many features of our results are robust. We have in fact indications that this is indeed the case from solving the Boltzmann equation with gluon elastic scattering in the small scattering angle approximation. Such results will be reported in a separate publication \cite{BlaizotTanji}.

 
\appendix

\section{Analysis of the function ${\cal F}_n(x)$}\label{App:functionFn}
The function ${\cal F}_n(x)$ is defined in Eq.~(\ref{calFdef}) of the main text, which we reproduce here for convenience: 
\be\label{calFnb}
\mathcal{F}_{n}(x)\equiv\int_{0}^1 \rmd y \left[1-(1-x^2)y^2\right]^{1/2}
P_{2n}\left(\frac{xy}{\left[1-(1-x^2)y^2\right]^{1/2}}\right),
\ee
where $P_{2n}$ is a Legendre polynomial. 

The first moment can be given a simple analytical expression in terms of elementary functions. For instance, 
\beq
\F_0(x)=\frac{1}{2} \left(x-\frac{i \cosh ^{-1}(x)}{\sqrt{1-x^2}}\right),
\eeq
with $\cosh^{-1}(z)$ defined with a branch cut on $(-\infty,1]$. We can also write this function as
\beq
x \F_0(x)=\frac{1}{2}\left[ x^2+\frac{\arctan\sqrt{\frac{1}{x^2}-1}}{ \frac{1}{x^2}-1}\right].
\eeq
The first few moments can then be conveniently expressed in terms of $\F_0$ and its successive derivatives. 
Thus, the first moment is given by      
\beq
\F_1(x)=\frac{3}{2} x \F_0'(x)-\frac{1}{2}  \F_0(x),
\eeq
and the second moment by
\beq
\F_2(x)=\frac{7}{12}\F_0(x)+\frac{5}{12}\F_1(x)-\frac{35}{8}x^2 \F_0''(x).
\eeq
      
\begin{figure}[h]
\begin{center}
\includegraphics[width=0.55\textwidth] {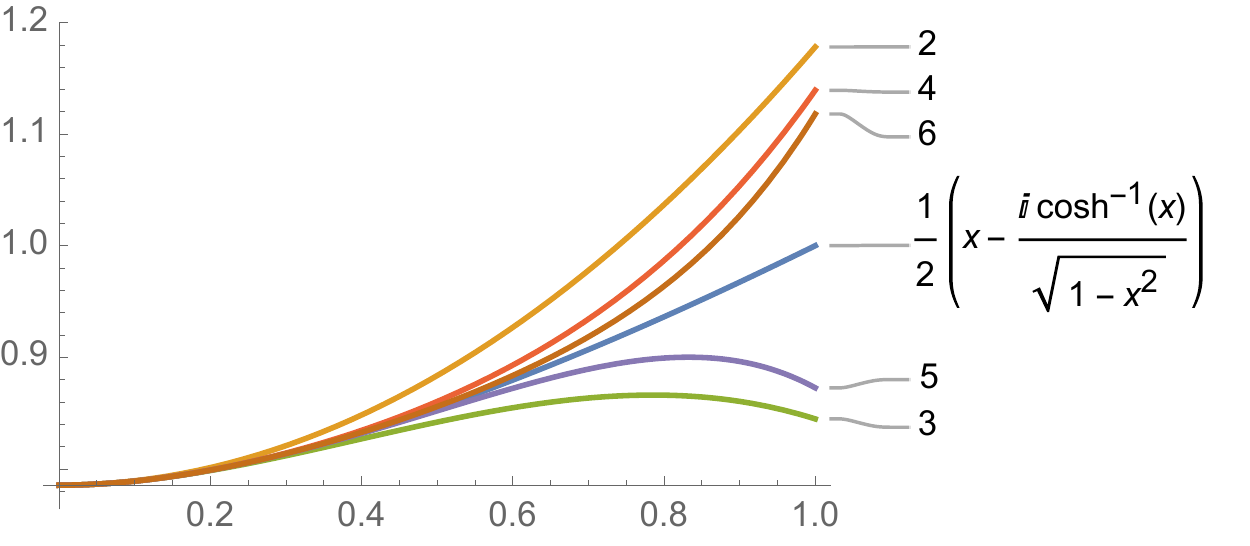} \includegraphics[width=0.40\textwidth] {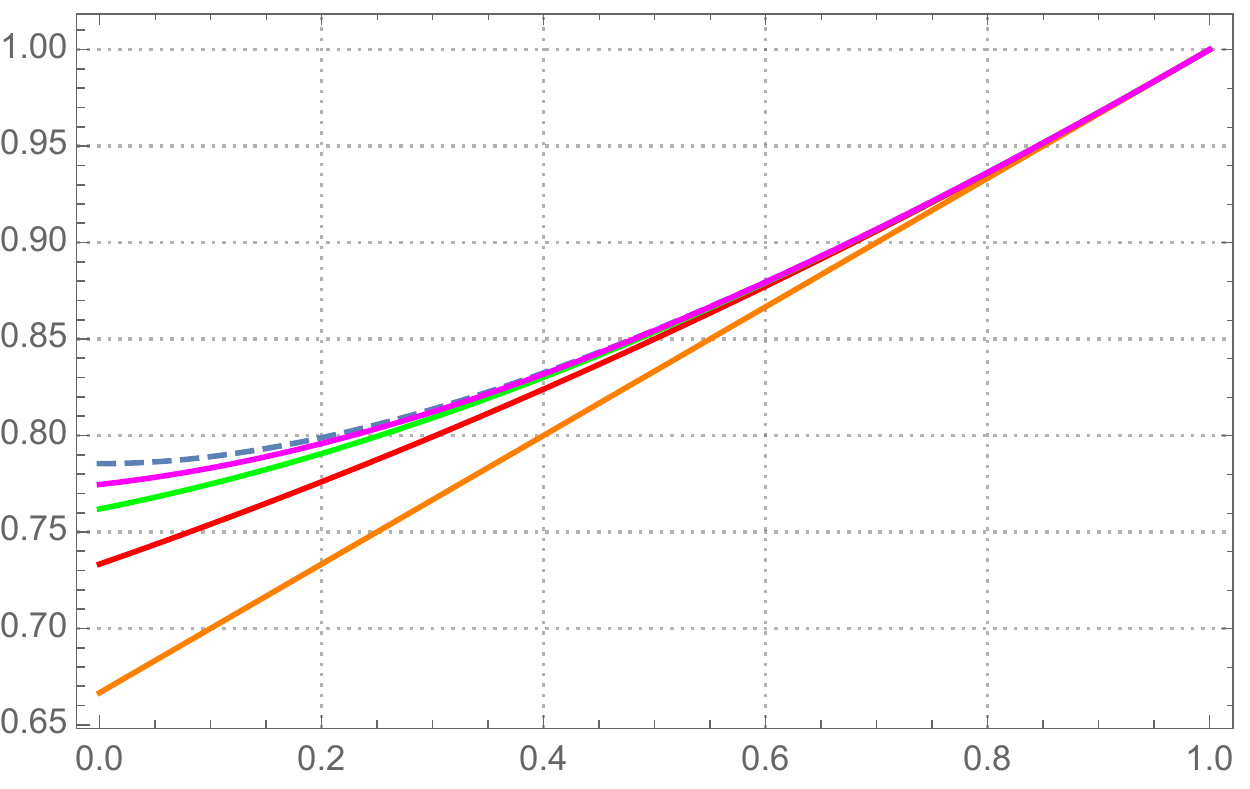}
\caption{(Color online.) The moment $\F_0(x)$ and its expansions to 6th order near $x=0$ (left) and to 4th order near $x=1$ (right). Near $x=1$ the convergence is smooth as the order of the expansion increases (the curves from bottom to top represent orders 1,2,3 and 4, while the dashed line is the exact result).
\label{fig:function_h}
}
\end{center}
\end{figure}

 Let us focus now on the zeroth moment $\F_0(x)$, in order to illustrate a generic properties of the functions $\F_n(x)$, and their expansions near $x=0$ and $x=1$ which exhibit very different convergence behaviors.  As shown in Fig.~\ref{fig:function_h}, while the expansion near $x=1$ exhibits apparent convergence even when extrapolated near $x=0$, this is not the case of the expansion near $x=0$ whose extrapolation near $x=1$ is not smootly convergent. As we shall see, this is related to the non uniform convergence of the integral in Eq.~(\ref{calFnb}). 

 Near $x=1$, the Taylor expansion of $\F_n(x)$ is indeed regular and can, in particular, be obtained simply by    differentiating with respect to $x$ under the integral sign in Eq.~(\ref{calFnb}). We get, for the first few $\F_n$, 
\beq\label{shorttimemoments}
{\cal F}_0(x)&=&1+\frac{1}{3}(x-1)+\frac{1}{15}(x-1)^2-\frac{1}{35} (x-1)^3+\frac{4}{315} (x-1)^4+O(x-1)^5\nn
{\cal F}_1(x)&=&\frac{8}{15}(x-1)+\frac{4}{105}(x-1)^2-\frac{4}{105} (x-1)^3+O(x-1)^4\nn
{\cal F}_2(x)&=&\frac{32}{105}(x-1)^2-\frac{128}{1155} (x-1)^3+\frac{184}{1505} (x-1)^4+O(x-1)^5\nn
{\cal F}_3(x)&=&-\frac{512}{3003} (x-1)^3+O(x-1)^4\nn
{\cal F}_4(x)&=&\frac{2048}{21879} (x-1)^4+O(x-1)^5.
\eeq
Note that the coefficients of the leading contributions are related to the coefficients $b_n$. We have indeed, modulo a sign, 
\beq
\left.\frac{\rmd {\cal F}_n(x)}{\rmd x^n}\right|_{x=1}=\frac{1}{n!}\,b_1\cdots b_n.
\eeq

When $x\to 0$,  the simple strategy of expanding the integrand with respect to $x$ in Eq.~(\ref{calFnb}) does not work, except for the first two coefficients (beyond these first two terms, one generates divergent integrals, and more elaborate techniques must be used). We obtain, for the first few moments (restricting ourselves, for $n\ge 3$, to the first two terms in the small $x$ expansion)
\begin{align}
\F_0(x)=&\frac{\pi}{4}+\frac{\pi x^2}{8}-\frac{x^3}{3}+\frac{3\pi x^4}{32} -\frac{4 x^5}{16} + O(x^6)\nn
\F_1(x)=&-\frac{\pi}{8}+\frac{5\pi x^2}{16}-\frac{4x^3}{3}+\frac{33\pi x^4}{64} -\frac{28 x^5}{15} + O(x^6)\nn
\F_2(x)=& \frac{3\pi}{32}-\frac{57\pi x^2}{64}+8x^3-\frac{1191\pi x^4}{256}+\frac{112 x^5}{5}+ O(x^6)\nn
\F_3(x)=&-\frac{5\pi}{64}+\frac{205\pi x^2}{128}+ O(x^3)\nn
\F_4(x)=&\frac{35\pi}{512}-\frac{2485\pi x^2}{1024}+ O(x^3). 
\end{align}
It is easily verified that the ratios $\F_n(0)/\F_0(0)$ coincide with the coefficients $A_n$ given in Eq.~(\ref{Ans}).

\section{Moment truncations in the free streaming regime}\label{truncationFS}

In this appendix, we complement the discussion in the main text by providing more details on the truncations of the moment equations in the free streaming regime. We first give below the explicit expressions for the moments $\L_n(t)$ corresponding to the truncations involving the first three and four moments, respectively. The results are expressed in terms of $t=\ln(\tau/\tau_0)$.

\begin{figure}
\begin{center}
\includegraphics[width=0.45\textwidth] {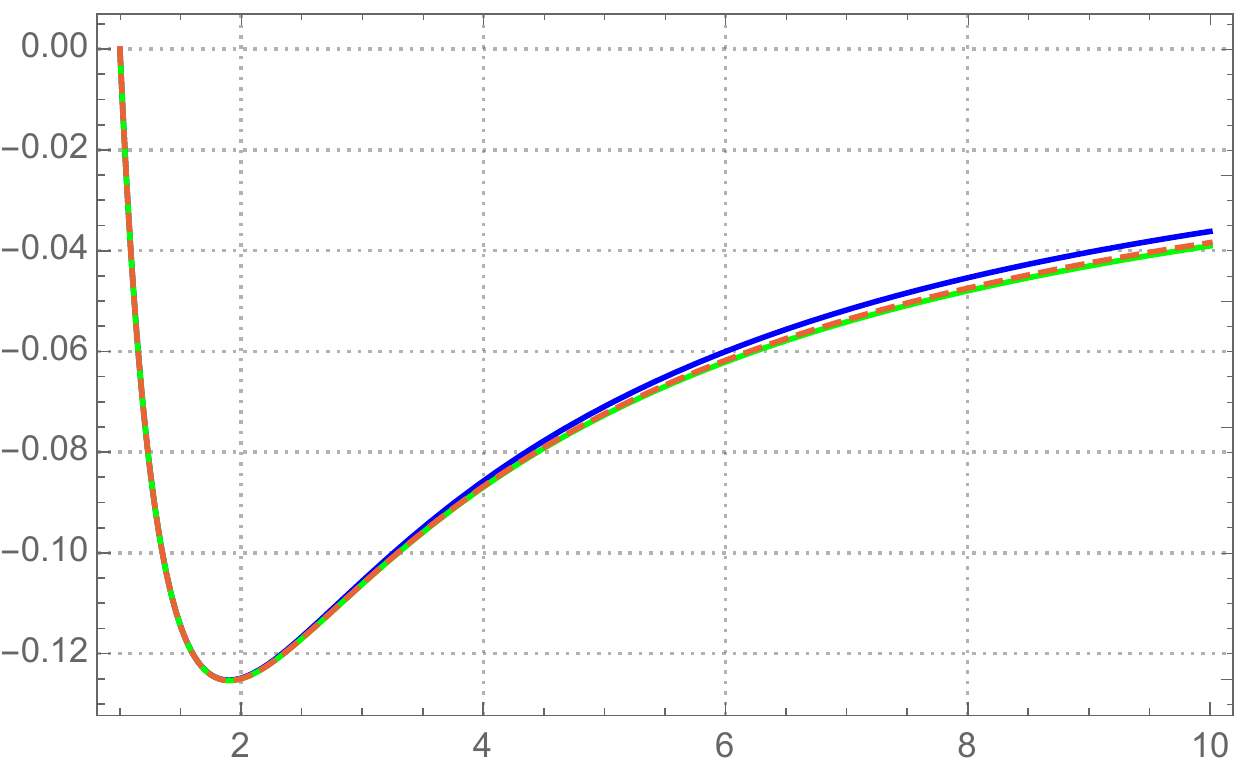}\includegraphics[width=0.45\textwidth] {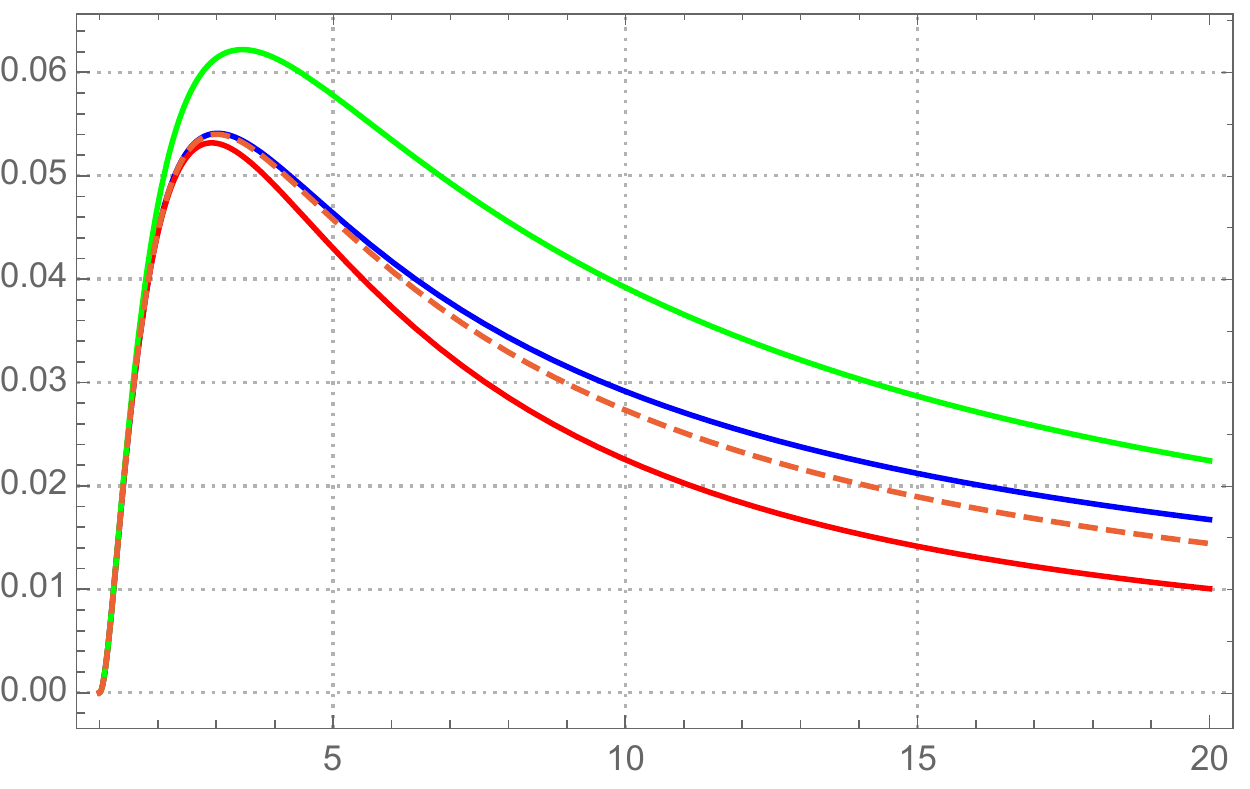}
\caption{(Color online.) Left: the moment $\L_1$ calculated  in the truncations with three (blue) and four (green) moments, compared to the exact result (dashed line). Right: the moment $\L_2$ calculated  in the truncations with three (green), four (red) and five (blue) moments, compared to the exact result (dashed line)
\label{fig:truncation4l1}
}
\end{center}
\end{figure}

For the truncation involving the first three moments $\L_0,\L_1,\L_2$, we  get 
\beq\label{truncation3}
&&\L_0(t)=0.96\, \rme^{-1.06 t} + 0.04\, \rme ^{-1.92 t}
   \cos[0.46 t] - 0.51  \rme^{-1.92 t} \sin[0.46  t],\nn
&&\L_1(t)=-0.39\, \rme^{-1.06 t} + 0.39\, \rme^{-1.92 t}
   \cos[0.46 t] - 0.42\, \rme^{-1.92 t} \sin[0.46  t],\nn
&&\L_2(t)= 0.63\, \rme^{-1.06 t} - 
 0.63 \, \rme^{-1.92 t}
   \cos[0.46  t] - 1.18\rme^{-1.92 t} \sin[0.46  t].
    \eeq   
Keeping the first four moments, one gets    
    \beq\label{truncation4}
&&\L_0(t)=    0.31\, \rme^{-2.15 t} + 0.71\,  \rme^{-0.97 t} - 
 0.02\, \rme^{-1.77 t} \cos[1.36 t] - 
 0.013\,\rme^{-1.77185 t} \sin[1.36  t],\nn
 &&\L_1(t)=  0.38\, \rme^{-2.15 t} - 0.39\, \rme^{-0.97 t} + 0.01\, \rme^{-1.77 t}
   \cos[1.36 t] - 0.06 \rme^{-1.77 t} \sin[1.36 t] ,\nn
&&\L_2(t)=    0.10 \,\rme^{-2.15 t} + 
 0.16\, \rme^{-0.97 t} - 0.26\,\rme^{-1.77 t} \cos[1.36  t] - 
 0.07\rme^{-1.77 t} \sin[1.36  t],\nn
&&\L_3(t)= 0.43 \,\rme^{-2.15 t} - 0.35\, \rme^{-0.97 t} - 
 0.08\,\rme^{-1.77 t} \cos[1.36 t] + 
 0.32 \,\rme^{-1.77 t} \sin[1.36  t].
 \eeq
 To illustrate the convergence of the truncation scheme, we compare  in Fig.~\ref{fig:truncation4l1} the moment $\L_1$ obtained from Eqs.~(\ref{truncation3}) and (\ref{truncation4}) with the exact result. The various curves are nearly indistinguishable. On the right panel, a similar comparison is made for the moment $\L_2$. Not surprinsingly in this case the convergence is slower, and it would be even slower if higher moments were considered.

\begin{figure}[h]
\begin{center}
\includegraphics[width=0.55\textwidth] {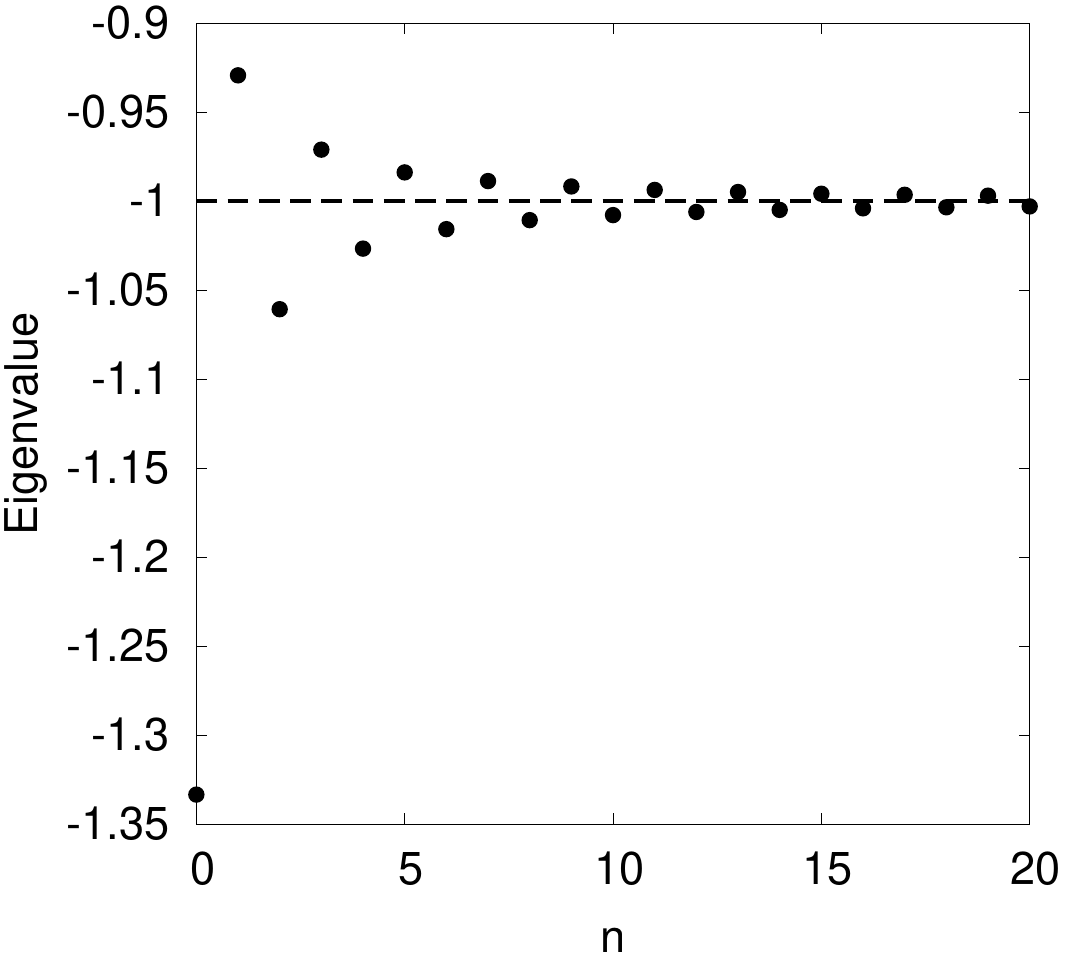}
\caption{The convergence of the lowest eigenvalue toward -1.
\label{fig:attractorg0}
}
\end{center}
\end{figure}

 Recall that in Eqs.~(\ref{truncation3}) and (\ref{truncation4}), the coefficients of $t$ in the exponents are the eigenvalues of the linear problem associated with the considered truncation. A pattern emerges as we continue exploring higher truncations, which is already visible on the expressions above. We note in particular the presence of two eigenvalues that are close to -1 and -2, respectively. These eigenvalues are already present in the two dimensional problem discussed in the main text (see Eqs.~(\ref{eigenmodes2})), and they are associated with the two fixed points of the free streaming, as discussed in Sec.~\ref{FSfpt}. The figure \ref{fig:attractorg0} shows how the eigenvalue corresponding to the stable fixed point converges to -1 as the order of the truncation increases. A similar convergence pattern is observed for the other eigenvalues.      

Starting with the three-moment truncation, one notes the appearance of a complex eigenvalue, 
whose real part is close to -1.75. As one considers higher truncations, 
such complex eigenvalues remain present and, as the order $n$ of the truncation increases,  their real part converges indeed to 7/4, while the imaginary part grows with $n$. Such a behavior can be easily understood from the following rough argument.  Let us just focus on the three modes, $\L_{n-1},\L_n,\L_{n+1}$ and ignore their coupling with the other modes. The matrix $M$ to be diagonalized is then
\beq
M=\left(
\begin{array}{ccc}
 a_{n-1} &c_{n-1} & 0 \\
 b_n & a_n & c_n \\
 0 & b_{n+1}& a_{n+1}\\
\end{array}
\right).
\eeq
Using the asymptotic values $a_n\simeq 7/4$, $b_n\simeq -c_n\simeq n/2$, one finds the eigenvalues: $7/4$, and $7/4\pm in/\sqrt{2}$.
This qualitatively corresponds to what one finds in the direct solution of the complete eigenvalue problem. The presence of complex eigenvalues signals oscillatory behavior in the time dependence of the moments. However, the oscillations are quickly damped and not visible in any of the plots shown in this paper.


\section{Gradient expansion from Chapman-Enskog expansion}\label{Chapman}

Our goal in this section is to show that the expansion (\ref{gradientL10}) of the moments $\L_n$, with $n\ge 1$ follow from the Chapman-Enskog expansion.\footnote{This expansion has been considered in greater generality in \cite{Jaiswal:2013vta}.}
 To derive this expansion, it is convenient to write the kinetic equation in  its covariant form
\beq\label{kincov}
p^\mu\del_\mu f=-\frac{u\cdot p}{\tau_R} (f-f_{\rm eq}),
\eeq
where $f_{\rm eq}$ is the local equilibrium distribution function, which is a function of $u\cdot p/T$. The temperature is determined by the Landau matching condition, Eq.~(\ref{Landaumatching}). 
Assuming that the ``true'' distribution  $f$ differs only slightly from the local equilibrium one, $f_{\rm eq}$, we write
\beq
f=f_{\rm eq}+\delta f, \qquad  \delta f=\delta f^{(1)}+\delta f^{(2)}+\cdots,
\eeq
where $\delta f^{(1)}$ is first order in gradient, $\delta f^{(2)}$ is second order, etc. We may then rewrite Eq.~(\ref{kincov}) as follows
\beq\label{CEeqn}
f=f_{\rm eq}-\frac{\tau_R}{u\cdot p} \,p^\mu\del_\mu f,
\eeq
and solve this equation  iteratively. The leading order is obtained by substituting $f_{\rm eq}$ for $f$ in the right hand side of Eq.~(\ref{CEeqn}). One gets 
\beq
f^{(1)}=f_{\rm eq}-\frac{\tau_R}{u\cdot p} \,p^\mu\del_\mu f_{\rm eq}=f_{\rm eq}+\delta f^{(1)}.
\eeq
By substituting $f^{(1)}$ in the right hand side of Eq.~(\ref{CEeqn}), one obtains the second order correction
\beq
\delta f^{(2)}=\left(\frac{\tau_R}{u\cdot p}\right) p^\mu\del_\mu\left(\frac{\tau_R}{u\cdot p}p^\nu\del_\nu \left( \frac{u\cdot p}{T}\right) f'_{\rm eq}\right),
\eeq
and repeating recursively the procedure, one obtains at order $n$ 
\beq
\delta f^{(n)}=(-\tau_R)^n  \left(\frac{1}{u\cdot p}\right)\left[p\cdot\del \frac{1}{p.u}  \right]^{n-1} p\cdot\del f_{\rm eq}.
\eeq

For the boost invariant system, the velocity field is entirely determined by the symmetry, $u^\mu=(t/\tau, 0,0,z/\tau)$. The correction $\delta f^{(n)}$ emerges then explicitly as a term of  order $(\tau_R/\tau)^n$. The angular dependence is also  made explicit in terms of Legendre polynomials. Thus, the leading order correction reads 
\beq
\delta f^{(1)}&=&-\frac{\tau_R}{u\cdot p}\,p^\mu\del_\mu\left( \frac{u\cdot p}{T}\right)f'_{\rm eq}\nn
&=&\frac{\tau_R}{\tau} \left( \cos^2\!\theta+\frac{\rmd \ln T}{\rmd \ln\tau} \right)\frac{p_0}{T}f'_{\rm eq}\nn
&=&  \frac{\tau_R}{\tau} \, \frac{2}{3} P_2(\cos\theta) \bar f'_{\rm eq},
\eeq
where the prime denotes the derivative with respect to $u\cdot p/T$, and in the last line we have defined $\bar f'_{\rm eq}\equiv (p/T)f'_{\rm eq}$. To obtain the last line, we have used the leading order relation ${\rmd \ln T}/{\rmd \ln \tau}=-1/3$. Note that this particular value of ${\rmd \ln T}/{\rmd \ln \tau}$ has the effect of canceling a potential scalar contribution to $\delta f^{(1)}$, leaving a contribution proportional to $P_2(\cos\theta)$. This cancelation is a consequence of the Landau matching condition which forces the temperature to drop as $\tau^{-1/3}$ in leading order. Note that this leading order correction is independent of whether $\tau_R$ is chosen constant, or time dependent (as in the conformal setting for instance).

In second order, we have, with $\bar f''_{\rm eq}\equiv (p/T)^2f''_{\rm eq}$ and constant $\tau_R$, 
\beq\label{deltaf2cste}
\delta f^{(2)}=\frac{\tau_R^2}{\tau^2}\left\{\frac{8 }{35}\left(\bar f''_{\rm eq}-\bar f'_{\rm eq} \right) P_4 +\frac{1}{63}\left( 62\bar f'_{\rm eq}+8\bar f''_{\rm eq} \right)P_2+\frac{4 }{45}\left(\bar f''_{\rm eq}+4 \bar f'_{\rm eq} \right) P_0\right\}.  
\eeq
To obtain this result, we have used  ${\del \ln T}/{\del\ln\tau}=-\frac{1}{3}$, and ignored the second derivative of $T$ (of higher order in $1/\tau$). 
The last term, proportional to $P_0$ ($P_0=1$), suggests a correction to the energy density, given by 
\beq
\delta_2\L_0&=&\int_\p \delta f^{(2)}(\p) p P_0\nn
&=&\frac{4}{45}\frac{\tau_R^2}{\tau^2}\int_\p\left(\frac{p^3}{T^2}f''_{\rm eq}+4 \frac{p^2}{T}f'_{\rm eq} \right)\nn
&=&\frac{16}{45}\frac{\tau_R^2}{\tau^2}\L_0,
\eeq
where we have used the following integrals
\beq\label{integralsfeq}
\int_\p p^2 \frac{\del f_{\rm eq}}{\del p}=-4\L_0,\qquad \int_\p p^3 \frac{\del^2 f_{\rm eq}}{\del p^2}=20\L_0.
\eeq
 However, the Landau matching condition forbids such a correction. There is in fact an additional contribution at this order, coming from the correction to $\delta f^{(1)}(\p)$. We have indeed
\beq
\delta f^{(1)}&=&\frac{\tau_R}{\tau} p_0\left( \frac{2}{3}P_2(\cos\theta)+\frac{1}{3}+\frac{g_0}{4} \right)\frac{1}{T}f'_{\rm eq}\nn
&\simeq& \frac{\tau_R}{\tau} p_0\left( \frac{2}{3}P_2(\cos\theta)+\frac{4}{45}\frac{\tau_R}{\tau} \right)\frac{1}{T}f'_{\rm eq}.
\eeq
where we have used (see Eqs.(\ref{dlnTdlntau})) 
\beq
\frac{\rmd \ln T}{\rmd \ln \tau}=\frac{g_0}{4}=-\frac{1}{3}+\frac{4}{45 w}+\cdots
\eeq
and the gradient expansion of $g_0$.
It can be verified that this is just what is needed to cancel $\delta_2\L_0$.  

We can repeat the calculation for the conformal case. We get, in place of (\ref{deltaf2cste}),     
\beq\label{deltaf2conf}
\delta f^{(2)}=\frac{\tau_R^2}{\tau^2}\left\{  \frac{8}{35} (\bar f''-\bar f')P_4+\left(\frac{8}{63}\bar f''+\frac{16}{21} \bar f'   \right) P_2+\frac{4}{45}\left( \bar f''+4 \bar f'  \right) P_0 \right\}.
\eeq
Clearly, the coefficients of $P_4$ and $P_0$ are the same as in the constant $\tau_R$ case. We then calculate 
\beq
\delta_2\L_1&=&\int_\p \delta f^{(2)}(\p) p P_2(\cos\theta)\nn
&=& \frac{\tau_R^2}{\tau^2} \int_\p p P_2^2(\cos\theta) \left(\frac{8}{63}\bar f''+\frac{16}{21} \bar f'   \right)\nn
&=& \frac{\tau_R^2}{\tau^2} \frac{1}{5}\left(\frac{8}{63} 20\L_0-\frac{16}{21} 4\L_0\right)\nn
&=&-\frac{32}{315} \frac{\tau_R^2}{\tau^2} \L_0 ,
\eeq
in agreement with the results obtained in Sec.~\ref{sec:asymptkinetic}. 

Finally let us remark that it is relatively easy to isolate the coefficient of the polynomial $P_{2n}$ in $f^{(n)}$,  as this is the term that involves the largest number of derivatives of $u.p$. For instance, keeping only the terms with the largest number of derivatives of $p\cdot u$, one gets
\beq
\delta f^{(3)}\leftarrow  -3\frac{\tau_R^3}{p^5}\left[p\cdot\del u\cdot p\right]^3 f'_{\rm eq}+3\frac{\tau_R^3}{p^4}\left[p\cdot\del u\cdot p\right]^3 f''_{\rm eq}-\frac{-\tau_R^3}{p^3}\left[p\cdot\del u\cdot p\right]^3 f'''_{\rm eq}.
\eeq
A simple calculation then yields  
\beq
\delta \L_3 =-\frac{\tau_R^3}{\tau^3}\frac{16}{3003}[-12-60-120]=-\frac{\tau_R^3}{\tau^3}\frac{1024}{1001}=-\frac{\tau_R^3}{\tau^3}\,b_1b_2b_3.
\eeq
This is the expected result. 

\section{Gradient expansions in the two-moment truncation}\label{gradientg0}

In this appendix we collect a few results on the gradient expansions of various quantities, within the two-moment truncation. These are most conveniently obtained from the differential equation obeyed by $g_0(w)$,  namely     
\beq\label{eqforbg02}
w\frac{\rmd g_0}{\rmd w}+g_0^2+\left(a_0+a_1+w\right)g_0+a_1a_0-c_0b_1+a_0 w =0
\eeq
for constant relaxation time, and, for the conformal case,
 \beq\label{eqfor g0}
w\left(1+\frac{g_0}{4}\right)\frac{\rmd g_0}{\rmd w} + g_0^2 +(a_0+a_1+w) g_0 +wa_0+a_0a_1-b_1c_0 =0.\label{eqfor g0}
\eeq
where we have used the fact that $\L_0(\tau)=\epsilon(\tau)\propto T^4(\tau)$ so that
\beq\label{dlnTdlntau}
\frac{\rmd \ln T}{\rmd \ln\tau}=\frac{g_0}{4}.
\eeq

\subsection{Gradient expansion of $g_0(w)$}

To derive the gradient expansion, we look for a solution of the form
\beq   
g_0(w)=\sum_{n=0}\frac{\gamma_n}{w^n}.    
\eeq
   By plugging this expansion into either one of the two equations above, we can determine the coefficients $\gamma_n$ for the corresponding choice of $\tau_R$. 
   We obtain then, for the case $\tau_R T={\rm Cste}$,    
\beq\label{gradientcoefconformal}
 &&  \gamma_0=-a_0=-\frac{4}{3},\quad \gamma_1=b_1 c_0=\frac{16}{45},\quad \gamma_2=\frac{b_1 c_0}{4} \left(3 a_0-4 a_1+4\right)=\frac{64}{945},\nn
 &&\gamma_3=-\frac{b_1 c_0}{8}  \left(-3 a_0^2+2 \left(5 a_1-8\right) a_0-8 a_1^2+24 a_1+6 b_1 c_0-16\right)=-\frac{1216}{33075}.
   \eeq
   For the case $\tau_R={\rm Cste}$,  we get 
\beq\label{gradientcoeffconstanttau}
 &&  \gamma_0=-a_0=-\frac{4}{3},\quad \gamma_1=b_1 c_0=\frac{16}{45},\quad \gamma_2=b_1 c_0\left(1+a_0-a_1\right)=\frac{176}{945},\nn
 &&\gamma_3=b_1 c_0 \left(a_0^2+\left(3-2 a_1\right) a_0+a_1^2-3 a_1-b_1 c_0+2\right)=-\frac{15616}{99225}.
   \eeq

\subsubsection{Gradient expansion of $g_1(w)$}

The expansion of $g_1[w]$ can be obtained from that of $g_0[w]$ by using the relation (\ref{relg0g1}) between $g_0$ and $g_1$. Keeping terms up to order $w^{-2}$, one gets
\beq\label{gradg1}
g_1[w]=-a_1-\frac{\gamma_2\, b_1 c_0}{\gamma_1^2}-\frac{\left(\gamma_1 \gamma_3-\gamma_2^2\right) b_1 c_0}{\gamma_1^3 \, w}-\frac{\left(\gamma_2^3-2 \gamma_1 \gamma_3 \gamma_2+\gamma_1^2 \gamma_4\right) b_1 c_0}{\gamma_1^4 \, w^2}
   \eeq    
 With the values of the coefficients $\gamma_i$ given above, one easily obtains, for the sum of the two constant terms, $
 g_1(\infty)= -1-\frac{3}{4}a_0$, and  $g_1(\infty)= -1-a_0$,
  for the cases $\tau_R T={\rm Cste}$ and $\tau_R={\rm Cste}$, respectievely. These values agree with the general result (\ref{eq:gninf}).\\

\subsection{Gradient expansion of $\L_0(w)$}

The expansion of the moment $\L_0$ can be easily obtained by integrating the expansion for $g_0$.  We do that first for the case of a constant $\tau_R$. Then we have
\beq
\frac{\rmd \ln\L_0(w)}{\rmd  \ln w}=g_0=\gamma _0+\frac{\gamma_1}{w}+\frac{\gamma_2}{w^2}+\frac{\gamma_3}{w^3}+\cdots
\eeq
which can be easily integrated to give (to order $w^{-3}$)
\beq
\L_0(w)\simeq w^{\gamma_0} \left(1-\frac{\gamma _1}{w}+\frac{\gamma _1^2-\gamma _2}{2
   w^2}+\frac{-\gamma _1^3+3 \gamma _2 \gamma _1-2 \gamma _3}{6
   w^3}\right)
   \eeq
    with (using the values of the coefficients $\gamma_i$ appropriate for constant $\tau_R$)
   \beq
   \gamma_0=-a_0,\quad \gamma_1=b_1 c_0,\quad \gamma_1^2-\gamma_2=(b_1c_0)^2-b_1c_0(1+a_0-a_1).
   \eeq
   Thus, up to order $w^{-2}$, the expansion of $\L_0(w)$ reads
  \beq
  \L_0(w)\simeq w^{\gamma _0} \left(1-\frac{b_1c_0}{w}+\frac{(b_1c_0)^2- b_1c_0(1+a_0-a_1)}{2
   w^2}\right),
   \eeq
  which agrees with the expression (\ref{L0B})) obtained using a different method.\\    
   
 In the conformal case, we use
 \beq  
\frac{\rmd\ln \L_0(w)}{\rmd  \ln w}=\frac{g_0}{1+g_0/4}
\eeq
together with the gradient expansion of $g_0(w)$. We get
\beq\label{gradL0w}
\L_0(w)\simeq w^{\frac{4 \gamma _0}{\gamma _0+4}}\left( 1-\frac{16 \gamma _1}{\left(\gamma _0+4\right){}^2
   w} + \frac{\frac{256 \gamma _1^2}{\left(\gamma _0+4\right){}^4}-\frac{16 \left(\left(\gamma _0+4\right) \gamma
   _2-\gamma _1^2\right)}{\left(\gamma _0+4\right){}^3}}{2 w^2}\right)
\eeq
Using the values of the coefficients $\gamma_i$ appropriate for the conformal case, we obtain
\beq
\frac{4 \gamma _0}{\gamma _0+4}=-2.
\eeq
Note that this corresponds indeed to the expected behavior of ideal hydrodynamics since $w^{_2}\sim \tau^{-2} (\tau_R T)^2 T^{-2}\sim \tau^{-4/3}$. 
  
\subsection{Gradient expansion of $\L_1(w)/\L_0(w)$}

 The ratio $\L_1/\L_0$ can be obtained 
directly in terms of $g_0$ by writing the first equation (\ref{eq:l0l1}) as follows   
   \beq
   \frac{\L_1(w)}{\L_0(w)}&=& -\frac{1}{c_0}\left( a_0+g_0 \right)\nn
   &=&-\frac{1}{c_0}\left(\frac{\gamma_1}{w}+\frac{\gamma_2}{w^2} +\frac{\gamma_3}{w^3}  \right)
   \eeq
which hold for any choice of the relaxation time. For the constant relaxation time we get
\beq
\frac{\L_1(w)}{\L_0(w)}=-\frac{b_1}{w}-\frac{b_1(1+a_0-a_1)}{w^2},
\eeq
 and for the conformal case
\beq
\frac{\L_1(w)}{\L_0(w)}=-\frac{b_1}{w}-\frac{b_1}{4w^2}(3a_0-4a_1+4).      
\eeq
These results are consistent with Eq.~(\ref{B1B2}).

\section{Stability analysis near the hydrodynamical fixed point}\label{stabilityanalysis}

To proceed with the stability analysis, we generalize the temperature dependence of the relaxation time $\tR$, and write, as in \cite{Heller:2018qvh},  
\beq
\tR\propto (1/T)^{\Delta}\,,
\eeq
via a constant
$\Delta$. The conformal case corresponds to $\Delta=1$, while a constant $\tau_R$ is obtained for $\Delta =0$.
This effectively changes $\rmd w/\rmd\tau$ into
\beq
\frac{dw}{d\tau} = 1+\frac{\Delta}{4} g_0,
\eeq
so that the equations for $g_0$ becomes
\beq\label{eqforg0conf}
w\left(1+\frac{\Delta }{4}g_0\right)g_0' + g_0^2 & = -a_0g_0 +b_1c_0 -(a_1+w)(a_0+g_0),
\eeq

\begin{figure}
\begin{center}
\includegraphics[width=0.75\textwidth] {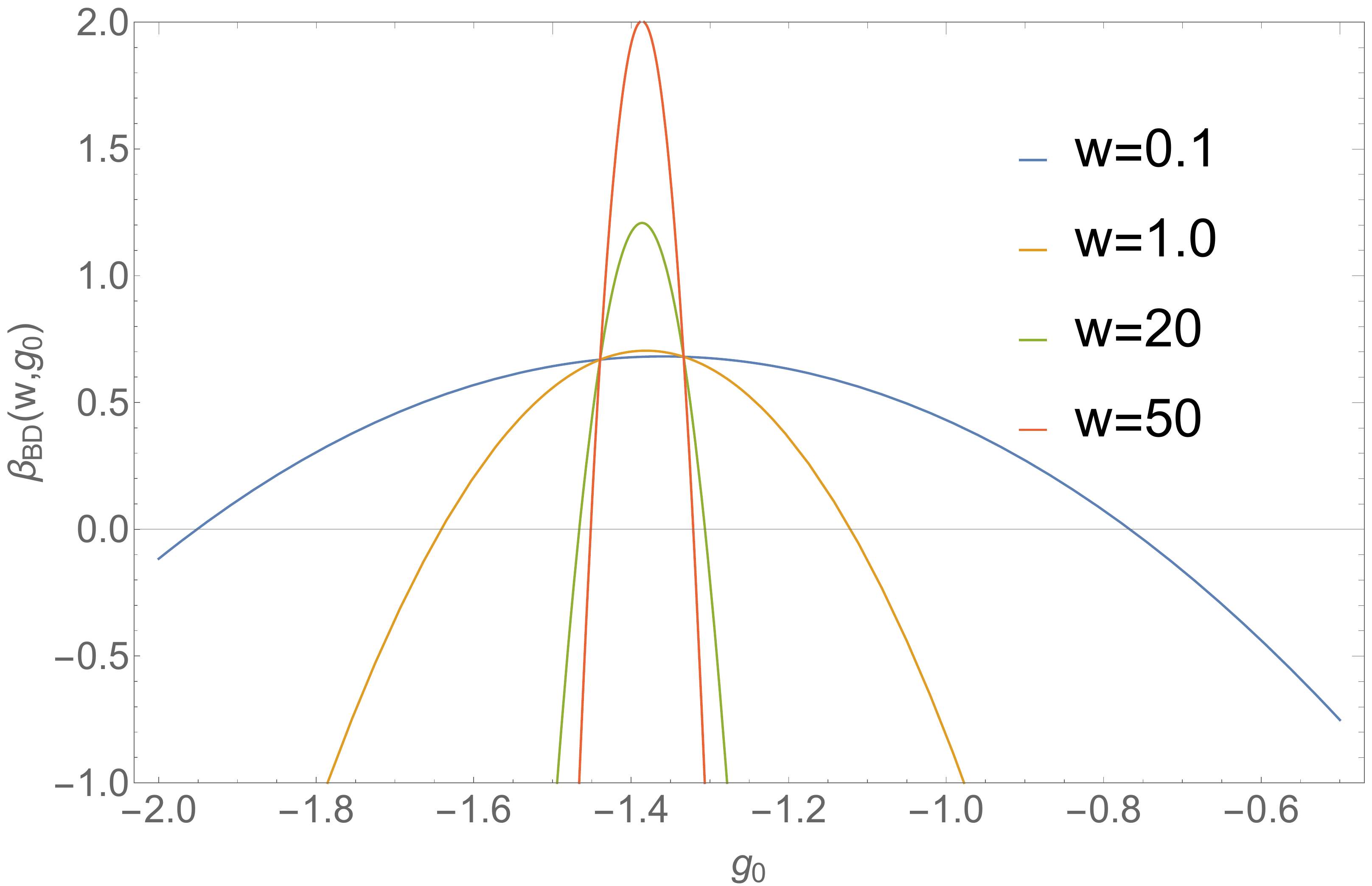}
\caption{(Color online.) The function $\beta_{BD}(w,g_0)$ (Eq.~(\ref{betag0BD})).  The curves corresponding to various values of $w$ cross at $g_0=-a_0$ corresponding to a stable pseudo fixed point and hydrodynamics, and at $g_0=-a_0-(8/3) C_\eta/C_{\lambda_1}$ which corresponds to an instable pseudo fixed point.  \label{fig:fixedptsBD}
}
\end{center}
\end{figure}

Our goal now is to linearize this equation near the hydrodynamical fixed point. To do so, we rewrite the equation above as follows 
\begin{align}
\label{eq:Geq}
G'\left( 1+\frac{\Delta}{4}G  \right)=-S G+\frac{\beta}{w} G-\frac{G^2}{w}+\frac{C}{w} S^2,
\end{align}
with   
$    
G=S(g_0+a_0)
$   
and 
\beq
S=\left(1-\frac{\Delta a_0}{4}\right)^{-1},\quad
\beta = -(a_1-a_0)\left(1-\frac{\Delta a_0}{4}\right)^{-1},\quad C=b_1c_0.    
\eeq
Note that at large $w$, $g_0+a_0\simeq b_1c_0/w$, independently of the value of $\Delta$. We then set $G=\bar G+\delta G$, with
\beq
\bar G=\frac{CS}{w}+O(1/w^2),
\eeq
 and substitute this in Eq.~(\ref{eq:Geq}). We get, after dropping the terms that are either quadratic in the fluctuation, or of order $1/w^2$, 
 \beq
 \delta G'\left(1+\frac{\Delta}{4}\frac{CS}{w}\right)=-S\delta G +\frac{\beta}{w}\delta G.
 \eeq
The solution of this equation at large $w$ behaves as   
\beq
\delta G(w)
\propto  \rme^{-Sw} w^{\beta+CS^2/4}.
\eeq
 This is compatible with the expression (41) in Ref.~\cite{Heller:2018qvh}, except for the value of  $\beta$. Note, however, that this result pertains to the two-moment truncation. Had we started from the BRSSS hydrodynamic equation (\ref{eq:l0l1_brsss}), and used the standard substitutions $\eta/s=C_\eta$, $\lambda_1=C_{\lambda_1}\eta/T$, $C_R=\tau_R T$, one would have obtained the following equation (for the case $\Delta=1$)
 \beq\label{eqforg0conf2}
 w g_0'\left(1+\frac{g_0}{4}\right)+\left(g_0+a_0\right)^2\left[1+\frac{3w}{8}\frac{C_{\lambda_1}}{C_\eta}\right] +w(g_0+a_0)-\frac{16}{9}\frac{C_\eta}{C_R}=0. 
\eeq   
  It is easily verified that the same stability analysis as that presented above yields a value of $\beta$ that agrees with that quoted in Ref.~\cite{Basar:2015ava}.   Note that, by construction, the two equations (\ref{eqforg0conf}) and (\ref{eqforg0conf2}) yield the same hydrodynamic behaviors in leading orders in the expansion in $1/w$. However, the pseudo fixed point structures of the two equations are different. For the sake of comparison, we show in Fig.~\ref{fig:fixedptsBD} the function 
    \beq\label{betag0BD}
  \beta_{BD}(g_0)=-\left(g_0+a_0\right)^2\left[1+\frac{3w}{8}\frac{C_{\lambda_1}}{C_\eta}\right] -w(g_0+a_0)+\frac{16}{9}\frac{C_\eta}{C_R},    
  \eeq
   corresponding to Eq.~(\ref{eqforg0conf2}). This is to be compared to Fig.~\ref{fig:fixedpoint1}. The stable pseudo fixed point corresponding to hydrodynamics behave in the same way in both cases, but   Eq.~(\ref{eqforg0conf2}) admits a second, unstable, pseudo fixed point located at a finite value of $g_0$, $g_0=-a_0-(8/3) C_\eta/C_{\lambda_1}$.

\section{Insights from the exact solution}

Our goal in this appendix is illustrate the very different natures of the expansions which are valid for small and large values of $r_0$, the ratio  of the collision to the expansion rates. We do this by starting directly from the exact equation for the energy density, or equivalently from the general equation for the moments, Eq.~(\ref{eq:rta_0a}), which we rewrite here for convenience:
\be
\label{eq:rta_0a1b}
\L_n(\tau)=D(\tau,\tau_0) \L_n^{(0)}(\tau) + \int_{\tau_0}^\tau 
\frac{\rmd\tau'}{\tR(\tau')} D(\tau,\tau') \L_0(\tau')(\tau'/\tau) \F_n(\tau'/\tau)\,,
\ee
Recall that this equation is valid for arbitrary initial conditions, isotropic or not,  these initial conditions being entirely coded in the free streaming moments  $\L_n^{(0)}(\tau)$  given by Eq.~(\ref{FSLnanisotrop}). In the course of this study, we shall also recover results that we have obtained by other means, although this is not our main goal here. 

\subsection{Expansion at early time}     

We consider first the small time behavior and calculate the time derivative of the first two moments.  By taking the derivative of Eq.~(\ref{eq:rta_0a1b}) with respect to $\tau$, we get
\beq
\tau_0\frac{\rmd \L_n}{\rmd \tau}=-\frac{\tau_0}{\tau_R(\tau_0)}\L_n^{(0)}(\tau_0)+\tau_0\left.\frac{\rmd \L_n^{(0)}}{\rmd\tau}\right|_{\tau_0}+ \frac{\tau_0}{\tau_R(\tau_0)}\L_0(\tau_0)\F_n(1).
\eeq      
Recall that for $n\ge 1$, $\F_n(1)=0$, while for $\F_0(1)=1$. Also, for $n=0$, $\L_0(\tau_0)=\L_0^{(0)}(\tau_0)=\varepsilon_0$. We get then
\beq\label{initialderivatives}
  &&\tau_0\frac{\rmd \L_0}{\rmd \tau}  =  \tau_0\left.\frac{\rmd \L_0^{(0)}}{\rmd\tau}\right|_{\tau_0}\nn
  && \tau_0\frac{\rmd \L_1}{\rmd \tau} =-\frac{\tau_0}{\tau_R(\tau_0)}\L_1^{(0)}(\tau_0)+\tau_0\left.\frac{\rmd \L_1^{(0)}}{\rmd\tau}\right|_{\tau_0}.
  \eeq   
  The first equation shows that the energy density decreases initially with time in the same way as in free streaming. The derivative of  $\L_1$ on the other hand depends explicitly on the relaxation time $\tau_R$, so that the moment $\L_1$ is immediately sensitive to the collisions. 
  To calculate $\tau_0{\rmd \L_0^{(0)}}/{\rmd \tau}$ we may use Eq.~(\ref{FSLnanisotrop}) and the relations given in Appendix~\ref{App:functionFn}. We then easily reproduce the results given in Sec.~\ref{equadiffL0L1}, Eqs.~(\ref{FSb2b}) and (\ref{dotL1}).

We could in principle continue and expand for $\tau\ll \tau_0$ by taking further derivatives, but we are in fact interested in the regime $\tau_0\ll\tau_R$, and we want an expansion valid for $\tau\lesssim \tau_R$, i.e. not limited to very small $\tau\ll\tau_0$. In line with the time-dependent perturbation theory that we have used in Sec.~\ref{sec:perturbation}, we look for an expansion that treats the effect of the collisions as a correction to free streaming. In more precise terms, we look for an expansion in powers of $r_0\equiv \tau_0/\tau_R$, the ratio of the collision rate to the expansion rate. 

\begin{figure}
\begin{center}
\includegraphics[width=0.75\textwidth] {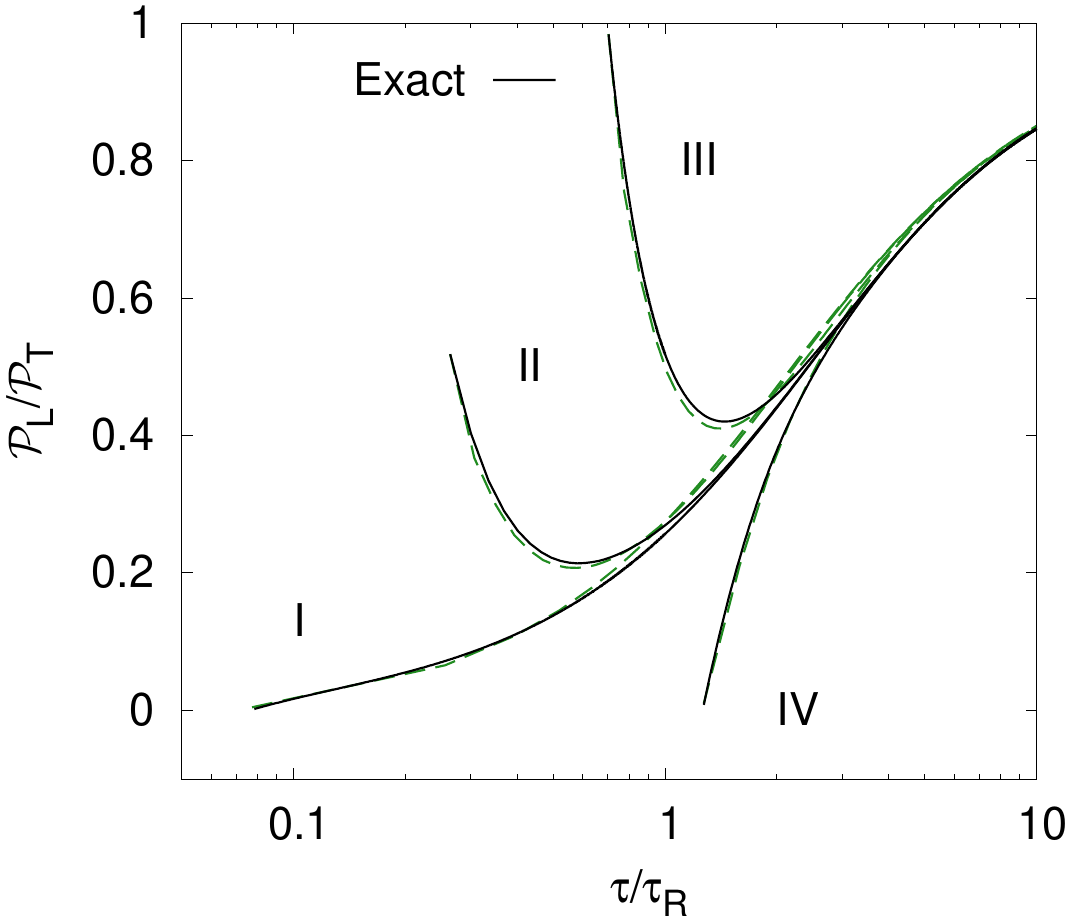}
\caption{(Color online.) The ratio $\P_L/\P_T$ as a function of $\tau/\tau_R$ for various initial conditions. The solid (black) lines represent the exact solution of the kinetic equation and the dashed (green) lines are the small $r_0$ expansion in leading order. \label{fig:exactsmallr0}
}
\end{center}
\end{figure}

Let us then rewrite here \Eq{eq:rta_0a1b} for the moment $\L_n$, for a constant $\tau_R$, and after performing the change of variables
$\tau=z\tau_0$:
\beq
\label{eq:rta_0a2c}
\L_n(\tau)=\rme^{-r_0(z-1)}  \L_n^{(0)}(\tau) + r_0 \,\int_{1}^z 
\rmd z' \rme^{-r_0(z-z')} \L_0(\tau')\left( \frac{z'}{z} \right) \F_n\left( \frac{z'}{z} \right).
\eeq
 We note that, in leading order in the expansion parameter $r_0$,     $\L_n(\tau)$ is given by the first term in Eq.~(\ref{eq:rta_0a2c}), which is the free streaming solution multiplied by the exponential factor   $\rme^{-r_0(z-1)}$. (This factor, for $\tau\lesssim \tau_R$ and $\tau_0\ll \tau_R$ is nearly equal to unity.) We then insert this whole fist term into the integral, and repeat iteratively the operation. We shall be satisfied here with the leading order. We get then           
\beq
\label{eq:rta_0a2d}
\L_n(\tau)=\rme^{-r_0(z-1)}  \L_n^{(0)}(\tau) + r_0 \rme^{-r_0(z-1)}\,\int_{1}^z 
\rmd z' \,\L_0^{(0)}(z')\left( \frac{z'}{z} \right) \F_n\left( \frac{z'}{z} \right),
\eeq
where we have used the fact that the free streaming solution is indeed a function of $\tau'/\tau_0=z'$. This solution captures the leading order of time-dependent perturbation theory, and it  can be easily checked that the slope at the origin is correctly reproduced, for both $\L_0$ and $\L_1$. The present expansion can be seen as a generalization of the perturbative approach of Sec.~\ref{sec:perturbation}, which is not limited here to the two-moment truncation, but includes implicitly the effects of all the moments (the perturbative approach of course assumes that $r_0(z-1)\ll 1$ so that the exponential factors are approximately equal to unity).   We show in Fig.~\ref{fig:exactsmallr0} the ratio $\P_L/\P_T$ obtained in this approximation (keeping the exponential factors in Eq.~(\ref{eq:rta_0a2d})). The small time behavior is clearly well reproduced. In fact, even the late time behavior is  well captured for all initial conditions. Indeed Eq.~(\ref{eq:rta_0a2d}) can be used as the starting point for a numerical iterative solution of the full kinetic equation \cite{Florkowski:2013lza}.

\subsection{Expansion in powers of $\tau_R$ at late times}     

In the small $r_0$ regime, as we have seen, the contribution of the integral in Eq.~(\ref{eq:rta_0a2c}) is subleading since it is proportional to $r_0$. In the hydrodynamic regime, when $\tau_R\to 0$, the integral plays an essential role and it is  the first term in Eq.~(\ref{eq:rta_0a1b}) which can be then neglected. Because of the exponential factor, the integrand is localized in the region $\tau'\lesssim \tau$. Set $\tau'=\tau-\tau''$. Then $\int_{\tau_0}^\tau \rmd\tau'\to \int^{\tau-\tau_0}_0\rmd\tau''$ and we get, for the case of a constant $\tau_R$,    
\beq\label{L0latetime}
\L_0(\tau)\simeq\int^{\tau-\tau_0}_0\frac{\rmd\tau''}{\tau_R}\rme^{-\tau''/\tau_R}\L_0(\tau-\tau'')((\tau-\tau'')/\tau) \F_0((\tau-\tau'')/\tau).
\eeq    
We can then extend the upper bound of the integration to infinity, and  expand the integrand for small $\tau''/\tau$ ($\tau''\lesssim\tau_R$, $\tau\gg \tau_R$). We get 
\beq
\L_0(\tau)\simeq\L_0(\tau) \left(1 -\frac{4}{3} \frac{\tau_R}{\tau}\right)-\tau_R\del_\tau\L_0\left(1 - \frac{8}{3}\frac{\tau_R}{\tau}  \right),
\eeq
that is, we recover the leading order relation 
\beq
\frac{4}{3} \frac{\tau_R}{\tau}\L_0(\tau)=-\tau_R\del_\tau\L_0.
\eeq

We can proceed systematically by using integrations by parts, noting that 
\beq
\rme^{\tau'/\tau_R}=\tau_R \frac{\rmd {\rme^{\tau'/\tau_R}}}{\rmd \tau'}.
\eeq
Inserting this relation into Eq.~(\ref{L0latetime}), integrating by parts, and 
 ignoring exponentially small contributions, one reproduces iteratively the gradient expansion that we have obtained by other means.       

\bibliographystyle{unsrt}
\bibliography{refsbib}

\end{document}